\documentclass[useAMS,usenatbib]{mn2e}
\pdfoutput = 1
\usepackage{graphicx}
\usepackage{color}
\usepackage{times}
\usepackage{multirow}
\title[The SAURON Project - XIV]
  {The SAURON Project - XIV. No escape from V$_{\mathrm{esc}}$: a global and local parameter in early-type galaxy evolution}
\author[N. Scott et al.]
  {Nicholas Scott,$^1$\thanks{E-mail: nscott@astro.ox.ac.uk} Michele Cappellari$^1$, Roger L. Davies$^1$, R. Bacon$^2$, P. T. de Zeeuw$^{3,4}$, \newauthor Eric Emsellem$^{2,3}$, J\'esus Falc\'on-Barroso$^5$, Davor Krajnovi\'c$^1$, Harald Kuntschner$^6$, \newauthor Richard M. McDermid$^7$, Reynier F. Peletier$^8$, Antonio Pipino$^9$, Marc Sarzi$^{10}$,\newauthor Remco C. E. van den Bosch$^{4,11}$, Glenn van de Ven$^{12}$\thanks{Hubble Fellow} and Eveline van Scherpenzeel$^4$
\\$^1$Sub-Department of Astrophysics, University of Oxford, Denys Wilkinson Building, Keble Road, Oxford, OX1 3RH
\\$^2$Universit\'e de Lyon, France; Universit\'e Lyon 1, F-69007; CRAL, Observatoire de Lyon, F-69230 Saint Genis Laval; CNRS, UMR 5574;\\ ENS de Lyon, France 
\\$^3$European Southern Observatory, Karl-Schwarzschild-Str 2, 85748 Garching, Germany 
\\$^4$Sterrewacht Leiden, Leiden University, Niels Bohrweg 2, 2333 CA Leiden, the Netherlands 
\\$^5$European Space and Technology Centre (ESTEC), Keplerlaan 1, Postbus 299, 2200 AG Noordwijk, the Netherlands 
\\$^6$Space Telescope European Coordinating Facility, European Southern Observatory, Karl-Schwarzschild-Str 2, 85748 Garching, Germany 
\\$^7$Gemini Observatory, Northern Operations Centre, 670 N. A'ohoku Place, Hilo, Hawaii 96720, USA 
\\$^8$Kapteyn Astronomical Institute, Postbus 800, 9700 AV Groningen, the Netherlands 
\\$^9$Department of Physics and Astronomy, University of Southern California, Los Angeles, CA 90089-0484
\\$^{10}$Centre for Astrophysics Research, University of Hertfordshire, Hatfield, Herts AL1 09AB 
\\$^{11}$McDonald Observatory, The  University of Texas at Austin, TX 78712, Austin, USA
\\$^{12}$Institute for Advanced Study, Peyton Hall, Princeton, NJ 08544, USA}
\date{\today}
\pagerange{\pageref{firstpage}--\pageref{lastpage}} \pubyear{2009}

\def\LaTeX{L\kern-.36em\raise.3ex\hbox{a}\kern-.15em
    T\kern-.1667em\lower.7ex\hbox{E}\kern-.125emX}

\begin{document}

\label{firstpage}

\maketitle

\begin{abstract}
 We present the results of an investigation of the local escape velocity (V$_{\mathrm{esc}}$) - line strength index relationship for 48 early type galaxies from the SAURON sample, the first such study based on a large sample of galaxies with both detailed integral field observations and extensive dynamical modelling. Values of V$_{\mathrm{esc}}$ are computed using Multi Gaussian Expansion (MGE) photometric fitting and axisymmetric, anisotropic Jeans' dynamical modelling simultaneously on HST and ground-based images. We determine line strengths and escape velocities at multiple radii within each galaxy, allowing an investigation of the correlation within individual galaxies as well as amongst galaxies. We find a tight correlation between V$_{\mathrm{esc}}$ and the line-strength indices. For Mgb we find that this correlation exists not only between different galaxies but also inside individual galaxies - it is both {\it a local and global} correlation. The Mgb-V$_\mathrm{esc}$ relation has the form: $\log (\mathrm{Mgb}/4\mathrm{\AA}) = (0.32 \pm 0.03) \log (\mathrm{V}_\mathrm{esc}/500\mathrm{km/s}) - (0.031 \pm 0.007)$ with an rms scatter $\sigma=0.033$. The relation within individual galaxies has the same slope and offset as the global relation to a good level of agreement, though there is significant intrinsic scatter in the local gradients. We transform our line strength index measurements to the single stellar population (SSP) equivalent ages (t), metallicity ([Z/H]) and enhancement ([$\alpha$/Fe]) and carry out a principal component analysis of our SSP and V$_{\mathrm{esc}}$ data. We find that in this four-dimensional parameter space the galaxies in our sample are to a good approximation confined to a plane, given by $\log \mathrm({V}_{\mathrm{esc}}/500\mathrm{km/s}) = 0.85 \mathrm{[Z/H]} + 0.43 \log (\mathrm{t}/\mathrm{Gyrs})$ - 0.20. It is surprising that a combination of age and metallicity is conserved; this may indicate a `conspiracy' between age and metallicity or a weakness in the SSP models. How the connection between stellar populations and the gravitational potential, both locally and globally, is preserved as galaxies assemble hierarchically may provide an important constraint on modelling.
\end{abstract}


\section{Introduction}
In the Hubble classification \citep{Hubble} scheme elliptical and lenticular (or S0) galaxies are collectively known as early-type galaxies, and are thought to represent the end-point of many billions of years of evolution. Early-type galaxies exhibit smooth morphologies, appearing as essentially featureless collections of stars on the sky. For many years this simple appearance was thought to reflect a straightforward and homogeneous behaviour, both dynamically and in terms of their stellar populations, across a broad range in luminosity and size. More recently observations have shown that while in many ways the structure of early-type galaxies is intrinsically simple there is a rich diversity in their properties that requires a more complex understanding of these objects. Such an understanding will yield important information about the formation and evolution of structure in the Universe.

Many different properties of early-type galaxies are found to be well correlated with their luminosities. The earliest correlations discovered were those relating global quantities of these galaxies. The most luminous galaxies were found to have large half-light radii R$_e$ \citep{Kormendyc}, low surface brightnesses within R$_e$, $\langle \mathrm{I}_e \rangle$ and large central velocity dispersions $\sigma_e$ \citep{FJ}. These correlations can be combined if we plot the measurements in $\log \sigma_e$, $\log$ R$_e$, $\log \langle \mathrm{I}_e \rangle$ space. In this variable space it is found that galaxies are confined to a tight plane, known as the Fundamental Plane \citep{DD,Dressler}. In this case the value of any one of the variables can be calculated once the other two are known - early-type galaxies are a two-parameter family. The most luminous galaxies were also found to be predominantly pressure-supported \citep[low V/$\sigma$,][]{B&C,Illingworth,Binney}, have core surface brightness profiles \citep{Kormendya,Lauer,Ferraresea,Ferrareseb,Faber,Kormendyb} and boxy isophotes \citep{Bendera}. In contrast the less luminous galaxies are predominantly rotationally supported \citep{Davies} with cuspy surface brightness profiles and discy isophotes. These observations hinted at a dichotomy in the early-type population \citep{Faber,KB96} but the inclusion of two-dimensional kinematics reveals a different, more marked separation into two distinct populations \citep[][hereafter Paper IX and Paper X]{Paper IX, Paper X}. Another quantity that is found to correlate well with $\sigma_e$ is the mass of the galaxy's central black hole M$_\bullet$ \citep{FM,BHsigma}, which also correlates with many other galaxy properties including bulge mass M$_{bulge}$ \citep{Magorrian,MD02,Marconi,Haring}.

As well as relationships between these global, predominantly dynamical quantities there are tight correlations relating stellar population parameters. The first known of these was the colour-magnitude relation relating the total luminosity of a galaxy to its {\it B-V} colour \citep{VS}. Global colours were also found to be well correlated with other galaxy properties, most notably central velocity dispersion $\sigma_e$ and central absorption line strengths for a number of commonly observed absorption indices \citep{Bender}. That colour and line strength should be tightly related is not entirely surprising. The fact that a quantity measuring a global property of the galaxy (in this case the global colour) which is dominated by light from the outer parts of the galaxy should be closely related to a quantity measured only in the very centre (true for both $\sigma$ and the absorption indices) suggests that the behaviour of these properties {\it within} a galaxy, as well as between different galaxies, must also be confined to a relatively narrow region in parameter-space. We shall explore further evidence for this idea and it's consequences later in this work.

\begin{table*}
\caption{Properties of the 48 E and S0 galaxies from the SAURON sample used in this paper}
\label{table:sample}
\begin{center}
\begin{tabular*}{\textwidth}{@{\extracolsep{\fill}}l l c c c c c c c c c c c c c}
\hline
Galaxy& Type & $R_e$ & Dist &Rotator & $i$&(M/L)$_{\it X}$&Band&M$_I$& {\it X $-$ I} &(M/L)$_{\it I}$  &HST& Quality&\multirow{2}{*}{$\frac{d(\log \mathrm{Mgb})}{d(\log \mathrm{V}_\mathrm{esc})}$}\\
Name& &  (arcsec) &(Mpc) & & ($^{\circ}$)&&{\it X}&(mag)& (mag) & &Imaging&of fit&\\
(1) & (2) & (3) & (4) & (5) & (6) & (7) & (8) & (9) &(10)&(11)&(12)&(13)&(14)\\
\hline
NGC 474		& $\mathrm{S0^0(s)}$		& 29		& 32.0   	&F	&37	&2.86	&I	&-21.94 	&--		& 2.86	& 814W & 3	& 0.66\\ 
NGC 524		& $\mathrm{S0^+(rs)}$		& 51		& 23.3   	&F	&20	&5.36	&I	&-22.99	&-		& 5.36	& 814W & 1	& 0.19\\
NGC 821		& $\mathrm{E6?}$			& 39		& 23.4   	&F	&79	&3.58	&I	&-22.32	&-		& 3.58	& 814W & 1	& 0.43\\ 
NGC 1023	& $\mathrm{SB0^-(rs)}$		& 48		& 11.1   	&F	&73	&2.90	&I	&-21.97	&-		& 2.90	& 814W & 2	& 0.09\\
NGC 2549	& $\mathrm{S0^0(r)sp}$		& 20		& 12.3   	&F	&90	&4.84	&R	&-20.49	&0.65*	& 3.64	& 702W & 1	& 0.41\\
NGC 2685	& $\mathrm{(R)SB0^+pec}$	& 20		& 15.0   	&F	&76	&1.74	&I	&-20.65	&-		& 1.74	& 814W & 2	& 0.47\\
NGC 2695	& $\mathrm{SAB0^0(s)}$		& 21		& 31.5  	&F	&48	&5.64	&V	&-21.63	&1.18	& 3.80	& - & 1		& 0.30\\
NGC 2699	& $\mathrm{E:}$			& 14		& 26.2  	&F	&46	&3.21	&R	&-20.81	&0.61	& 2.48	& 702W & 2	& 0.51\\
NGC 2768	& $\mathrm{E6:}$			& 71		& 21.8  	&F	&90	&5.32	&I	&-22.73	&-		& 5.32	& 814W & 1	& 0.39\\
NGC 2974	& $\mathrm{E4}$			& 24		& 20.9  	&F	&56	&4.79	&I	&-21.94	&-		& 4.79	& 814W & 1	& 0.39\\
NGC 3032	& $\mathrm{SAB0^0(r)}$		& 17		& 21.4  	&F	&38	&1.99	&I	&-20.41	&-		& 1.99	& 814W & 1	& -1.46\\
NGC 3156	& $\mathrm{S0:}$			& 25		& 21.8  	&F	&67	&1.46	&I	&-20.43	&-		& 1.46	& 814W & 1	& -0.07\\
NGC 3377	& $\mathrm{E5-6}$			& 38		& 10.9  	&F	&90	&2.31	&I	&-21.19	&-		& 2.31	& 814W & 1	& 0.74\\
NGC 3379	& $\mathrm{E1}$			& 42		& 10.3  	&F	&68	&3.43	&I	&-22.11	&-		& 3.43	& 814W & 1	& 0.32\\
NGC 3384	& $\mathrm{SB0^-(s):}$		& 27		& 11.3  	&F	&66	&1.89	&I	&-21.45	&-		& 1.89	& 814W & 3	& 0.31\\
NGC 3414	& $\mathrm{S0 pec}$		& 33		& 24.5  	&S	&60	&4.23	&I	&-22.23	&-		& 4.23	& 814W & 2	& 0.75\\
NGC 3489	& $\mathrm{SAB0+(rs)}$		& 19		& 11.8  	&F	&60	&0.99	&I	&-20.99	&-		& 0.99	& 814W & 2	& 0.25\\
NGC 3608	& $\mathrm{E2}$			& 41		& 22.3  	&S	&60	&3.73	&I	&-22.22	&-		& 3.73	& 814W & 1	& 0.47\\
NGC 4150	& $\mathrm{S0^0(r)?}$		& 15		& 13.4  	&F	&51	&1.43	&I	&-19.91	&-		& 1.43	& 814W & 1	& -0.37\\
NGC 4262	& $\mathrm{SB0^-(s)}$		& 10		& 15.4  	&F	&26	&8.84	&B	&-20.51	&2.24	& 4.08	& ACS/475W & 3	& 0.52\\
NGC 4270	& $\mathrm{S0}$			& 18		& 33.1  	&F	&70	&4.01	&V	&-21.27	&1.06	& 3.01	& 606W & 1	& 0.02\\
NGC 4278	& $\mathrm{E1-2}$			& 32		& 15.6  	&F	&40	&4.86	&I	&-22.08	&-		& 4.86	& 814W & 1	& 0.47\\
NGC 4374	& $\mathrm{E1}$			& 71		& 18.5  	&S	&60	&4.08	&I	&-23.46	&-		& 4.08	& 814W & 1	& 0.34\\
NGC 4382	& $\mathrm{S0^+(s)pec}$		& 67		& 17.9  	&F	&78	&2.58	&I	&-23.39	&-		& 2.58	& 814W & 3	& -0.29\\
NGC 4387	& $\mathrm{E}$			& 17		& 17.9  	&F	&65	&5.23	&B	&-20.19	&2.22	& 2.46	& ACS/475W & 1	& 0.21\\
NGC 4458 	& $\mathrm{E0-1}$			& 27		& 16.4  	&S	&76	&2.43	&I	&-20.42	&-		& 2.43	& 814W & 1	& 0.44\\
NGC 4459	& $\mathrm{S0^+(r)}$		& 38		& 16.1  	&F	&46	&2.86	&I	&-21.86	&-		& 2.86	& 814W & 1	& 0.31\\
NGC 4473	& $\mathrm{E5}$			& 27		& 15.3  	&F	&73	&3.38	&I	&-21.87	&-		& 3.38	& 814W & 1	& 0.42\\
NGC 4477	& $\mathrm{SB0(s):?}$		& 47		& 16.5  	&F	&26	&6.66	&V	&-21.66	&1.28	& 4.09	& 606W & 3	& 0.08\\
NGC 4486	& $\mathrm{E0-1^+pec}$		& 105	& 17.2  	&S	&60	&4.90	&I	&-23.43	&-		& 4.90	& 814W & 1	& 0.50\\
NGC 4526	& $\mathrm{SAB0^0(s)}$		& 40		& 16.4  	&F	&78	&3.54	&I	&-22.54	&-		& 3.54	& 814W & 1	& 0.09\\
NGC 4546	& $\mathrm{SB0^-(s):}$		& 22		& 13.7  	&F	&70	&6.12	&V	&-21.17	&1.15	& 4.23	& 606W & 1	& 0.46\\
NGC 4550	& $\mathrm{SB0^0:sp}$		& 14		& 15.5  	&S	&82	&3.38	&I	&-20.50	&-		& 3.38	& 814W & 1	& 0.41\\
NGC 4552	& $\mathrm{E0-1}$			& 32		& 15.8  	&S	&60	&4.33	&I	&-22.45	&-		& 4.33	& 814W & 1	& 0.28\\
NGC 4564	& $\mathrm{E}$			& 21		& 15.8  	&F	&74	&4.29	&R	&-21.03	&0.65*	& 3.22	& 702W & 1	& 0.48\\
NGC 4570	& $\mathrm{S0 sp}$			& 14		& 17.1  	&F	&90	&3.39	&I	&-21.42	&-		& 3.39	& 814W & 1	& 0.40\\
NGC 4621	& $\mathrm{E5}$			& 46		& 14.9  	&F	&90	&3.95	&I	&-22.80	&-		& 3.95	& 814W & 1	& 0.42\\
NGC 4660	& $\mathrm{E}$			& 11		& 15.0  	&F	&68	&3.16	&I	&-20.20	&-		& 3.16	& 814W & 1	& 0.56\\
NGC 5198	& $\mathrm{E1-2:}$			& 25		& 37.4  	&S	&60	&6.48	&R	&-22.04	&0.65*	& 4.87	& 702W & 1	& 0.41\\
NGC 5308	& $\mathrm{S0^-sp}$		& 10		& 34.1  	&F	&90	&3.56	&I	&-22.13	&-		& 3.56	& 814W & 1	& 0.28\\
NGC 5813	& $\mathrm{E1-2}$			& 52		& 31.3  	&S	&60	&4.23	&I	&-23.28	&-		& 4.23	& 814W & 1	& 0.19\\
NGC 5831	& $\mathrm{E3}$			& 35		& 26.4  	&S	&60	&3.94	&R	&-21.85	&0.58	& 3.16	& 702W & 3	& 0.64\\
NGC 5838	& $\mathrm{S0^-}$			& 23		& 19.8  	&F	&70	&5.22	&I	&-21.73	&-		& 5.22	& 814W & 1	& 0.26\\
NGC 5845	& $\mathrm{E:}$			& 4.6		& 25.2  	&F	&75	&3.04	&I	&-20.72	&-		& 3.04	& 814W & 1	& 0.39\\
NGC 5846	& $\mathrm{E0-1}$			& 81		& 24.2  	&S	&60	&5.09	&I	&-23.34	&-		& 5.09	& 814W & 1	& 0.30\\
NGC 5982	& $\mathrm{E3}$			& 27		& 46.4  	&S	&60	&3.51	&I	&-23.18	&-		& 3.51	& 814W & 1	& 0.30\\
NGC 7332	& $\mathrm{S0\ pec\ sp}$		& 11		& 22.4  	&F	&83	&2.78	&V	&-21.44	&1.11	& 2.00	& WF1/555W & 1	& 0.16\\
NGC 7457	& $\mathrm{S0^-(rs)?}$		& 65		& 12.9 	&F	&64	&1.89	&I	&-20.73	&-		& 1.89	& 814W & 1	& 0.07\\
\hline
\end{tabular*}
\end{center}
\begin{minipage}{17.8cm}
Notes: Column (1): NGC number. Column (2): Morphological type from RC3. Column (3): Effective (half-light) radius $R_e$ measured in the I band (see Paper IV). Column (4): Distances  were taken from preferentially from \citet{Mei} or \citet{Tonry}. Virgo galaxies without other distance determinations were assigned the mean Virgo distance of 16.5 Mpc from \citet{Mei}. Distances for other galaxies were taken from \citet{Paturel}. Column (5): Galaxy classification from Paper IX: F = fast rotator ($\lambda_R > 0.1$), S = slow rotator ($\lambda_R \leq 0.1$). Column (6): The best-fitting inclination determined from axisymmetric Jeans Anisotropic MGE (JAM) modelling. For slow rotators an inclination of 60$^{\circ}$ is assumed (except NGCs 4458 and 4550 which have independent determinations of their inclination). Column (7):  The M/L of the best-fitting JAM model, in the given band. Column (8): Photometric band the M/L was determined in. Column (9): Total magnitude determined from the MGE models and converted to {\it I}-band using colours from the literature. Column (10): Galaxy colour, where {\it X} is the band of the HST imaging. Colours were taken preferentially from \citet{Tonry} or from \citet{Prugniel}. For those galaxies marked with a * no colour was available and an average early-type colour from \citet{Prugniel} was used. Column (11):  The M/L of the best-fitting JAM model, converted to {\it I}-band. Column (12): The instrument and filter on the HST from which the photometry is taken. Unless otherwise stated data was taken using HST/WFPC2. Column (13): The quality of fit of the MGE model. For galaxies ranked 1 a good fit was achieved. For galaxies ranked 2 the fit achieved was still good apart from minor discrepancies or dust absorption. For those ranked 3 significant discrepancies had to be taken into account in the fitting process and the MGE model does not closely follow the isophotes (mostly due to bars) but still reproduces reasonable second-moment velocity maps. Column (14): Mgb - V$_\mathrm{esc}$ gradient determined as described in Section \ref{Sec:Grads}.
\end{minipage}
\end{table*}

There is one further relation most closely related to this work, that linking $\sigma$ and the magnesium line strength index (either Mgb or Mg$_2$) measured in a central aperture \citep{Burstein}. This is the tightest and best-studied relation linking a dynamical quantity $\sigma$ with a quantity depending only on the stellar population, the Mg index. Many authors have measured this relation for many hundreds of early-type galaxies \citep[see e.g.][]{Colless}, and while the precise gradient and zero-point found for the relation vary from author to author the tightness of the relation is common to all studies. Any successful model of early-type galaxy formation must explain this and the other relations discussed above before it can be accepted as accurately describing the formation histories of these objects.

The previously discussed relations are all global ones; some local relations have also been studied but only with small samples. \citet{FI} found that the local colour in elliptical galaxies correlated well with the local escape velocity, V$_\mathrm{esc}$, whereas the local colour - local $\sigma$ relation has significantly larger scatter. A similar result for the local Mg$_2$-V$_\mathrm{esc}$ relation was found by \citet{DSP93}, hereafter DSP93. This dependence on local parameters, which holds both within a single galaxy and between a sample of early-type galaxies begins to suggest that a key parameter in the formation and evolution of early-type galaxies is the gravitational potential $\Phi$, for which V$_\mathrm{esc}$ is a proxy.

In this work we explore the line strength - V$_\mathrm{esc}$ relation for the SAURON sample of galaxies \citep{Paper II} for which integral-field spectroscopy obtained on the SAURON integral-field unit \citep{Paper I} and extensive photometry are available. In Section 2 we give details of the SAURON sample, the observations and the data reduction process for the photometry and the spectroscopy. In Section 3 we discuss the dynamical modelling of the sample from which we derive the potential $\Phi$ and the escape velocity V$_\mathrm{esc}$. In Section 4 we present the resulting line strength-V$_\mathrm{esc}$ relations and translate these to the physical properties age, metallicity and alpha enhancement. In Section 5 we consider the implications of our results in the context of galaxy formation scenarios. Finally, our conclusions are summarised in Section 6.
\section{Sample and Data}
\begin{figure}
\includegraphics[width=3in]{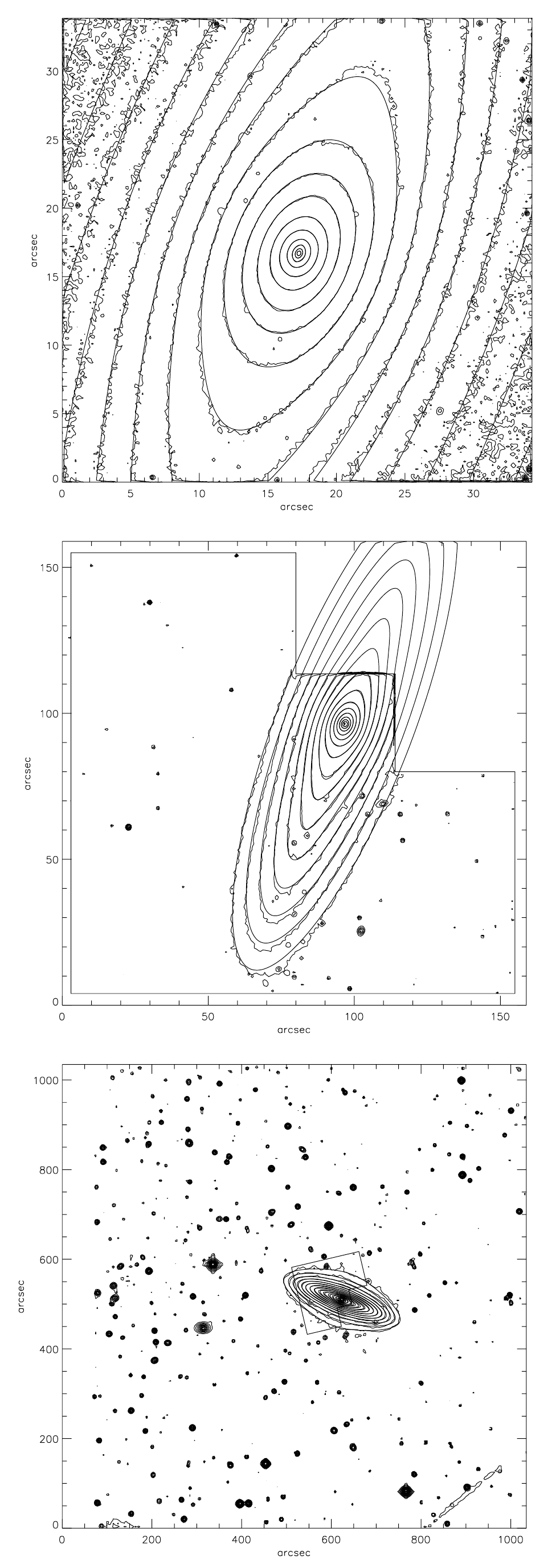}
\caption{Photometry of NGC4570 with the MGE model contours overplotted. From top to bottom the figures are: HST/PC1 field, HST/WFPC2  mosaic image, wide-field MDM image. The MDM image is oriented with north to the top and east to the left. The WFPC2 mosaic field is shown on the MDM image. As can be seen the MGE contours closely follow the isophotes at all radii.}
\label{Fig:NGC4570}
\end{figure}
\subsection{Selection}
The sample of galaxies used in this investigation is the SAURON sample of 48 early type galaxies \citep{Paper II}, divided equally between E and S0 morphologies \citep[where the classification is taken from][]{RC3}. This sample is representative of nearby  bright early-type galaxies ($cz \leqslant 3000\ \mathrm{kms}^{-1}$; $M_B \leqslant -18\ \mathrm{mag}$) and is fully described in \citet{Paper II}. The sample consists of an equal number of `cluster' and `field' objects (where cluster objects are defined as those belonging to the Virgo cluster, the Coma I cloud and the Leo group) uniformly covering the ellipticity-$M_B$ plane. 

\subsection{Data} \label{Photometry}
The photometric data consists of ground-based photometry obtained in the F555W filter on the 1.3-m McGraw-Hill Telescope at the MDM observatory on Kitt Peak \citep{MDM}, supplemented by HST observations where available (see Table \ref{table:sample} for the complete list.) A relatively large field of view of 17.1 $\times$ 17.1 $\mathrm{arcmin}^2$ was used for the MDM observations in order to provide accurate sky-subtraction from the images. The space-based observations consist primarily of HST/WFPC2 imaging or imaging from ACS or WPFC where WFPC2 data was not available. 

The HST data were used as reference for the photometric calibration and the MDM images were rescaled to the same level. The method of photometric calibration is described in detail in \citet[][hereafter Paper IV]{Paper IV}, but in summary we measured logarithmically sampled photometric profiles using circular apertures for each image after masking bright stars or galaxies. We do not expect or observe strong colour gradients between the F555W, F814W and intermediate filters, allowing us to match the MDM and HST photometry. The photometric profiles were then fitted by minimising the relative error between the two profiles in the region of overlap. The HST and MDM images were then merged to form a single photometric profile for each object. 

The spectroscopic information was obtained using the SAURON integral-field unit on the 4.2-m William Herschel Telescope at the Roque de los Muchachos observatory on La Palma. For details of the instrument and the data reduction pipeline see \cite{Paper I}. The SAURON field of view covers objects out to typically 1 R$_e$ and at least $0.5$ R$_e$. The data reduction steps include bias and dark subtraction, extraction of the spectra using a fitted mask model, wavelength calibration, flat fielding, cosmic-ray removal, sky subtraction and flux calibration. The flux calibration is described in detail in \citet[][hereafter Paper VI]{Paper VI}. The stellar absorption lines are also properly corrected for nebular emission. The SAURON wavelength range allows us to measure four Lick indices \citep{Lick II}: $\mathrm{H}\beta$, Fe5015, Fe5270 and Mgb \citep[see][for a full definition of these indices]{Lick}. In this work we consider only $\mathrm{H}\beta$, Fe5015 and Mgb as the Fe5270 line lies at the edge of SAURON's spectral range and so has incomplete spatial coverage in some objects. The measurement of the line strength indices from the final data cubes is described in Paper VI where line strength maps for the whole sample are presented. The stellar kinematics\footnote{Available from http://www.strw.leidenuniv.nl/sauron/} we use in this paper is the same that was used in Paper IV which was presented in Emsellem et at. (2004). This makes our $M/L$ values directly comparable with those of paper IV, when scaled to the same distances.

\section{Dynamical modelling}

\subsection{Multi-Gaussian Expansion mass models}
\label{Sec:MGEs}
Photometric models for all 48 galaxies in the sample were constructed using the Multi-Gaussian Expansion (MGE) parametrization of \cite{MGE I}. The observed surface brightness profile is described in terms of the sum of two-dimensional Gaussians, which allows the reproduction of ellipticity variations and strongly non-elliptical isophotes. The MGE fitting method of \cite{MGE II} was used to facilitate fitting of a large sample of galaxies. The MGE models were constrained to have constant position angle (PA) to enable axisymmetric Jeans modelling to be used in determining the underlying potential.

Twenty four of the MGE models used in this investigation were discussed in Paper IV and will not be discussed further here. The remaining 24 early-type galaxies of the SAURON sample are those for which either accurate Surface Brightness Fluctuation (SBF) distances were unavailable, WFPC2/F814W data was unavailable or the objects show strong non-axisymmetric features. While accurate distances are required to estimate mass to light ratios they are not required for this investigation and by relaxing the requirement for WFPC2/F814W data we can now model the entire SAURON early-type sample. MGEs for those galaxies not already presented in Paper IV are listed in the appendix. Note that several of the galaxies already modeled in Paper IV are triaxial objects.

The MGE models were fitted simultaneously to the wide-field MDM images and the higher resolution HST images. Where WFPC2 imaging was available the models were fitted simultaneously to the ground based, lower resolution mosaic and higher resolution WFPC2/PC1 images. The MGE fits were performed by keeping the position angle (PA) of the Gaussians constant and also taking the point spread function (PSF) into account. PSFs were calculated using TinyTim \citep{Tinytim} and modeled using the above MGE fitting method. The PSFs used are presented in Table \ref{Tab:PSFs}. The resulting analytically deconvolved MGE models are all corrected for galactic extinction following \citet{Extinct}, as given by the NASA/IPAC Extragalactic Database (NED). They are then converted to a surface density in solar units in the Johnson-Cousins magnitude system using the calibration relevant to each instrument (WFPC1 - \citet{WFPC1}; WFPC2 - \citet{WFPC2}; ACS - \citet{ACS}.) Absolute magnitudes for the Sun (M$_\mathrm{B} = 5.48$,M$_\mathrm{V} = 4.83$, M$_\mathrm{R} = 4.42$, M$_\mathrm{I} = 4.08$,) are taken from Table 2.1 of \citet{Gal Ast}. The values of the MGE parameterizations are presented in the Appendix in Table \ref{Tab:MGEs} and \ref{Tab:MGEs2}.

Because the SAURON field-of-view is relatively small when compared to our imaging we are principally interested in fitting the central regions of each galaxy, while the MDM imaging is used to provide additional constraints on the MGE model. We do not attempt to accurately model structure such as shells or isophotal twists in the outer parts of these galaxies but we do seek to reproduce the overall shape of the object. As discussed in \citet{MGE II} the models were regularized by requiring the axial ratio of the flattest Gaussian to be as round as possible while still reproducing the observations. This is important so as not to artificially constrain the possible inclinations of the models and to reproduce realistic densities. We also masked dusty regions in the images; in the small number of galaxies in our sample that exhibit dust, the dust is only visible in one half of the galaxy image and so does not reduce the quality of our MGE fits.

The quality of the resulting models with respect to the photometry was visually inspected for all galaxies to ensure a reasonable fit had been achieved. We also compared the resulting kinematics (see Section \ref{Sec:Jeans}) to the SAURON kinematics presented in \citet{Paper III} and adapted the MGEs (within the rules outlined below) where necessary to obtain a match to the SAURON observations. The models were refined until a satisfactory qualitative fit was achieved for all galaxies. For those galaxies with regular photometry this was easily achieved. An example of the model and data photometry for such a galaxy, NGC4570 is given in Fig. \ref{Fig:NGC4570}. 

\subsubsection{Bars, twists and non-axisymmetry}
For those galaxies with non-axisymmetric features such as bars or isophotal twists achieving a good match proved more challenging. This is because by simply fitting the two-dimensional isophotes we infer a three-dimensional distribution of matter than does not reflect that in the real object. In these cases a simple prescription was followed to produce our MGE models. 

\begin{figure}
\begin{center}
\includegraphics[width=3.3in,trim = 0 0 0 3in, clip]{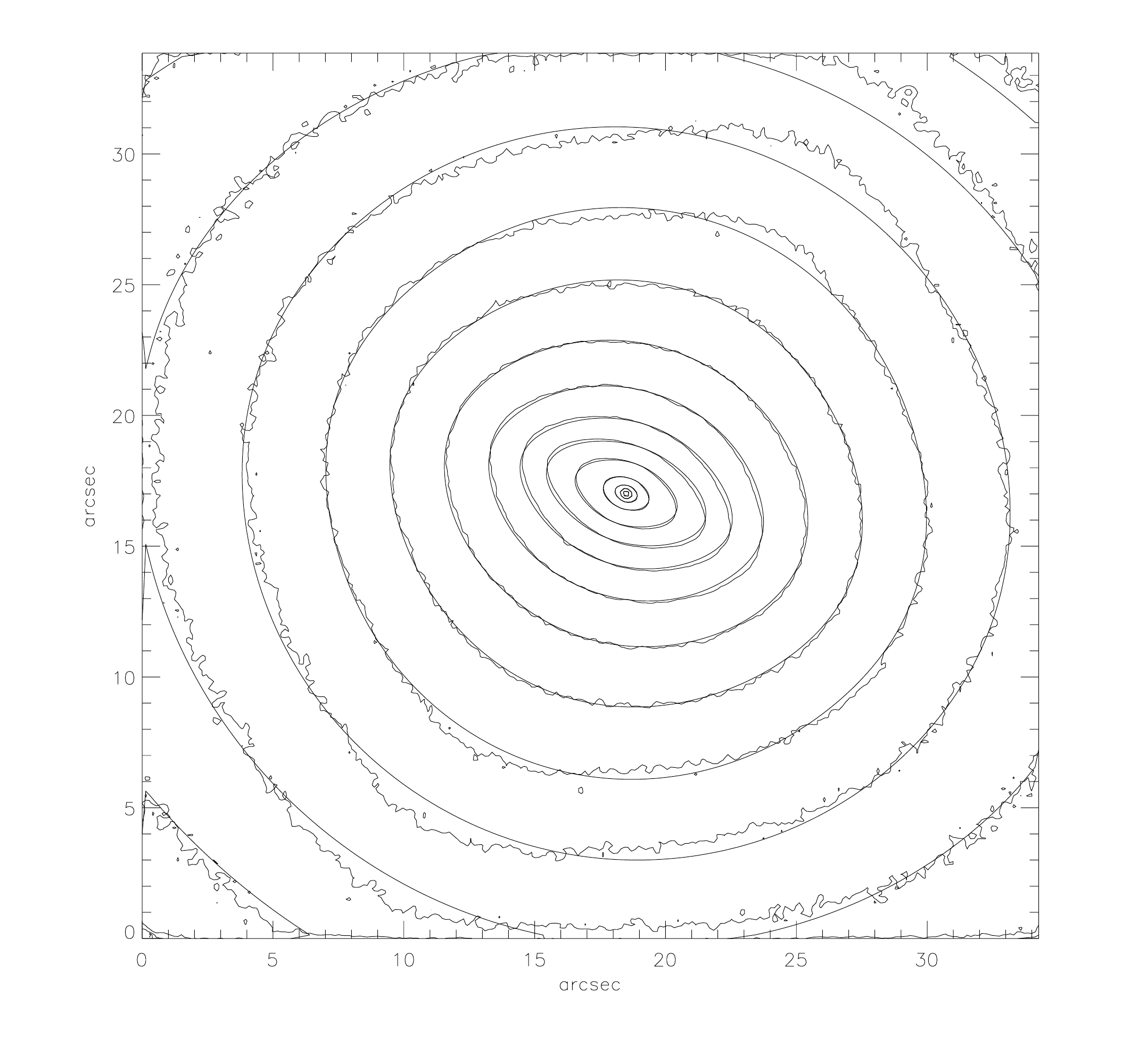}
\includegraphics[width=3.3in,trim = 0 0 0 3in, clip]{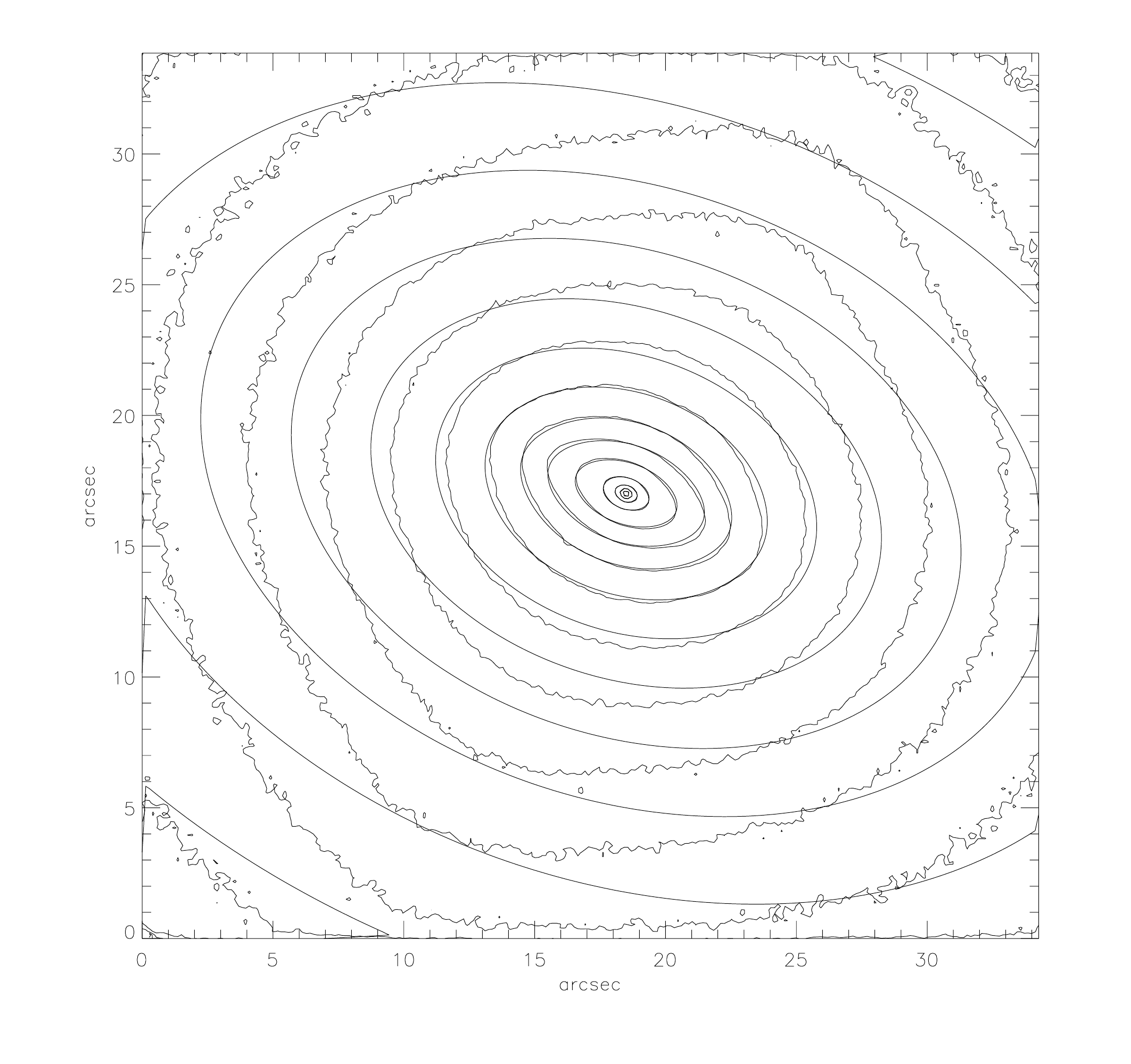}
\end{center}
\caption{HST/PC1 photometry on NGC3384 showing two different MGE model contours overplotted. In the upper figure we have allowed the model to follow the isophotes such that the effect of the bar is included in the MGE model. In the lower image we have constrained the axial ratios of the Gaussians as described in the text in order to reduce the influence of the bar on the MGE model. Despite being a superficially poorer fit to the photometry the MGE model from the lower figure reproduces a better fit to the observed kinematics than that from the upper figure (see Fig. \ref{Fig:NGC3384kin}).}
\label{Fig:NGC3384phot}
\end{figure}

\begin{figure}
\begin{minipage}{3.5in}
\begin{center}
\includegraphics[width=1.7in]{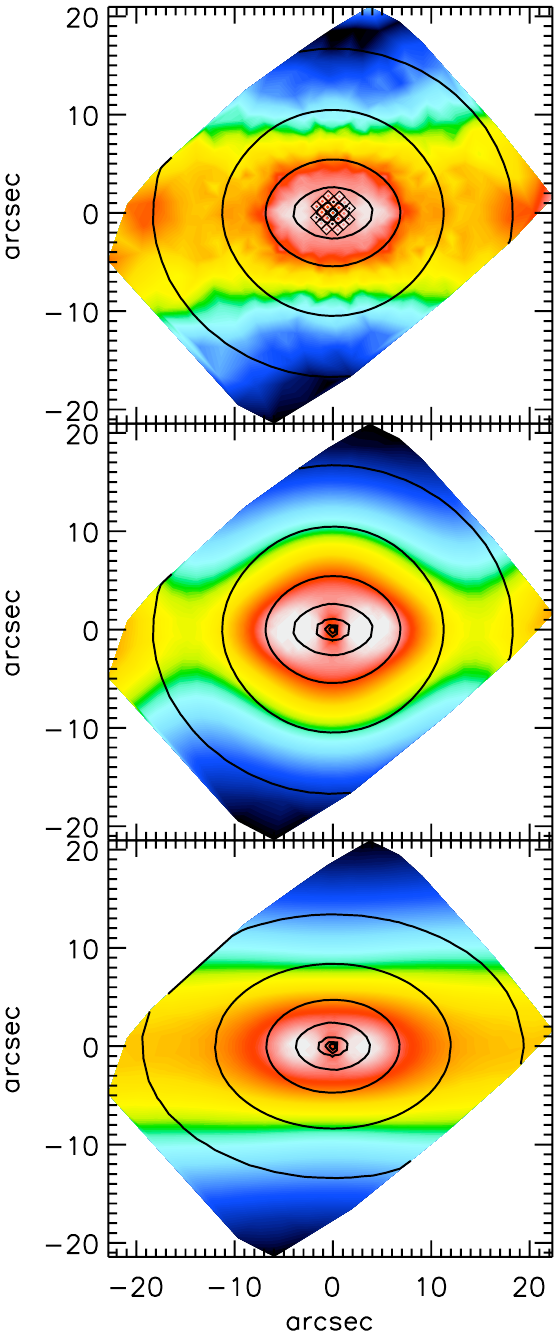}
\includegraphics[width=1.7in]{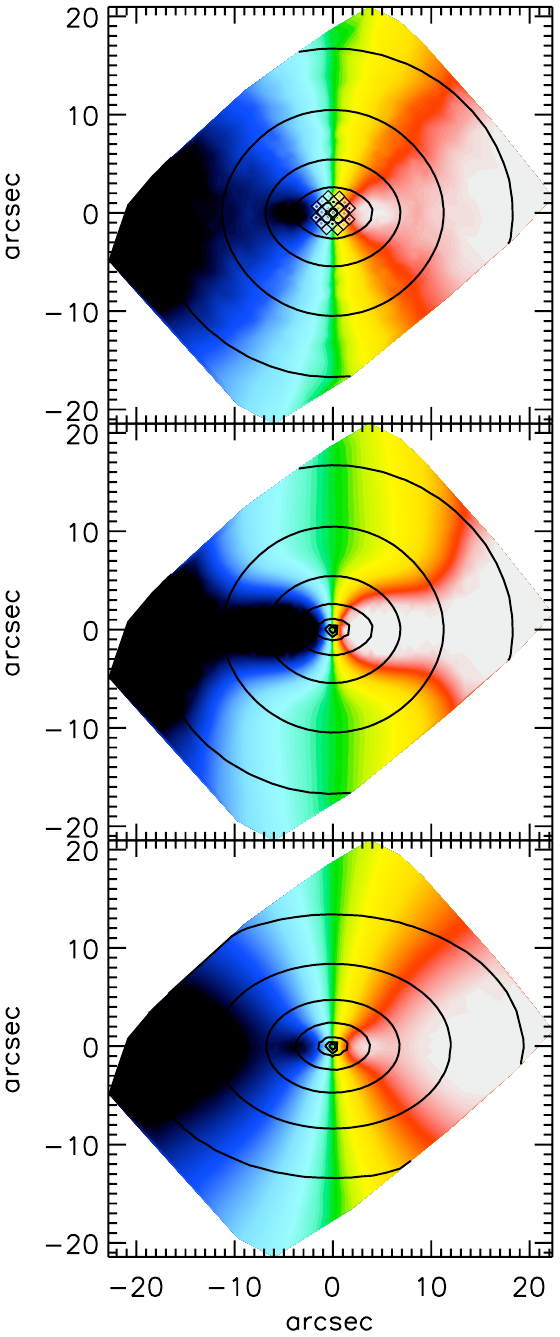}
\caption{Bi-symmetrized and linearly interpolated maps of the second velocity moment $\mu'_2 \equiv \sqrt{V^2 + \sigma^2}$ (left) and velocity field V (right) of NGC3384. The top image shows the observed $\mu'_2$ and V extracted from the SAURON stellar kinematics. The centre and bottom images show the predicted $\mu'_2$ and V from the two MGE models described in the caption to Fig. \ref{Fig:NGC3384phot}. The centre image shows the predictions for an MGE that follows the isophotes of the bar whereas the bottom image shows the predictions for an MGE that attempts to avoid the effect of the bar. As can be clearly seen the `bar-less' MGE is a much better fit to the observed $\mu'_2$ and V. The contours shown are the isophotes from the reconstructed SAURON image.}
\label{Fig:NGC3384kin}
\end{center}
\end{minipage}
\end{figure}

We note that bars are always associated with a discy structure which is to first order axisymmetric and of which the bar represents a perturbation. Because of this we assume that a reasonable axisymmetric mass model of a barred galaxy is found when the ellipticity over the barred region is fixed at the global ellipticity. In galaxies where there is an obvious bar we constrained the MGE model such that the axial ratio of the Gaussians over the barred region was consistent with the axial ratios of the inner and outer regions where the impact of the bar on the photometry was negligible. An example of a model for a barred galaxy is given in Figs. \ref{Fig:NGC3384phot} and \ref{Fig:NGC3384kin}. With the barred MGE the axisymmetric JAM model fails to reproduce the kinematics, we find a significant improvement when using our bar-less MGE. Moreover, the recovered inclination of the model becomes closer to the one inferred from the disc. This suggests that our bar-less MGEs is a better approximation to the global galaxy structure.

For galaxies where the photometric and kinematic PA (see table 1, Paper X) differ significantly the kinematic PA was used for the MGE models as being more representative of the galaxy over the SAURON field of view. Although in this way we do not represent the isophotal twists in the photometry the stellar kinematics are fitted better leading to a significant improvement in the $\chi^2$ value of the fit. We checked the kinematics produced by the axisymmetric Jeans modelling and in all cases a reasonable agreement was found with the observed kinematics. Further examples of the observed and modelled first and second velocity moments are shown in the Appendix in Fig. \ref{Fig:JAM_examples}1.

\subsection{Jeans Anisotropic MGE (\textsc{JAM}) axisymmetric modelling}
\label{Sec:Jeans}
In order to compute V$_\mathrm{esc}$ we constructed Jeans Anisotropic MGE (\textsc{JAM}) axisymmetric models\footnote{Available from http://www-astro.physics.ox.ac.uk/$\sim$mxc/idl/} \citep{newJeans} of all the galaxies in the sample. For a given inclination $i$, the MGE surface density can be deprojected analytically \citep{MBE} to obtain the intrinsic density $\rho (R,z)$ in the galaxy meridional plane, still expressed in terms of Gaussians. This deprojection is non-unique but represents a reasonable choice as the resulting intrinsic density resembles observed galaxies for all lines of sight. We then apply \textsc{JAM} modelling to the resulting deprojected densities to determine the underlying potential.

The method is described fully in \citet{newJeans} but we will briefly summarise the key points here. The positions $\mathbf{x}$ and velocities $\mathbf{v}$ of a large system of stars can be described by a distribution function $f(\mathbf{x},\mathbf{v})$ which in a steady state must satisfy the collisionless Boltzmann equation. In order to make use of this equation further simplifying assumptions must be made. A typical first choice is to assume axial symmetry, which leads to the two Jeans equations \citep{Jeans} , but this is not sufficient to specify a unique solution. We make the further assumptions that the velocity dispersion ellipsoid is aligned with the cylindrical coordinate  system $(R, z, \phi)$ and that the anisotropy is constant. We also assume that mass follows light, but allow for a constant dark matter fraction. Under these assumptions the Jeans equations reduce to:
\begin{equation}
\frac{b \nu \overline{v_z^2} - \nu \overline{v^2_{\phi}}}{R} + \frac{\partial(bv\overline{v_z^2})}{\partial R} = -\nu\frac{\partial \Phi}{\partial R}
\end{equation}
\begin{equation}
\frac{\partial ( \nu \overline{v_z^2})}{\partial z} = -\nu \frac{\partial \Phi}{\partial z}
\end{equation}
where b quantifies the anisotropy, $\overline{v_R^2} = b \overline{v_z^2}$, $\nu$ is the luminosity density and $v_i$ the components of the velocity. They provide the second moments of the line-of-sight velocity $\overline{v_{z'}^2} \equiv \overline{v_{los}^2}$, which are generally considered to be good approximations to the observed quantity $V^2_{rms} = V^2 + \sigma^2$. By comparing the observed and Jeans modelled second moments we determined the best fitting inclination, anisotropy and constant mass-to-light ratio $M/L$ for all 48 galaxies in our sample. 

\begin{figure*}
\centering
\includegraphics[width=7in]{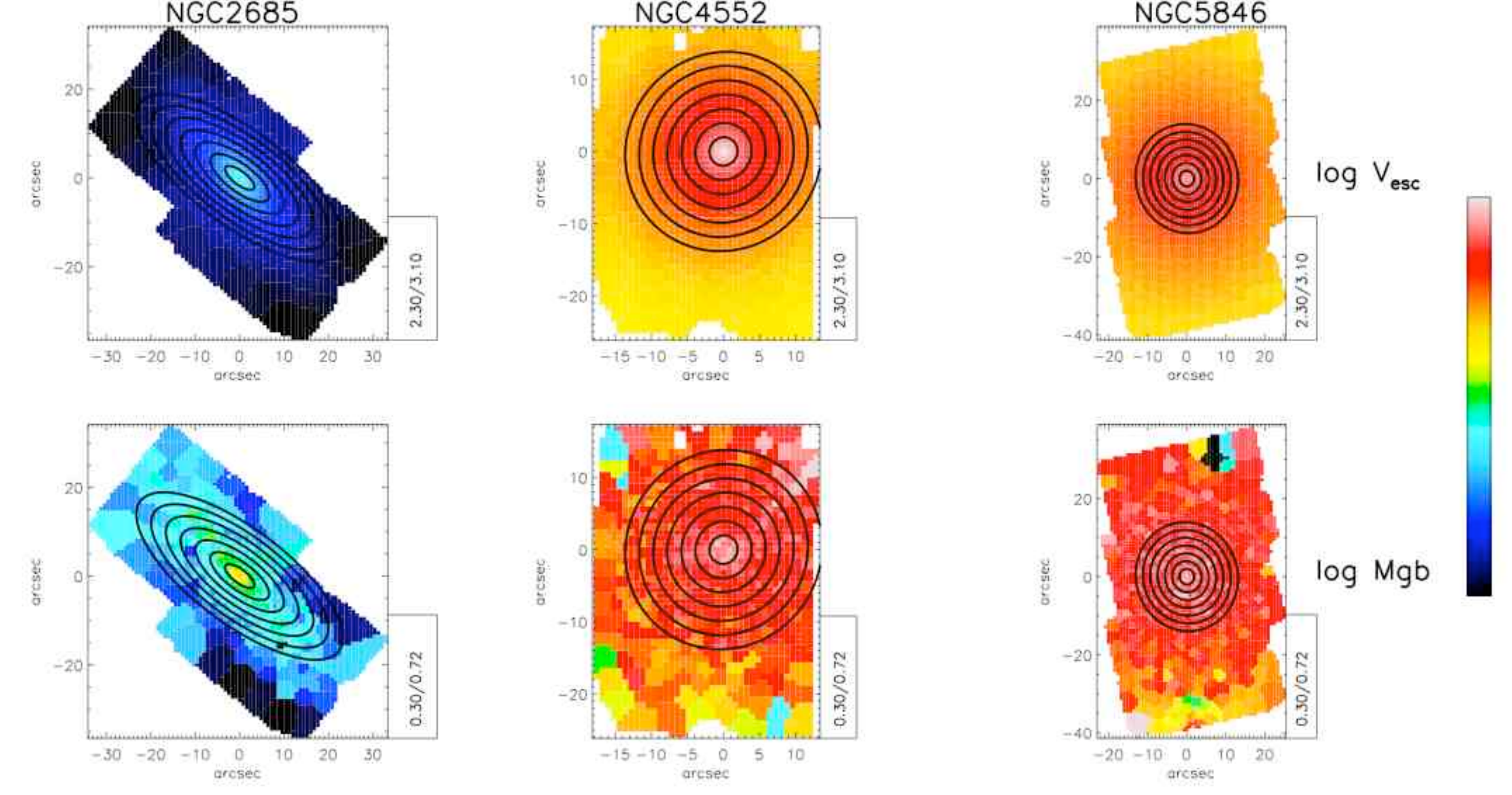}
\caption{V$_\mathrm{esc}$ and Mgb maps for several galaxies with the elliptical annuli used to extract the profiles shown. Here we show only a few of the ellipses to avoid cluttering the plot. The numbers to the bottom right of each plot indicate the range of values displayed. As can be seen the modelled V$_\mathrm{esc}$ field (or equally the potential $\Phi$) closely traces the ellipses but the observed Mgb maps are less regular.}
\label{Fig:Apertures}
\end{figure*}

Determining the inclination is difficult but we apply several independent checks which validate our fitted results. For highly flattened objects they must be close to edge on (16 objects). Six galaxies have an obvious embedded gas disk or dust lane, inclinations were estimated assuming these are thin discs. In all cases our inclinations were consistent with the independent determination. For the remaining 26 galaxies the inclinations are determined based purely on the model and may not always be accurate. The accuracy of these inclinations is discussed further in \citet{newJeans}. Our V$_\mathrm{esc}$ is only weakly dependent on the inclination used and so this uncertainty does not significantly affect the conclusions of this work. As an extreme test of the dependence of our modeled V$_\mathrm{esc}$ on inclination we artificially set  the inclinations for the models of all our galaxies to 90$^{\circ}$ (i.e. edge-on) and re-calculated the V$_\mathrm{esc}$. The effect on V$_\mathrm{esc}$ was small, only a 5 percent change in the most extreme cases..

As mentioned above some of the objects in our sample are clearly not axisymmetric systems and so the use of axisymmetric models requires some justification. The alternative would be to use the more general \citet{Schwarzschild} models. In Paper IV we compared the mass-to-light ratios (M/L) derived from axisymmetric Jeans and Schwarzschild modelling and find an excellent agreement between the two (see particularly Fig. 7 from that paper.) Additionally, the slow-rotators, while likely to be triaxial objects are also very round \citep[see Paper X,][]{KDC} and so any deviations from axisymmetry in their intrinsic shapes are relatively small. From this one should not expect major biases in the M/L we derive with axisymmetric models. We explicitly tested whether this is the case, using the M/L derived via more general triaxial models by \citet{Remco} for eight slow-rotators in common with our sample. We found good agreement between our M/L determinations with JAM and the triaxial models. A more detailed comparison will be presented elsewhere. As for the M/L, we expect V$_{\mathrm{esc}}$ to be only weakly affected by the assumption of axisymmetry and the use of Jeans models. This can be understood by noting that while there are many more possible orbits in a triaxial system than in an axisymmetric one this does not affect the potential and hence V$_{\mathrm{esc}}$. It is the distribution of the mass, not the structure of the orbits, that affects $\Phi$,  and this is largely unchanged between triaxial and axisymmetric systems, apart from a small geometric factor.

\subsection{Extraction of line strength indices and V$_\mathrm{esc}$}
In order to study the index-V$_\mathrm{esc}$ relations we must extract the intrinsic line strength indices from the SAURON maps and the V$_\mathrm{esc}$ from our \textsc{JAM} models in a consistent fashion. The potential $\Phi$ is calculated as in \citet{MGE I} and the V$_\mathrm{esc}$ is simply related to this by the expression:
\begin{equation}
	\mathrm{V}_\mathrm{esc} = \sqrt{2|\Phi(R,z)|}
\end{equation}

The observed indices on the sky plane are the luminosity-weighted average of the local values in the galaxy along the line-of-sight. We make the quite general assumption that the indices are related to V$_\mathrm{esc}$ by a simple power-law relationship of the form:

\begin{equation}
	\mathrm{Index} \propto \mathrm{V}_\mathrm{esc}^{\gamma}
\end{equation}
This leads to:
\begin{eqnarray}
\Sigma(x^\prime,y^\prime) \mathrm{Mgb}_p & \propto & \Sigma(x^\prime,y^\prime)\mathrm{V}_\mathrm{esc,p}^\gamma(x^\prime,y^\prime) \nonumber \\
& = & \int^{\infty}_{-\infty}\rho(R,z) \mathrm{V}_\mathrm{esc}^\gamma(R,z) dz^\prime \nonumber \\  
& =  &  \int^{\infty}_{-\infty}\rho(R,z) | 2 \Phi(R,z)|^{\gamma/2} dz^\prime
\end{eqnarray}
With this assumption it is possible to extract the luminosity-weighted average, V$_\mathrm{esc,p}$ of the local V$_\mathrm{esc}$ along the line-of-sight. In practice the V$_\mathrm{esc}$ values depend only weakly on the parameter $\gamma$ and our conclusions hold for any reasonable choice of the parameter.

To form profiles we sum the local line-of-sight integrated values over elliptical annuli aligned with the kinematic major axis of the galaxy and evenly spaced in radii over the entire SAURON field, where the ellipticity used is the global ellipticity as given in table 1 of Paper X. The noise in each profile is minimised by choosing the photometric ellipticity $\epsilon$ for the profile extraction ellipses. Several examples of the Mgb and V$_\mathrm{esc}$ maps with the elliptical annuli used to extract the profiles overplotted are shown in Fig. \ref{Fig:Apertures}. Before this was done the individual SAURON line strength maps were inspected for irregular bins. These occasionally occur in the outer parts of the SAURON field due to the continuum effects described in Paper VI. Masks were constructed for several of the most affected maps. Only the outer few elliptical annuli are affected by this issue and the use of un-masked maps does not significantly effect the Index-V$_\mathrm{esc}$ profiles. Bright stars and obvious dust features were also masked on the line strength maps. The error in the line strength for each data point is the rms sum of the measurement error from Paper VI and the rms scatter within an annulus. We adopted an error of 5 per cent in V$_\mathrm{esc}$ (see Paper IV).

\begin{figure*}
\begin{minipage}{1in}
\includegraphics[width=1in,height=6in]{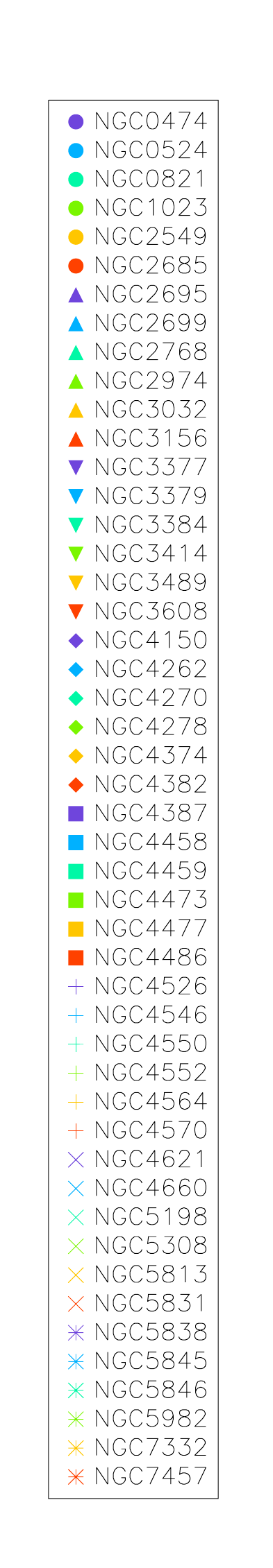}
\end{minipage}
\begin{minipage}{4in}
\includegraphics[height=2.95in]{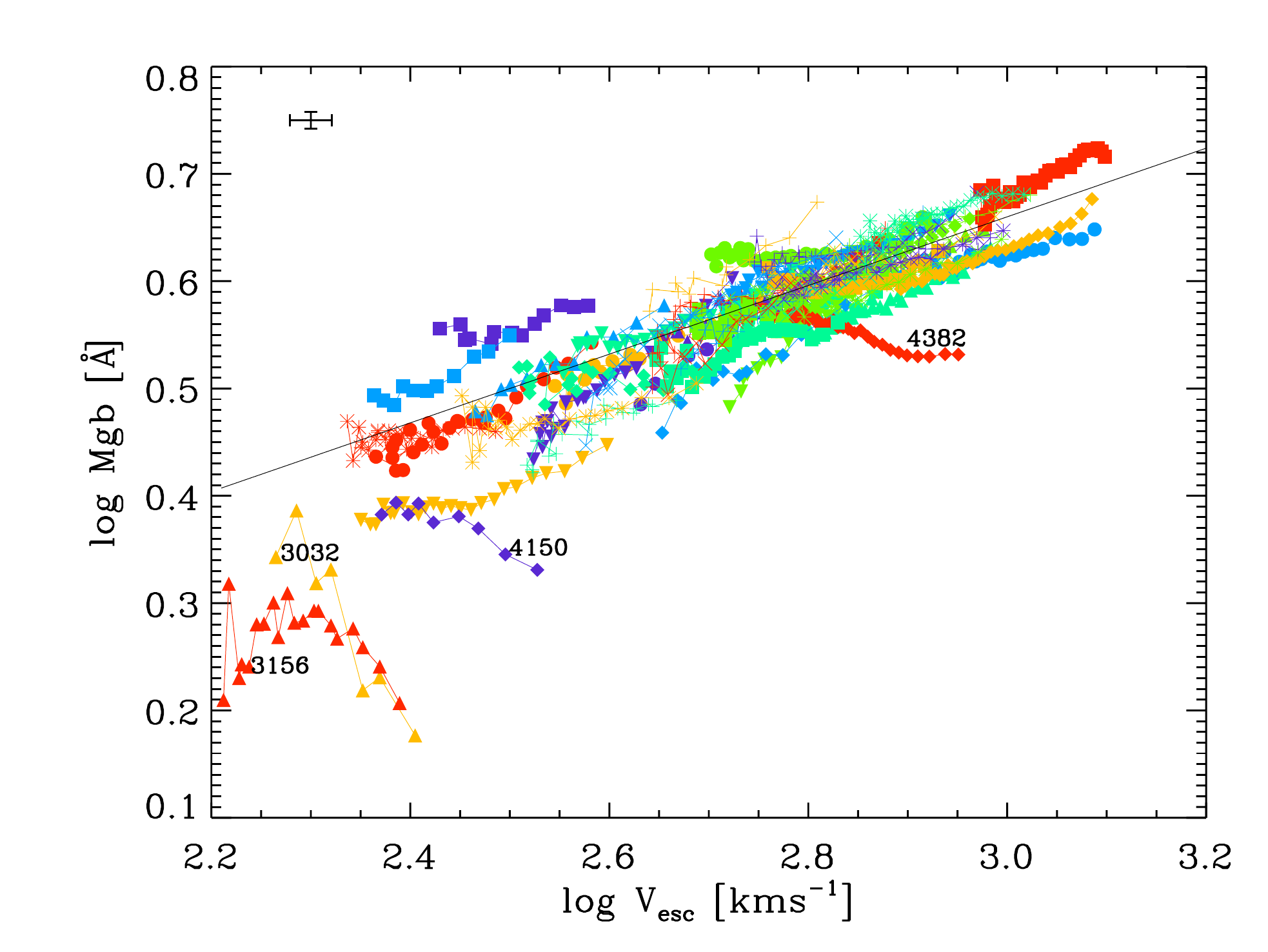}
\includegraphics[height=2.95in]{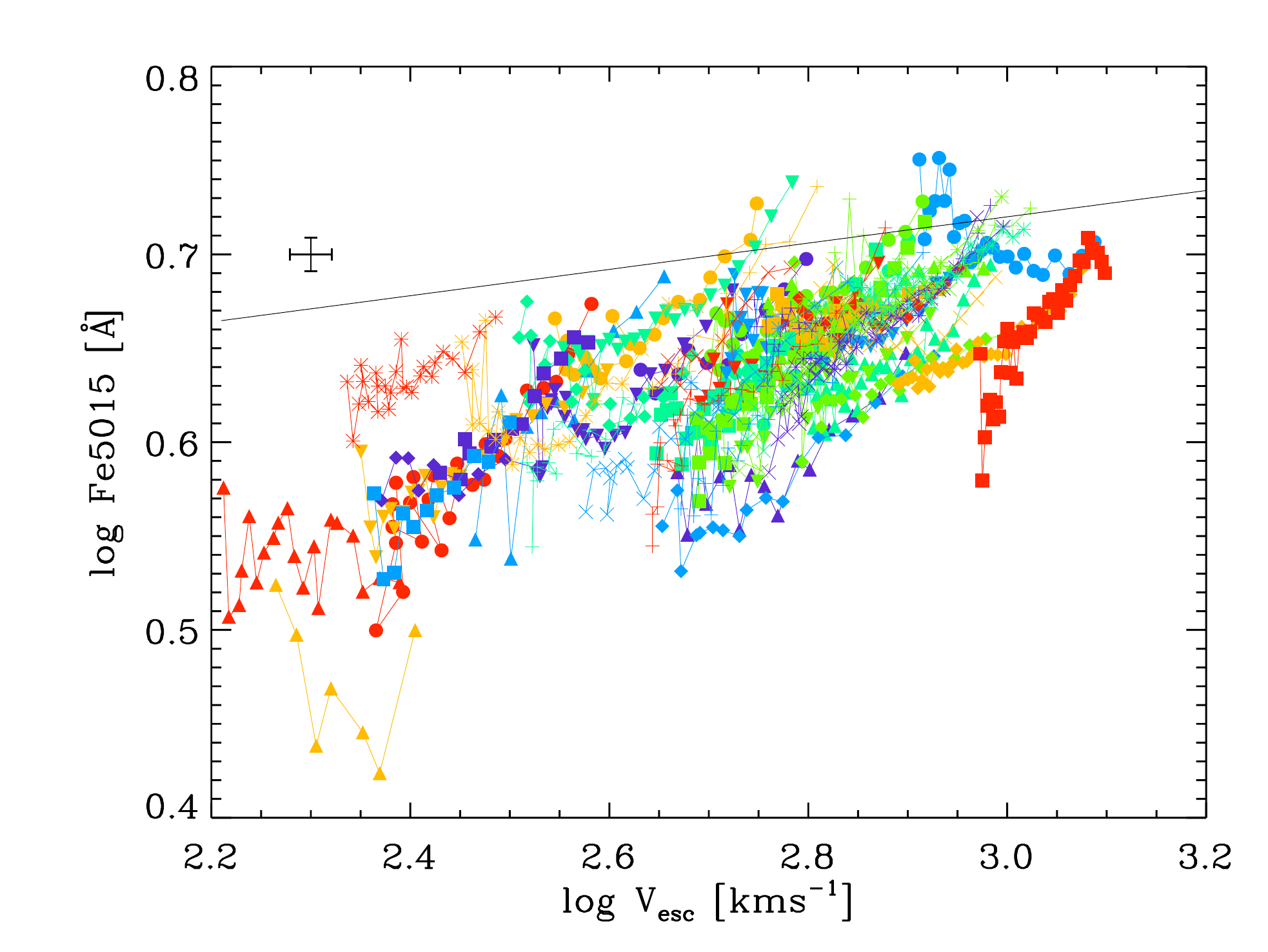}
\includegraphics[height=2.95in]{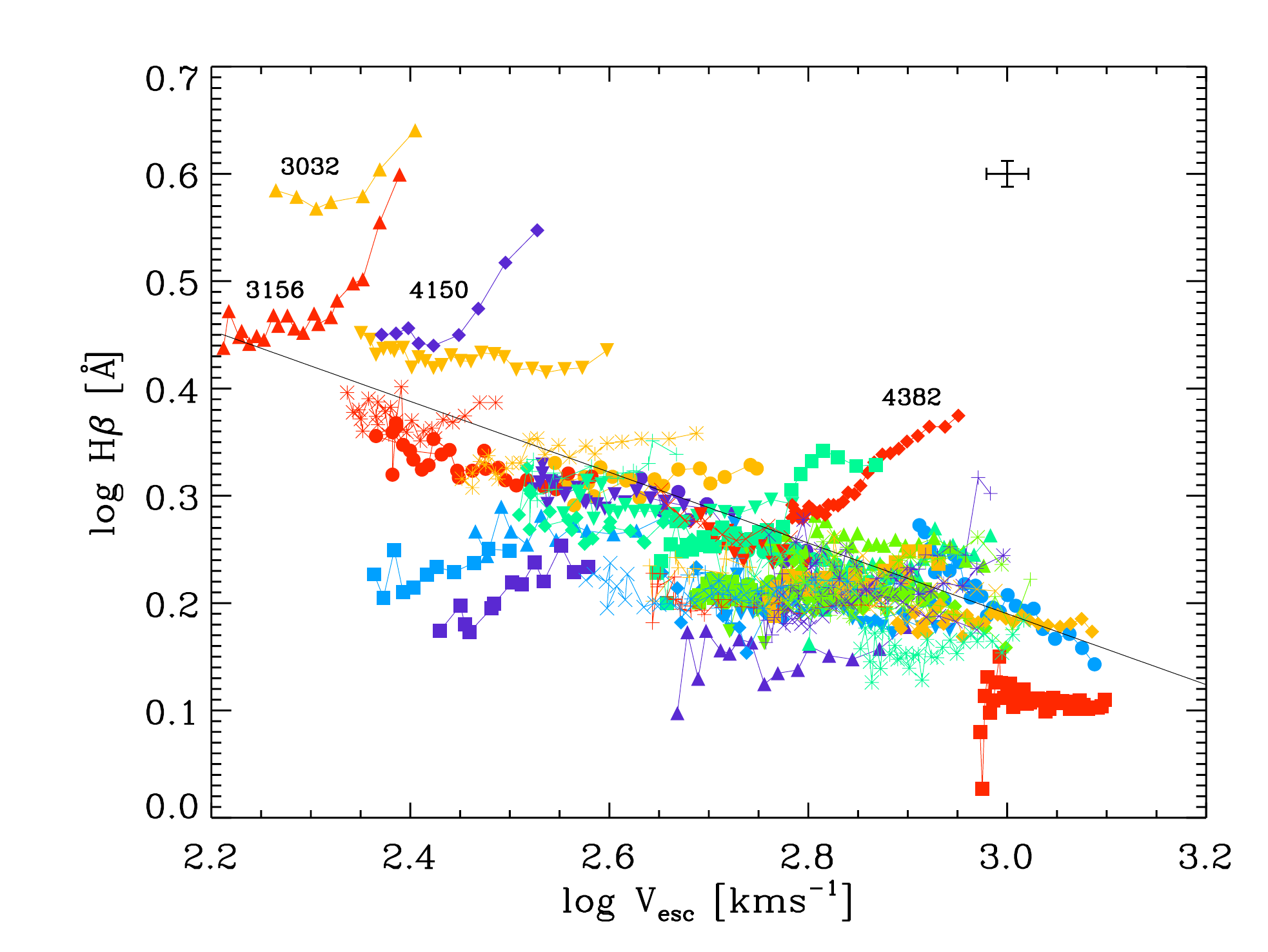}
\end{minipage}
\caption{The line strength index versus V$_\mathrm{esc}$ relations. All indices are measured in \AA \ of equivalent width. In each figure all the points for an individual galaxy are represented by a single combination of colour and symbol. In all three cases the correlation is remarkably tight. The black line in each panel is a fit to the central Re/8 aperture values, excluding outlying galaxies. The profiles with their NGC numbers shown are those galaxies showing evidence of recent star formation. Typical error bars are shown in each figure.} 
\label{Fig:Index_Vesc}
\end{figure*}

\begin{table*}
\caption{Linear fit parameters for the Index-V$_\mathrm{esc}$ relations from Re/8 circular aperture values.}
\label{Tab:Vesc1}
\begin{center}
\begin{tabular}{c l c c c}
Sample & Index & $a$ & $b$ & Standard\\
& & & & Deviation\\
\hline
& Mgb & -0.30$\pm$0.09 & 0.32$\pm$0.03 & 0.033\\
All (44) &Fe5015 & 0.51$\pm$0.09 & 0.07$\pm$0.03 & 0.033\\
& H$\beta$ & 1.18$\pm$0.15 & -0.33$\pm$0.05 & 0.049\\
\hline
&Mgb & -0.27$\pm$0.10 & 0.31$\pm$0.04 & 0.030\\
`Clean' (34) &Fe5015 & 0.45$\pm$0.10 & 0.09$\pm$0.04 & 0.031\\
&H$\beta$ & 1.04$\pm$0.16 & -0.28$\pm$0.06 & 0.046\\
\hline
&Mgb & -0.31$\pm$0.13 & 0.33$\pm$0.04 & 0.034\\
Fast Rotator (32)&Fe5015 & 0.60$\pm$0.11 & 0.04$\pm$0.04 & 0.039\\
&H$\beta$ & 1.27$\pm$0.18 & -0.36$\pm$0.06& 0.047\\
\hline
&Mgb & -0.28$\pm$0.16 & 0.31$\pm$0.05 & 0.030\\
Slow Rotators (12)&Fe5015 & 0.24$\pm$0.13 & 0.15$\pm$0.05 & 0.035\\
&H$\beta$ & 0.91$\pm$0.24 & -0.24$\pm$0.08 & 0.052\\
\end{tabular}
\end{center}
Notes: We fit a straight line of the form $\log \mathrm{Index} = a + b\times\log \mathrm{V}_\mathrm{esc}$ . The linear fit parameters were calculated using a $\chi^2$ minimisation technique as described in the text.
\end{table*}

\section{Results}
\label{Sec:Results}
In this section we present the Index-V$_\mathrm{esc}$ relations determined using the method described above. The profiles are presented in Fig. \ref{Fig:Index_Vesc}. Mgb and Fe5015 show a remarkably tight correlation, with the Mgb-V$_\mathrm{esc}$ relation having the tightest correlation. The H$\beta$-V$_\mathrm{esc}$ correlation is less tight. In this and all following linear fits we fitted a linear relation to the Re/8 circular aperture values for each of the observed correlations by minimizing the $\chi^2$ parameter using the {\sc FITEXY} routine taken from the IDL\footnote{http://www.ittvis.com/ProductServices/IDL.aspx} Astro-Library \citep{Landsman}, which is based on a similar routine by \citet{Press} and adding quadratically the intrinsic scatter to make $\chi^2/\nu = 1$, where $\nu$ is the number of degrees of freedom. For a discussion of the technique and its merits see \citet{Tremaine}. Four outlying galaxies were excluded from these fits (see Section \ref{Sec:Grads} for a description of how outliers were selected). The results of these fits are plotted as the straight line on each figure, and the zero-point, slope and rms scatter for each relation are shown in Table \ref{Tab:Vesc1}. Mgb and Fe5015 rise with increasing V$_\mathrm{esc}$, with Mgb having the steeper slope. In contrast H$\beta$ shows the opposite trend, in the sense that the areas of deepest potential $\Phi$ have the weakest H$\beta$ absorption.

It is known that some early type galaxies may be weakly triaxial \citep[][Paper IX; Paper X]{KB96} and an axisymmetric model may not reliably reproduce the intrinsic kinematics and potential. Galaxies where this is the case typically show a large kinematic and photometric misalignment; barred galaxies are also not fully axisymmetric systems. Some of the scatter observed in the Index-V$_\mathrm{esc}$ relations may be due to axisymmetric models not properly reproducing the intrinsic V$_\mathrm{esc}$. In order to investigate this we define a `clean' sample in which all galaxies that show evidence for a non-axisymmetric distribution have been removed (see Table \ref{table:sample}, column (13), only galaxies graded 1 were included in this clean sample.) The best-fitting linear fit parameters for the clean sample are given in Table \ref{Tab:Vesc1}. There is little change in the gradients between the full and axisymmetric samples, though the scatter is slightly reduced by $\sim$ 10 per cent. This suggests that while imperfect fitting of $\Phi$ due to the assumption of axisymmetry accounts for some of the scatter observed in the relations it is only a small effect.

In \citet{Paper IX} a classification scheme was described for galaxies based on a parameter $\lambda_R \equiv \langle R|V|\rangle / \langle R \sqrt{V^2 + \sigma^2}\rangle$ which is related to the angular momentum per unit mass of the stars integrated within 1 R$_e$. Within this classification galaxies with $\lambda_R \leq 0.1$ are described as slow rotators and those with $\lambda_R > 0.1$ as fast rotators. \citet{Paper IX} and \citet{Paper X} speculated that slow and fast rotators represent two different families of galaxies with significantly different formation histories (in terms of interaction or merger events, cold gas accretion episodes, secular evolution etc.). If this is the case we might expect the stellar populations of the two galaxies to have experienced different histories and for signatures of this to show up in the line strength-V$_\mathrm{esc}$ relations. Definite predictions of these differences are beyond the scope of this work but it is thought that dry merger processes are more important in the formation of slow rotators whereas the role of gas is more prominent in fast rotators. Mergers would tend to alter V$_\mathrm{esc}$ while leaving the stellar population (and therefore the line strengths) unchanged whereas gaseous processes will alter the line strengths. In order to explore this we separate our sample into fast- and slow-rotators and again perform linear fits to the two sub-populations; the best fitting parameters are again shown in Table \ref{Tab:Vesc1}. For the case of Mgb the fast- and slow-rotators follow the same relationship. In the case of H$\beta$ and Fe5015 there is some suggestion that fast- and slow-rotators follow different relationships, though when only `clean' galaxies are considered this disappears.

We also explored the dependence on the traditional division into ellipticals/S0s based on their RC3 classifications (though \citet{Paper IX} and \citet{Paper X} suggest the slow/fast rotator classification is more physically relevant) with 24 of each in our sample and found no significant difference in the Index-V$_\mathrm{esc}$ relations between the two sub-samples. We also split our sample into field galaxies and those belong to a cluster or group (again with 24 galaxies lying in each sub-sample) and again found no significant differences but it important to remember that the environmental classification used here is a simple one.

\subsection{Comparison to previous work}
The number of studies that have looked at the local Mg-$\sigma$ and Mg-V$_\mathrm{esc}$ relations is surprisingly small given the well known tightness of the global relation. The two main studies in the area are DSP93 and \citet{CD94}. Both have much smaller samples than our current work (8 galaxies and 5 galaxies with V$_\mathrm{esc}$ respectively). The DSP93 sample has four galaxies in common with the SAURON sample ( NGC 3379, 4278, 4374 and 4486) for three of which DSP93 have V$_\mathrm{esc}$ (the exception is NGC 4374) whereas we have no galaxies in common with the CD94 sample.

\begin{figure}
\centering
\includegraphics[width=3.5in]{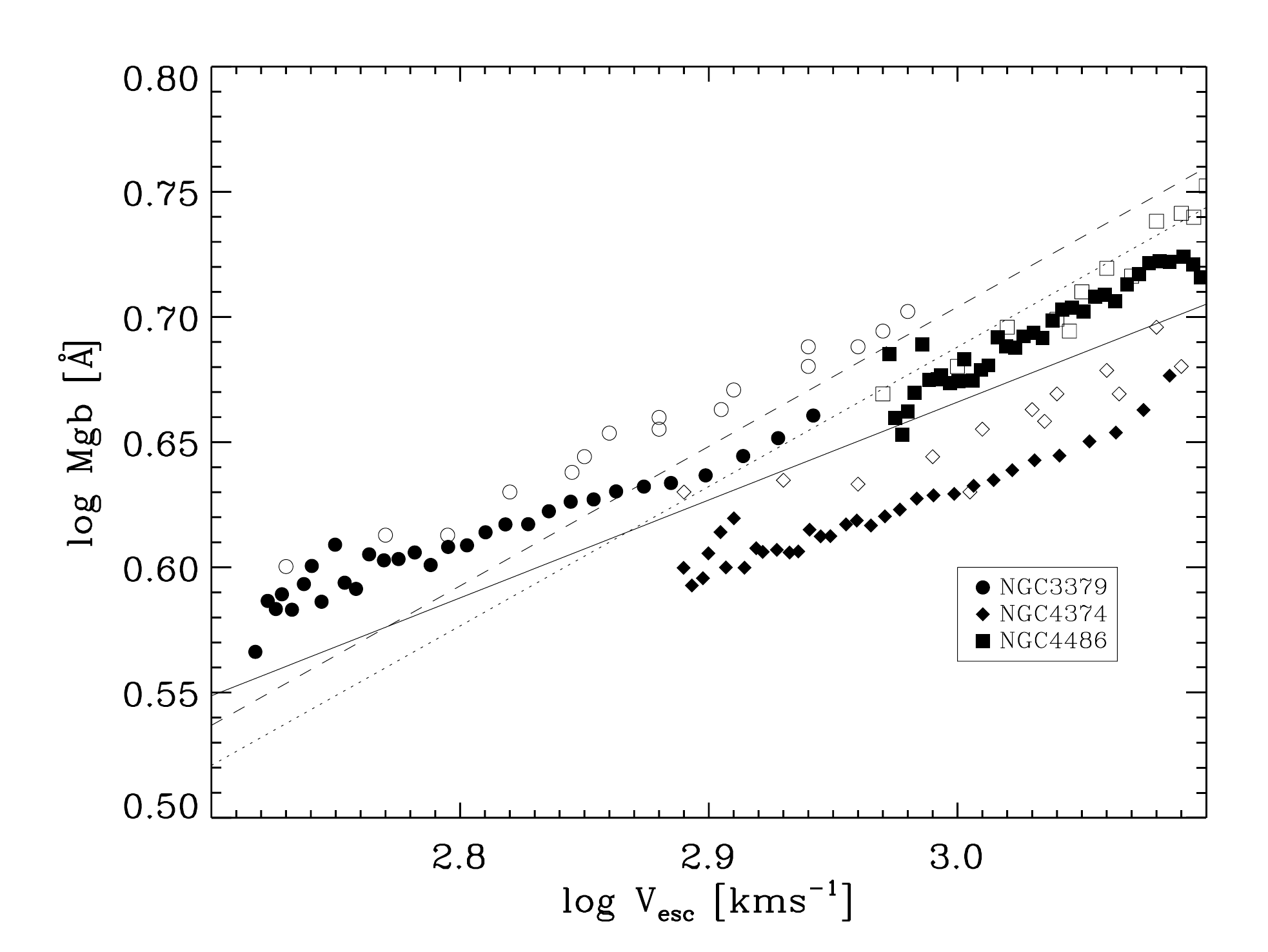}
\caption{Comparison of the results of our sample with those galaxies we have in common with DSP93. The Mg$_2$ index values of DSP93 and CD94 were converted to Mgb index values using Equation \ref{Eq:MgbtoMg2}. The open symbols are the values from DSP93 and the closed symbols are the results from this work. The solid line is the fit to the SAURON sample, the dotted line to the DSP93 sample and the dashed line the CD94 sample.}
\label{Fig:DSP93comp}
\end{figure}

\begin{figure}
\centering
\includegraphics[width=3.5in]{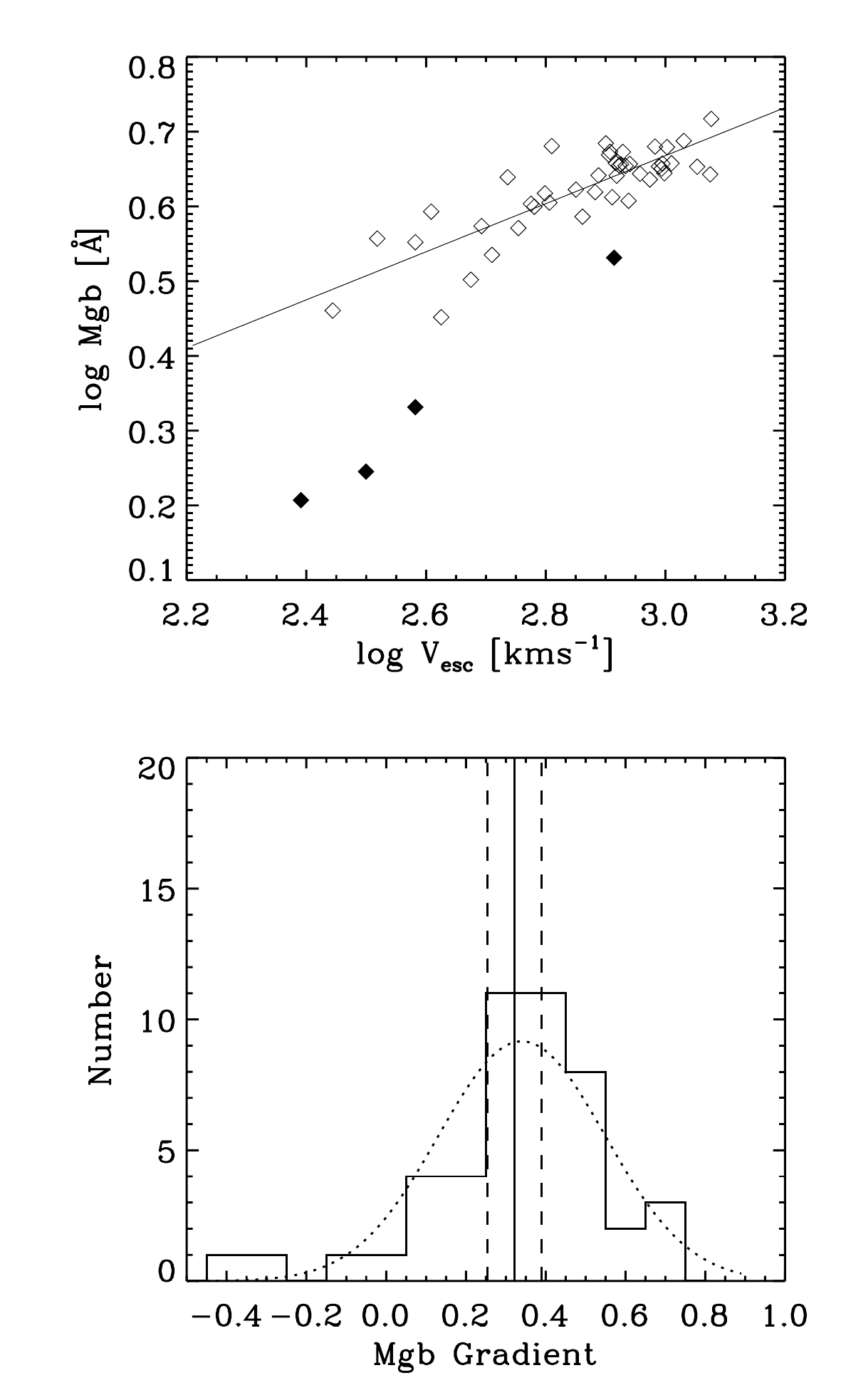}
\caption{ Upper panel: Mgb vs. V$_\mathrm{esc}$ measured within circular Re/8 apertures. The solid line is a linear fit to the data. The solid symbols mark the outlying galaxies which have been excluded from the fit. Lower panel: The gradients of the Mgb-V$_\mathrm{esc}$ relations for each individual galaxy profile, determined by fitting a straight line to the galaxy profiles in the same way as for the global relation. The histogram shows the individual galaxy gradients. The dotted line is a continuous distribution with the same mean, $\sigma$ and total area as the individual gradients. The vertical solid line shows the global gradient determined from fitting to Re/8 values only, with the dashed lines indicating the 2 $\sigma$ error.}
\label{Fig:Vesc_grads}
\end{figure}

Both DSP93 and CD94 looked at the $\mathrm{Mg}_2$-V$_\mathrm{esc}$ relation, rather than Mgb-V$_\mathrm{esc}$ relation. $\mathrm{Mg}_2$ is a broader molecular index but is tightly correlated with the Mgb index \citep{Mg2Mgb}. We convert their Mg$_2$ index values to Mgb using:

\begin{equation}
\mathrm{log\ Mgb}\ =\ 1.57\ \mathrm{Mg}_2 + 0.208
\label{Eq:MgbtoMg2}
\end{equation}

DSP93 and CD94 are also based upon long-slit spectroscopy rather than integral-field data. In order to fairly compare our data to the previous work we re-extract our V$_\mathrm{esc}$, $\sigma$ and Mgb profiles using a rectangular aperture 2.5 arcseconds wide (as used in the DSP93 observations) aligned with the major axis of the SAURON maps, and sampling linearly in distance along the slit from the centre of each galaxy. 

For the three galaxies in common with the DSP93 sample we find reasonable agreement with our Mgb-V$_\mathrm{esc}$ result (see Fig. \ref{Fig:DSP93comp}). While the individual measurements are in broad agreement the slope for our sample is significantly different to that found by DSP93 and CD94. This is largely because we sample a much broader range in $\mathrm{V}_\mathrm{esc}$, approximately twice that in DSP93 and CD94.

\subsection{The gradients within galaxies}
\label{Sec:Grads}
The Mgb-V$_\mathrm{esc}$ relation is particularly interesting, partly because it is the tightest correlation but also because the profiles for individual galaxies follow the global relation remarkably closely. This agrees with the results found by DSP93 and CD94. To better illustrate this important point we performed a linear fit to each of the individual galaxy profiles.  In Fig. \ref{Fig:Vesc_grads} we show the gradients determined by a linear fit for each galaxy, along with the global gradient determined from a linear fit to the Re/8 circular aperture value for all the galaxies. The global gradient is $0.32 \pm 0.03$. The distribution of the individual galaxy gradients has a biweight mean of 0.34 and robust $\sigma$ of 0.2 \citep[see][for a description of robust statistics]{Biweight}. The typical error in the individual galaxy gradients is 0.04. The mean of the individual gradients is consistent with the global gradient within the errors, but the distribution is considerably broader. The additional observed scatter in the individual gradients implies an intrinsic scatter of 0.16. The Fe5015 and H$\beta$ relations behaves quite differently. In the case of Fe5015 the local gradients are typically steeper than the global gradient. The local gradients in Mgb and Fe5015 vs V$_\mathrm{esc}$ appear the same, whereas the global gradient is significantly flatter in the case of Fe5015. Galaxies typically show H$\beta$ to be flat or slightly rising with V$_\mathrm{esc}$, whereas (as mentioned above) the global trend is that H$\beta$ falls with increasing V$_\mathrm{esc}$. This is in line with studies of radial gradients in early-type galaxies \citep[e.g.][]{Mehlert} which find galaxies have typically very uniform H$\beta$ indices and hence characteristic ages.

Taking a 2$\sigma$ cut in Fig. \ref{Fig:Vesc_grads} we note that 3 galaxies have significantly different gradients from the mean: NGCs 3032, 4150 and 4382. These three galaxies also stand out in the Mgb-V$_\mathrm{esc}$ relation, deviating significantly from the mean relation. NGC 3156 also deviates significantly from this relation, and even though its local gradient lies within our 2 $\sigma$ cut it has the 2nd largest error on it's local gradient due to it's U-shaped profile. For this reason we also consider NGC 3156 as an outlier. These four galaxies are labelled in Fig. \ref{Fig:Index_Vesc} with their NGC numbers. Three of these galaxies have the highest values of H$\beta$ in our sample, indicative of recent star formation (see Paper VI). The fourth,  NGC4382, is a peculiar galaxy in both the kinematic and line strength maps, showing a central dip in $\sigma$ and Mgb as well as a disturbed morphology. It also has significantly higher H$\beta$ than galaxies of a similar V$_\mathrm{esc}$, associated with star formation in a central disc. While these galaxies stand out noticeably in Mgb and H$\beta$ they lie much closer to the Fe5015 relation, with only NGC3156 showing a significant deviation.

Three of these outliers also have the lowest values of V$_\mathrm{esc}$ in our sample. It is possible that there is a break in the Mgb-V$_\mathrm{esc}$ relation at these low values of V$_\mathrm{esc}$ but we cannot make a definitive judgement on this, given the limited number of galaxies in our sample that occupy this regime. Either low-V$_\mathrm{esc}$ galaxies are simply the least massive galaxies, expected to have experienced more recent star formation in a downsizing scenario, in which case we might expect them to return to the observed relations, or the Mgb-V$_\mathrm{esc}$ relation breaks down at these low values of V$_\mathrm{esc}$, suggesting different processes determine the stellar population characteristics of galaxies in this regime. A sample with more galaxies in this regime would be required to decide between these two hypotheses.

\begin{figure}
\centering
\includegraphics[width=3.5in]{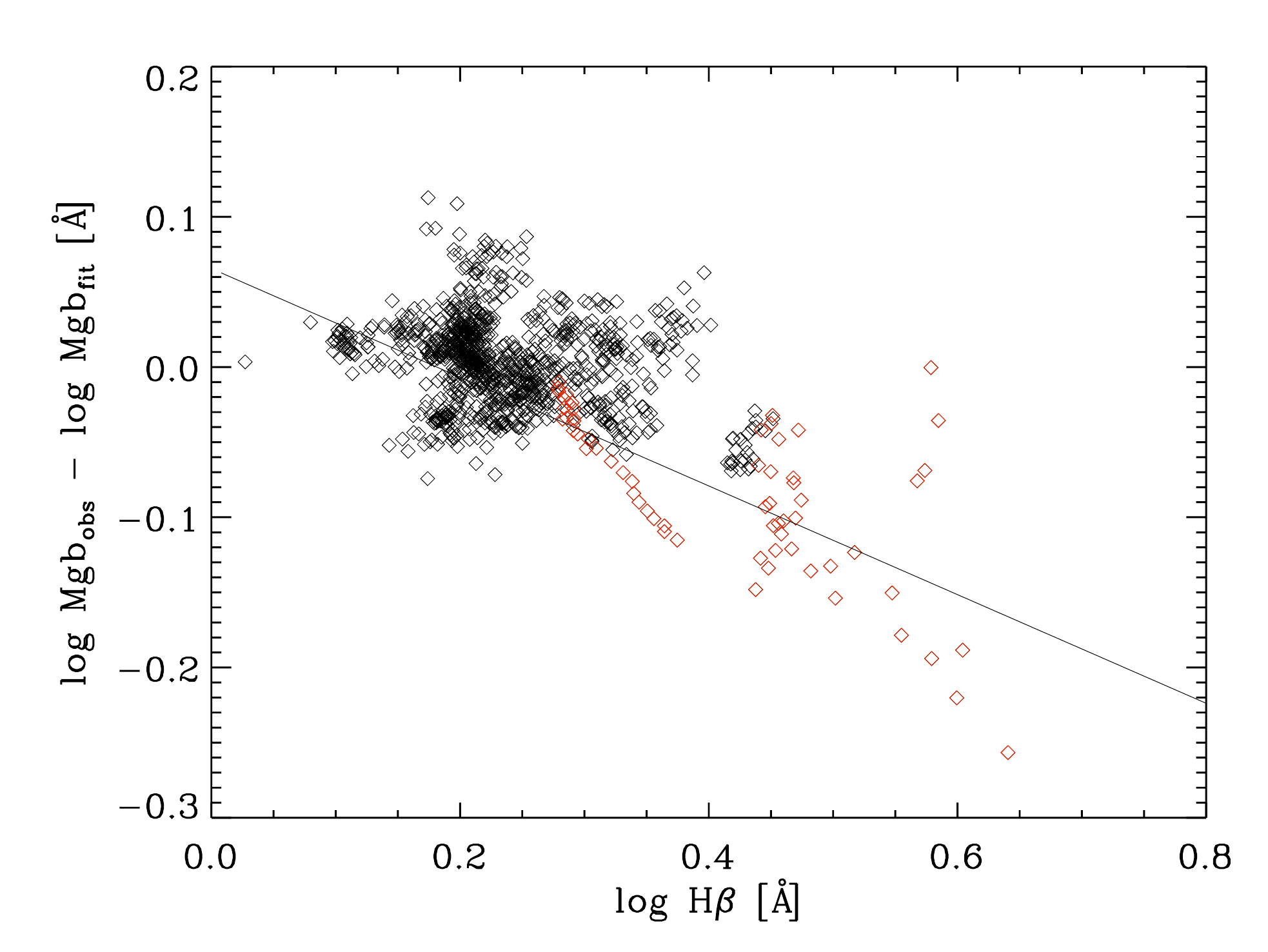}
\caption{The residuals of the Mgb-V$_\mathrm{esc}$ relation plotted against H$\beta$. There is a clear trend with H$\beta$, in the sense that the residuals are larger (in the absolute sense) as H$\beta$ increases. This trend is largely driven by those galaxies that stand out in the Index-V$\mathrm{esc}$ relations, shown as red symbols. The majority of galaxies cluster around a small region centred on zero residual. The straight line is a fit to the red points only.}
\label{Fig:Residuals}
\end{figure}
\begin{figure}
\centering
\includegraphics[width=3.5in]{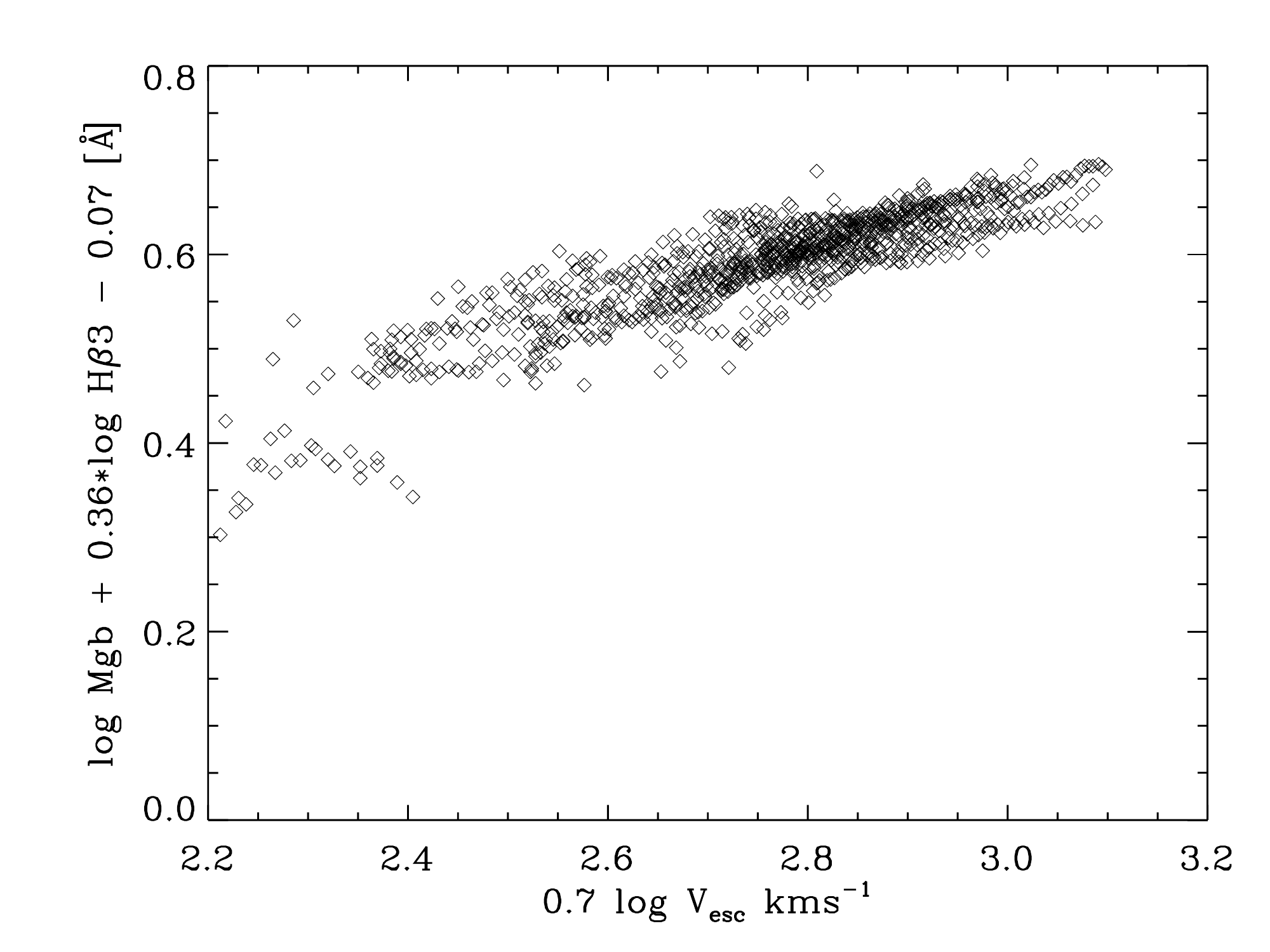}
\caption{The Index-V$_\mathrm{esc}$ relation corrected using the relationship between the Mgb residuals and H$\beta$ shown in Fig. \ref{Fig:Residuals}. The remaining scatter in this relation is consistent with the measurement errors.}
\label{Fig:mgb_corrected}
\end{figure}

\subsection{Accounting for H$\beta$-strong galaxies}
We mentioned above that those galaxies that deviate significantly from the Mgb-V$_\mathrm{esc}$ relation also have high values of H$\beta$. We quantify this in Fig. \ref{Fig:Residuals} by plotting the Mgb residuals, $\Delta$Mgb = Mgb(observed) - Mgb(fitted), against H$\beta$. While most galaxies cluster in a large clump centred on $\Delta$Mgb = 0 there is a significant tail of points with large residuals which appear to correlate with H$\beta$. A linear fit to only those galaxies with high values of H$\beta$ gives the relationship:
\begin{equation}
\Delta\log\mathrm{Mgb} = (-0.36\pm0.05)\log\mathrm{H}\beta + (0.07\pm0.03)
\end{equation}
We can use this relationship to modify our Mgb-V$_\mathrm{esc}$ relation for H$\beta$-strong galaxies. In Fig. \ref{Fig:mgb_corrected} we plot the `corrected' index given by: 
\begin{equation}
\mathrm{Index} = \log \mathrm{Mgb} + 0.36 \log \mathrm{H}\beta - 0.07
\end{equation}
against V$_\mathrm{esc}$ and again perform a linear fit to these data. The resulting fit is given by:
\begin{eqnarray}
\log (\mathrm{V}_{\mathrm{esc}}/500 \mathrm{kms}^{-1}) = &0.16 + 3.57\log(\mathrm{Mgb}/4\mathrm{\AA}) \nonumber \\ & + 1.29 \log(\mathrm{H}\beta/ 1.6\mathrm{\AA})
\end{eqnarray}
We find that this fit has a scatter of only $\sigma =  0.025$ in $\log$ V$_\mathrm{esc}$, reducing the scatter by 22 percent compared to the uncorrected Mgb-V$_\mathrm{esc}$ relation. This scatter is now consistent with the measurement errors.  Even with the four galaxies with strongest H$\beta$ removed the reduction in scatter is significant, $\sigma$ changes from 0.033 to 0.026, a reduction of 20 percent. As a specific example the two galaxies at low V$_\mathrm{esc}$ lying above the Mgb-V$_\mathrm{esc}$ relation fall on the relation after this correction is applied.
\subsection{V$_\mathrm{esc}$ as an alternative to $\sigma$ and $\Sigma_M$}

While $\sigma$ is related to the depth of the potential it is also dependent on the details of the orbital structure of the galaxy; in galaxies with significant rotation or other anisotropy $\sigma$ is a poor tracer of $\Phi$, whereas the true line-of-sight V$_\mathrm{esc}$ is always a reliable measure. As an example to support the idea that V$_\mathrm{esc}$ is a better predictor of local galaxy properties than $\sigma$ we need only look at \citet{Sombrero}. Here the authors study NGC 4594, the Sombrero galaxy, a discy edge-on galaxy. In their Fig 21. they show both Mgb vs $\log \sigma$ and Mgb vs $\log \mathrm{V}_\mathrm{esc}$. They clearly demonstrate that the local values of Mgb are not significantly correlated with $\sigma$ but are tightly correlated with V$_\mathrm{esc}$.

\begin{figure}
\includegraphics[width=3.5in]{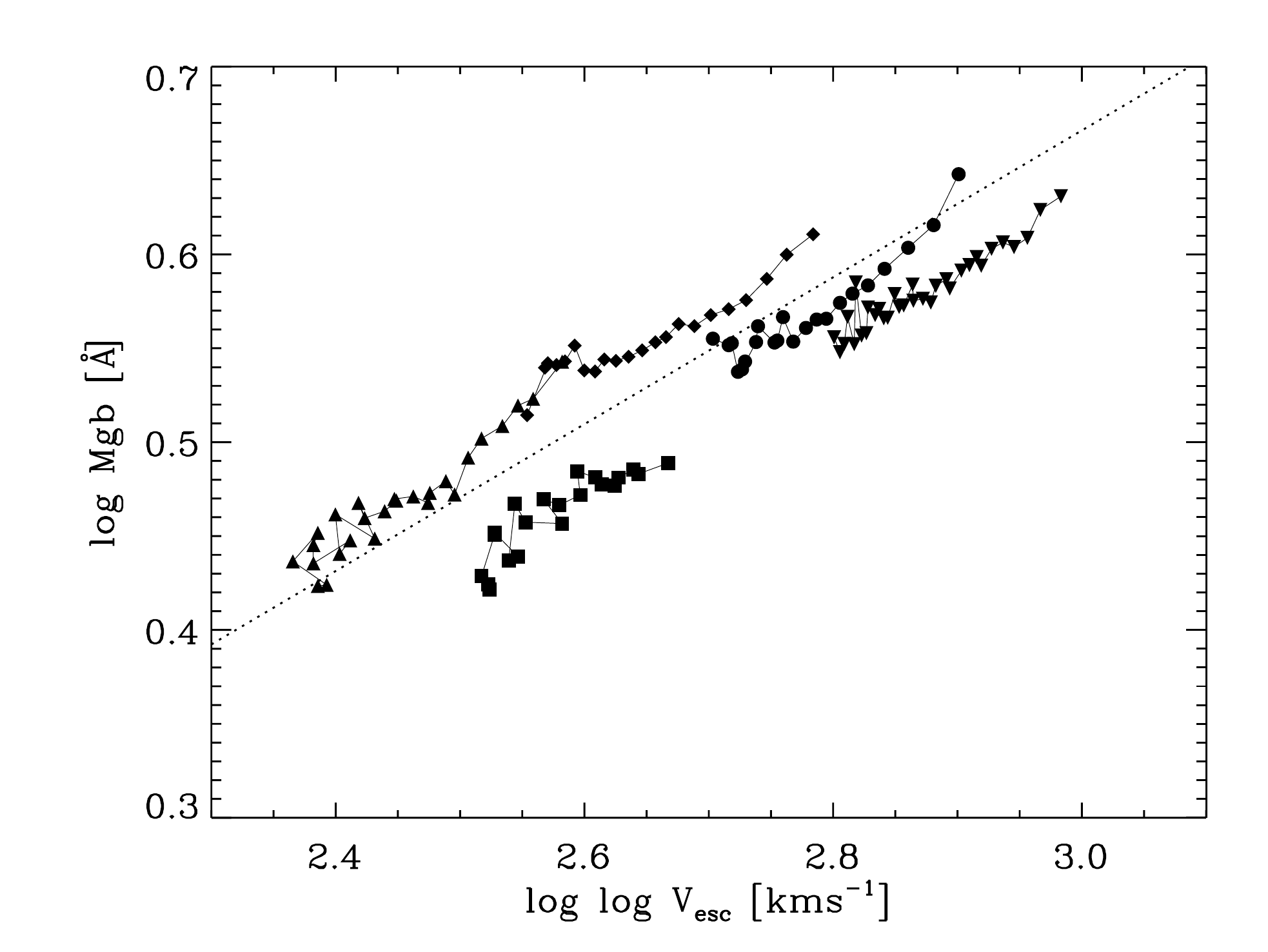}
\includegraphics[width=3.5in]{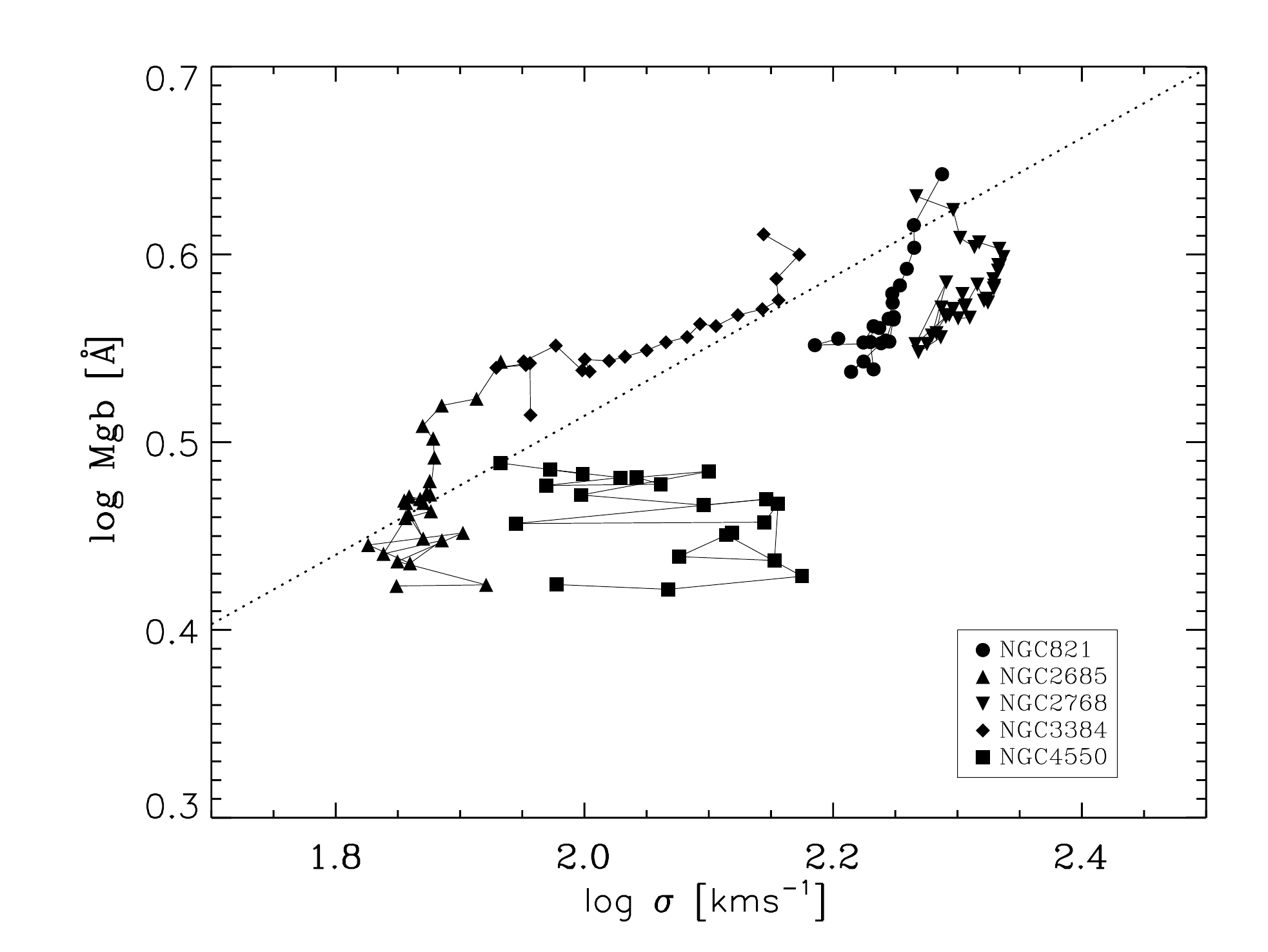}
\caption{Upper figure: Mgb - V$_\mathrm{esc}$ profiles for five galaxies from the sample. The individual galaxy profiles closely follow the global relation (shown as the dotted line). Lower figure: Mgb - $\sigma$ profiles for the same four galaxies. The individual profiles show little resemblance to the global relation (again shown as the dotted line), and in some cases exhibit essentially no trend with $\sigma$ at all. This illustrates the improvement in using V$_{esc}$ instead of $\sigma$ as a tracer of a galaxies dynamical properties.}
\label{Fig:profiles}
\end{figure}

\begin{figure}
\includegraphics[width=3.5in,height=2.75in]{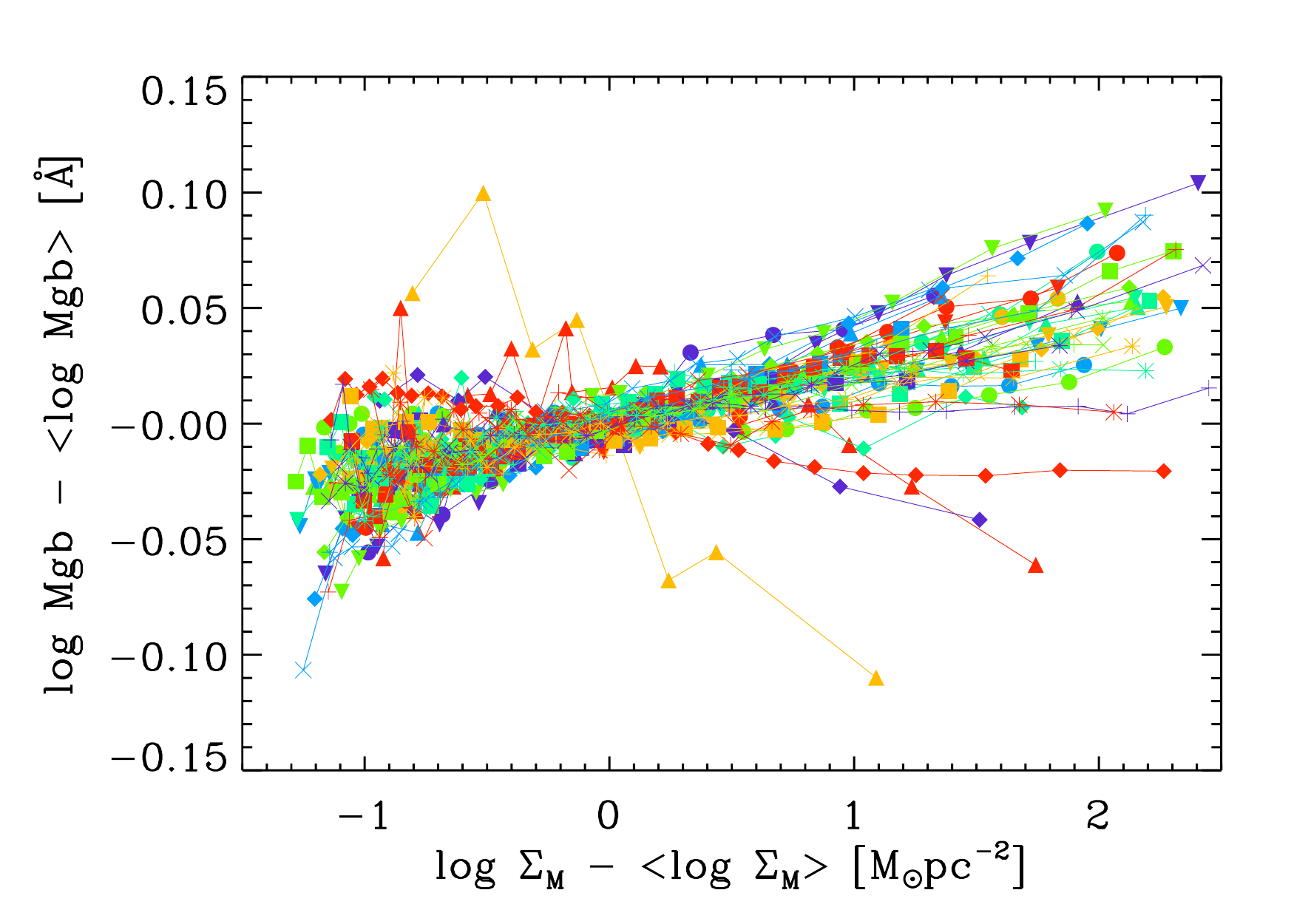}
\includegraphics[width=3.5in,height=2.75in]{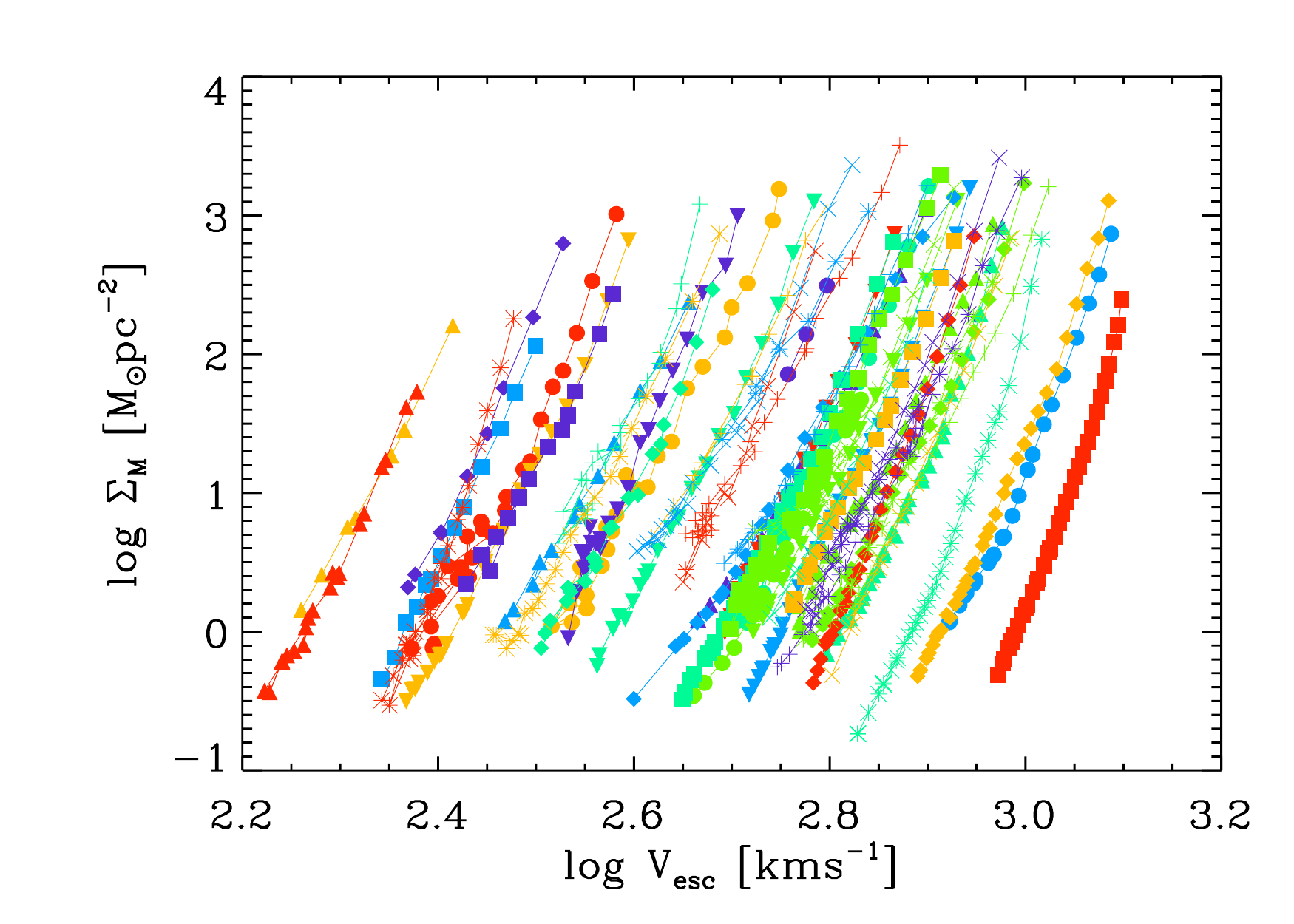}
\caption{Upper panel: the Mgb - $\Sigma_M$ relation for all 48 galaxies. For each galaxy the mean Mgb and mean $\Sigma_M$ of that galaxy has been subtracted from each profile.  The colour and symbol combination is the same as in Fig. \ref{Fig:Index_Vesc}. Galaxies have the same internal gradients but have very different offsets - this relation is local rather than global. The lower panel shows the tight relation between V$_\mathrm{esc}$ and $\Sigma_M$ within a galaxy and the lack of a connection between different galaxies.}
\label{Fig:Mass}
\end{figure}

In our own work we find a similar result. While $\sigma$ is a reasonable predictor for some galaxies it is generally worse than V$_\mathrm{esc}$, and in some cases fails spectacularly to reproduce the line strength trends observed with V$_\mathrm{esc}$. This is particularly true of galaxies with atypical $\sigma$ maps (central dips in $\sigma$, counter-rotating discs etc.) In Fig. \ref{Fig:profiles} we show the Mgb-V$_\mathrm{esc}$ and Mgb-$\sigma$ relations for a selection of galaxies from our sample. As can be clearly seen, while $\sigma$ produces a reasonable trend for some galaxies, in others there is no observable trend whatsoever yet in these cases the trend with V$_\mathrm{esc}$ is still quite obvious. We choose to use V$_\mathrm{esc}$ because it is a direct measure of the potential.

\begin{figure}
\includegraphics{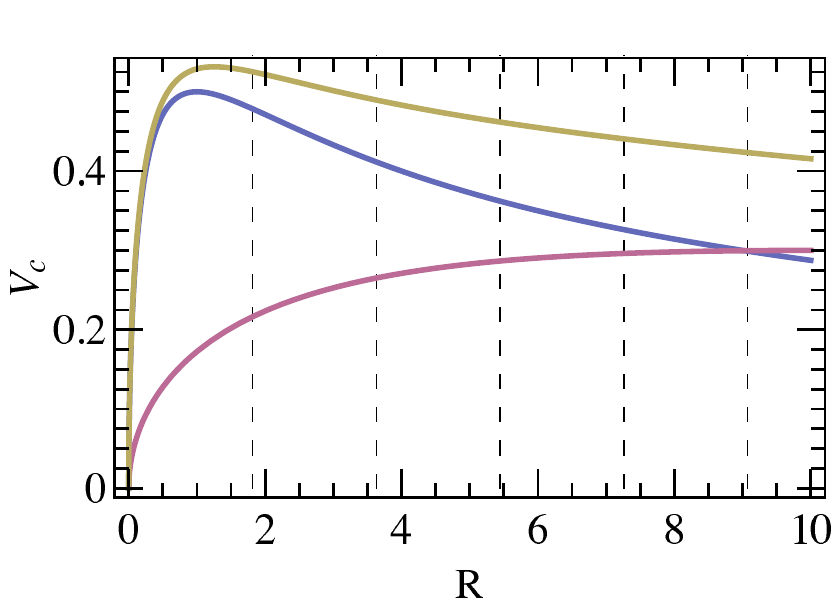}
\includegraphics{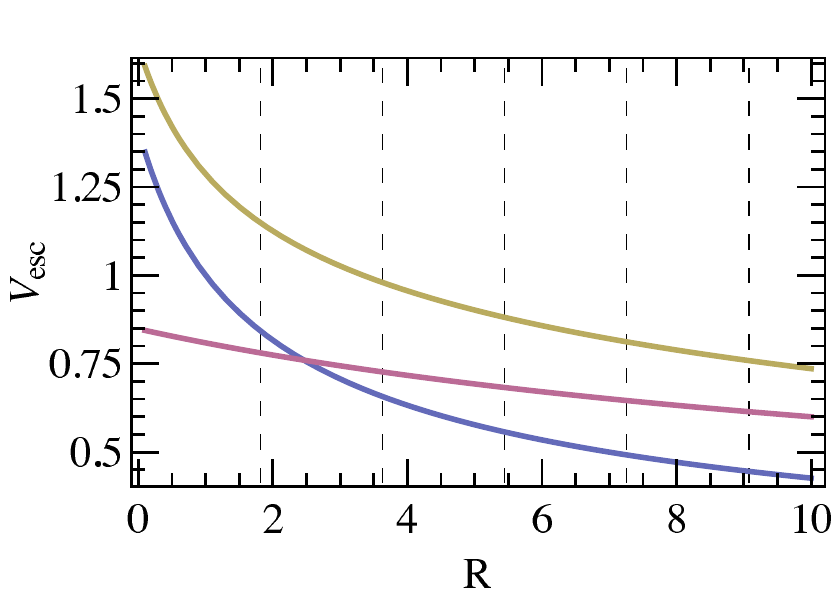}
\caption{Circular velocity (top panel) and escape velocity (bottom panel) as a function of radius. The yellow line indicated the total value while the red and blue lines represent the contributions due to dark and stellar mass respectively. The dashed vertical lines represent the position of 1, 2, 3, 4 and 5 R$_e$. The change in the gradient of V$_\mathrm{esc}$, between total and stellar contribution only, over the region 0-1 R$_e$ is 0.07 dex.}
\label{Fig:DM}
\end{figure}

\begin{figure*}
\includegraphics[width=7in,clip,trim=0 0 100 0]{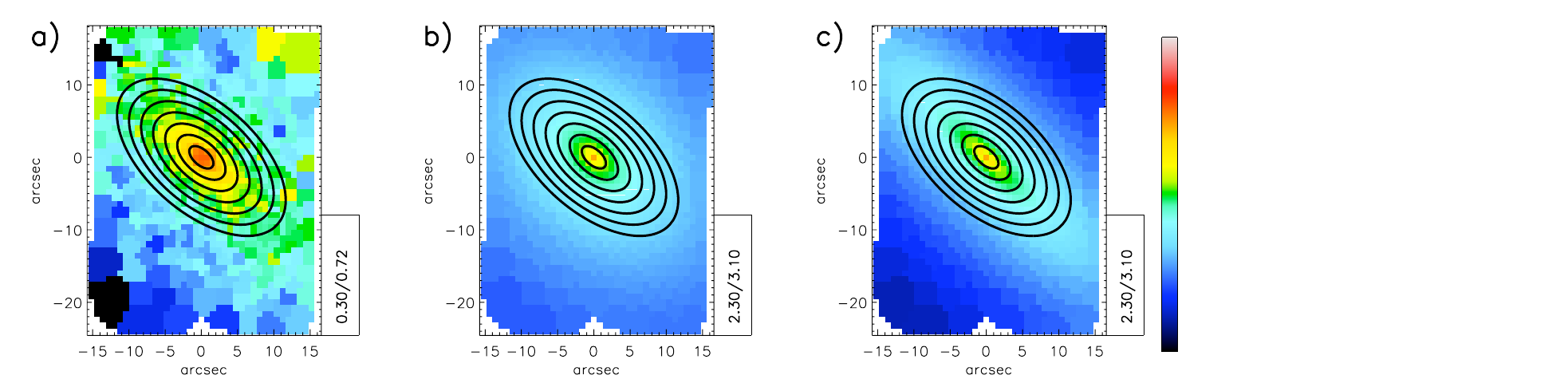}
\caption{Mgb and V$_\mathrm{esc}$ maps for NGC 3377. The solid lines show the ellipses used for extraction of the profiles. Panel a) log Mgb. Panel b) log V$_\mathrm{esc}$ derived from our best-fitting models. Panel c) log V$_\mathrm{esc}$ derived from our thin disc model. As can be seen the isocontours of V$_\mathrm{esc}$ produced from the thin disc model are significantly flatter than the isophotes or the isocontours of Mgb shown in panel a).}.
\label{Fig:edgeon_map}
\end{figure*}

We also investigated whether Mgb correlates with the local surface mass density $\Sigma_M$ (upper panel, Fig. \ref{Fig:Mass}). $\Sigma_M$ was calculated directly from the MGE models and then scaled using the M/L derived from the JAM models. We find that within each galaxy there is a tight relationship between Mgb and $\Sigma_M$ with consistent gradients between galaxies, however, there is no relation between the central $\Sigma_M$ and central Mgb. In this case we find only a local relation; no global relation is apparent. It is interesting to note that this implies a constant gradient for  V$_\mathrm{esc}$ vs $\Sigma_M$, but the offset between galaxies shows no such correlation (lower panel, Fig. \ref{Fig:Mass}). This suggests that $\Sigma_M$ is not physically related to Mgb; the local correlation arising simply because both $\Sigma_M$ and Mgb decrease with galactic radius. Again V$_\mathrm{esc}$ appears to be a more significant parameter because it shows both a local and global correlation.

\subsection{Influence of a dark matter halo}
\label{Sec:DM}
As mentioned above we made the assumption that mass follows light in order to calculate our V$_\mathrm{esc}$. If the dark matter distribution follows that of the stellar mass then this assumption is valid, but in general the dark matter profile may be different. The effect of a dark halo on the local potential was first explored by \citet{FI}. In order to investigate this issue we consider the effect of a dark matter halo on a simple galaxy model. We take the stellar density to be given by a \citet{Hernquist} profile with scale radius $a = 1$ and embed this in a dark halo also represented by a Hernquist profile with $a = 10$:
\begin{equation}
\rho(r) = \frac{M}{2\pi}\frac{a}{r}\frac{1}{(r + a)^3}
\end{equation}
The mass of the dark matter halo was fixed to give a dark matter fraction of 50 percent within 5 R$_e$ which is consistent with the measurements from dynamical studies \citep[see Paper IV,][]{DM1, Thomas} and from lensing \citep{Rusin,DM2}. This is similar to the NFW profile \citep{NFW} in that it has a slope of $ \rho \sim r^{-1}$ for $r < a$, but the Hernquist halo has finite total mass. Recalculating our V$_\mathrm{esc}$ with this new halo we find that the V$_\mathrm{esc}$ gradient decreases by 0.07 dex in the interval 0-1 R$_e$ (see Fig. \ref{Fig:DM}) between the model with a dark halo and that with only a stellar contribution. This shows that with these assumptions the dark matter halo produces a small but detectable change in V$_\mathrm{esc}$. Over the limited radial range covered by our SAURON observations a reasonable halo model produces only a modest change of slope. Over a larger radial range the dark matter halo can significantly change the slope as illustrated for NGC821 and NGC3379 by \citet{Anne-Marie}.

\subsection{Mgb and V$_\mathrm{esc}$ maps}
\label{Sec:Mgb_discs}
While there is a clear correlation between Mgb and V$_\mathrm{esc}$ in both a global and local sense this is not as obvious when comparing the SAURON and V$_\mathrm{esc}$ maps. In 18 of the 48 SAURON galaxies the isocontours of Mgb are more flattened than those of V$_\mathrm{esc}$ (see Paper VI. Note, this mostly applies to the Mgb maps. For Fe5015 the isocontours are typically rounder than the Mgb contours and so the problem is less pronounced if present at all, while for H$\beta$ the maps are essentially flat and so not affected by the choice of aperture.) Moreover, the Mgb maps show some structure whereas the V$_\mathrm{esc}$ maps, which trace the potential, are smooth by construction. A lot of this difference comes down to rms scatter in the observational data which is absent in the modeled V$_\mathrm{esc}$, but the flattening of the Mgb isocontours is a significant effect. In particular some galaxies (for example NGC3377, see Fig. \ref{Fig:edgeon_map}) exhibit a pronounced Mgb disc.

\begin{figure}
\includegraphics[width=3.5in,height=6in]{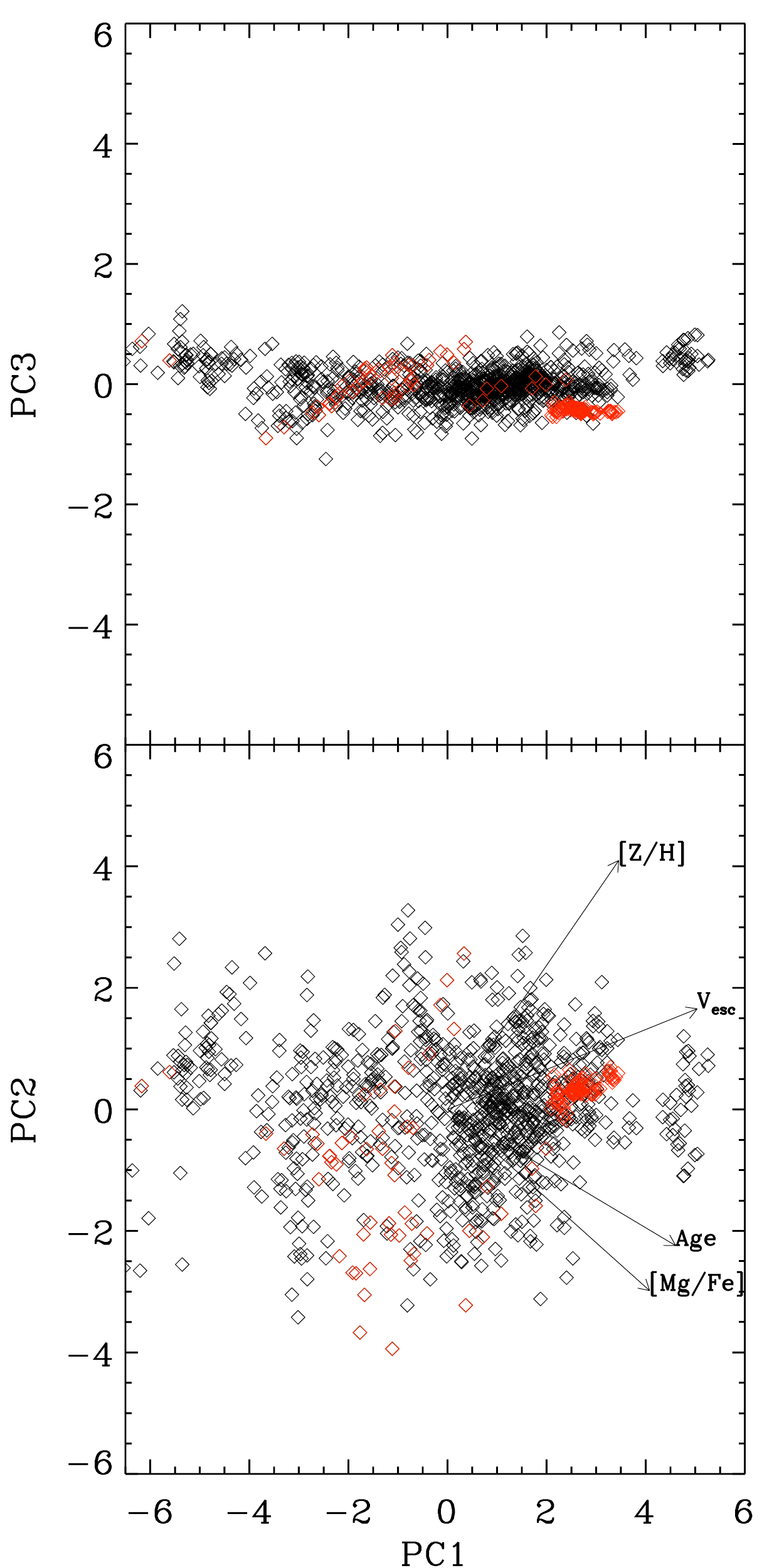}
\caption{Edge-on and face-on views of the hyperplane. The directions of V$_{\mathrm{esc}}$ and the SSP parameters are shown in the face on view. The red symbols on the lower panel show the four galaxies singled-out earlier as lying off the Mgb-V$_{\mathrm{esc}}$ relation. The arrows indicate the directions of the V$_\mathrm{esc}$ and the SSP parameters within the hyperplane.}
\label{Fig:PCA_Plane}
\end{figure}

\begin{table}
\caption{Principle components analysis}
\label{Tab:PCA}
\begin{center}
\begin{tabular}{c c c c c c c}
\hline
& V$_{esc}$' & Age' & [Z/H]' & [$\alpha$/Fe]' & Eigen- &\% of \\
& & & & & value &variance \\
\hline
PC1 & 0.581 & 0.286 & -0.214 & -0.731 & 2.49 & 62\\
PC2 & 0.531 & -0.388 & -0.606 & 0.447 & 1.10 & 28\\
PC3 & 0.399 & 0.708 & 0.275 & 0.514 & 0.30 & 8\\
PC4 & 0.471 & -0.516 & 0.715 & -0.037 & 0.10 & 2\\
\hline
\end{tabular}
\end{center}
Notes: The primed variables are standardised versions of the corresponding variables with zero mean and unit variance. The coefficients of the principal components are scaled to the variance and sensitive to the range of each variable, in the sense that vairables that only vary by a small amount tend to have a large coefficient.
\end{table}

We considered whether the galaxies which exhibit these Mgb disc structures would be better fitted by assuming all the Mgb comes from a thin disc rather than being uniformly distributed for each galaxy. This resulted in only a small change in the determined V$_\mathrm{esc}$ for these galaxies. The Mgb-V$_\mathrm{esc}$ relation derived from the disc-based models has a slightly larger scatter than for our best-fitting models ($\sigma \sim 0.036$ for the disc-based values compared to $\sigma\sim0.032$ for the best-fitting models) but the relation is still a tight one (see Fig. \ref{Fig:Caveats}). We also considered the effect of extracting our Mgb index and V$_\mathrm{esc}$ from the maps using a long slit aperture - again we recover the tight local and global correlation with similar scatter as we found using elliptical apertures.

While a pure-disc model is clearly unrealistic even this extreme assumption does not change our main results. We favour a scenario of a disc-like structure embedded in a spheroid (see Paper VI for further discussion of this idea) to account for the flattened Mgb contours and the structure observed in the Mgb maps. Still, the key issue here is the link between line strengths and the local V$_\mathrm{esc}$, which appears robust against the differing assumptions tested above.

\subsection{V$_\mathrm{esc}$ and single stellar population (SSP) parameters}
While the line strength-V$_\mathrm{esc}$ relations are interesting it is not entirely clear what they tell us about the formation of local early-type galaxies. The measured line strengths are a synthesis of the age, metallicity and chemical abundance distributions of the stellar population. In order to study these more fundamental properties of the stellar populations we transform our line strengths into the physical parameters, age (t), metallicity ([Z/H]), and alpha enhancement ([$\alpha$/Fe]) using the single stellar population models of \citet{Schiavon}. These are not true ages, metallicities and abundances but SSP-equivalent parameters assuming each galaxy formed its stars in a single burst. While this assumption is clearly unrealistic and we should not believe the precise values returned by the model it still allows us to make comparisons between the SSP-equivalent values for our galaxies. 

The models predict the Lick line strength indices for a wide range in age, [Z/H] and [$\alpha$/Fe] based upon accurate stellar parameters from library stars and fitting functions describing the response of the Lick indices to changes in stellar effective temperature, surface gravity and iron abundance. The models produce a grid of age, [Z/H] and [$\alpha$/Fe] iso-contours in the Mgb-Fe5015-H$\beta$ space of our data. For each data point we find the nearest point on the model grids, which gives us the best-fitting age, [Z/H] and [$\alpha$/Fe] and from this we construct age, [Z/H] and [$\alpha$/Fe] maps. These maps are then used to produce the age-, metallicity- and alpha enhancement- V$_\mathrm{esc}$ profiles using the same method used to construct the line strength index-V$_\mathrm{esc}$ profiles. This process, along with a more general discussion of the results is presented in \citet{Sauron_SSP}. Here we confine ourselves to a discussion of the SSP parameters in the context of V$_{\mathrm{esc}}$.

\begin{figure*}
\begin{minipage}{7in}
\includegraphics[width=3.5in]{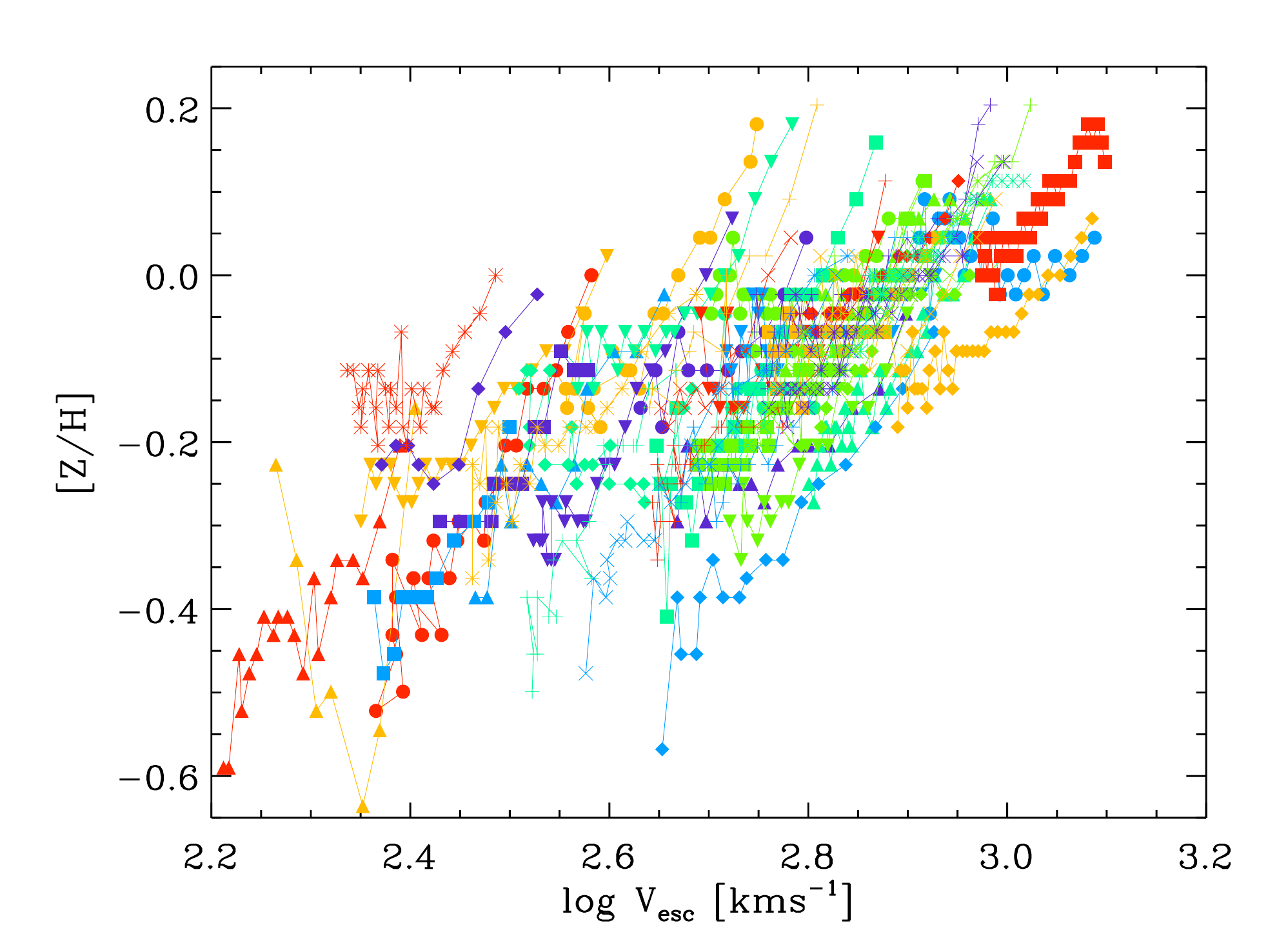}
\includegraphics[width=3.5in]{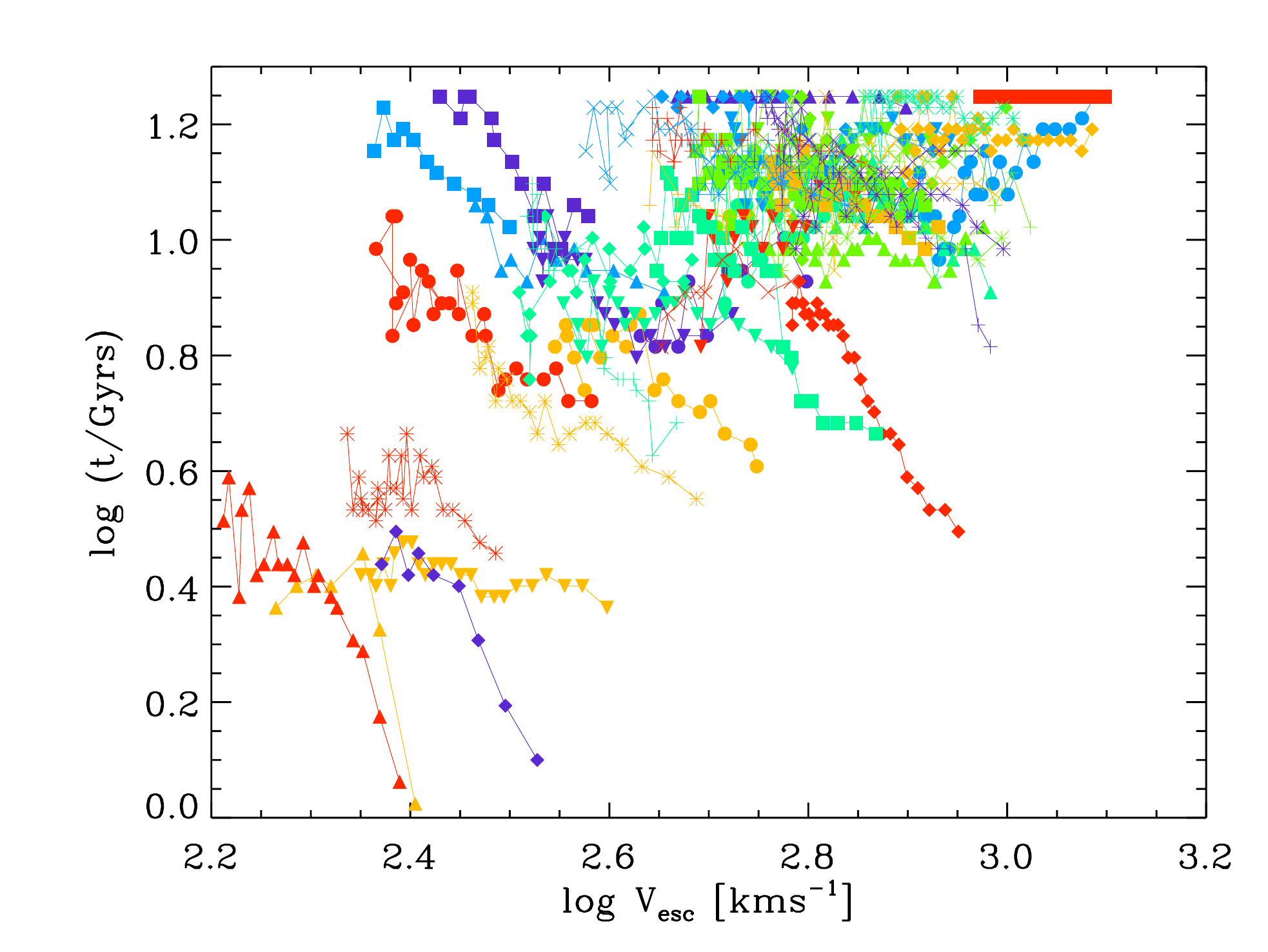}
\includegraphics[width=3.5in]{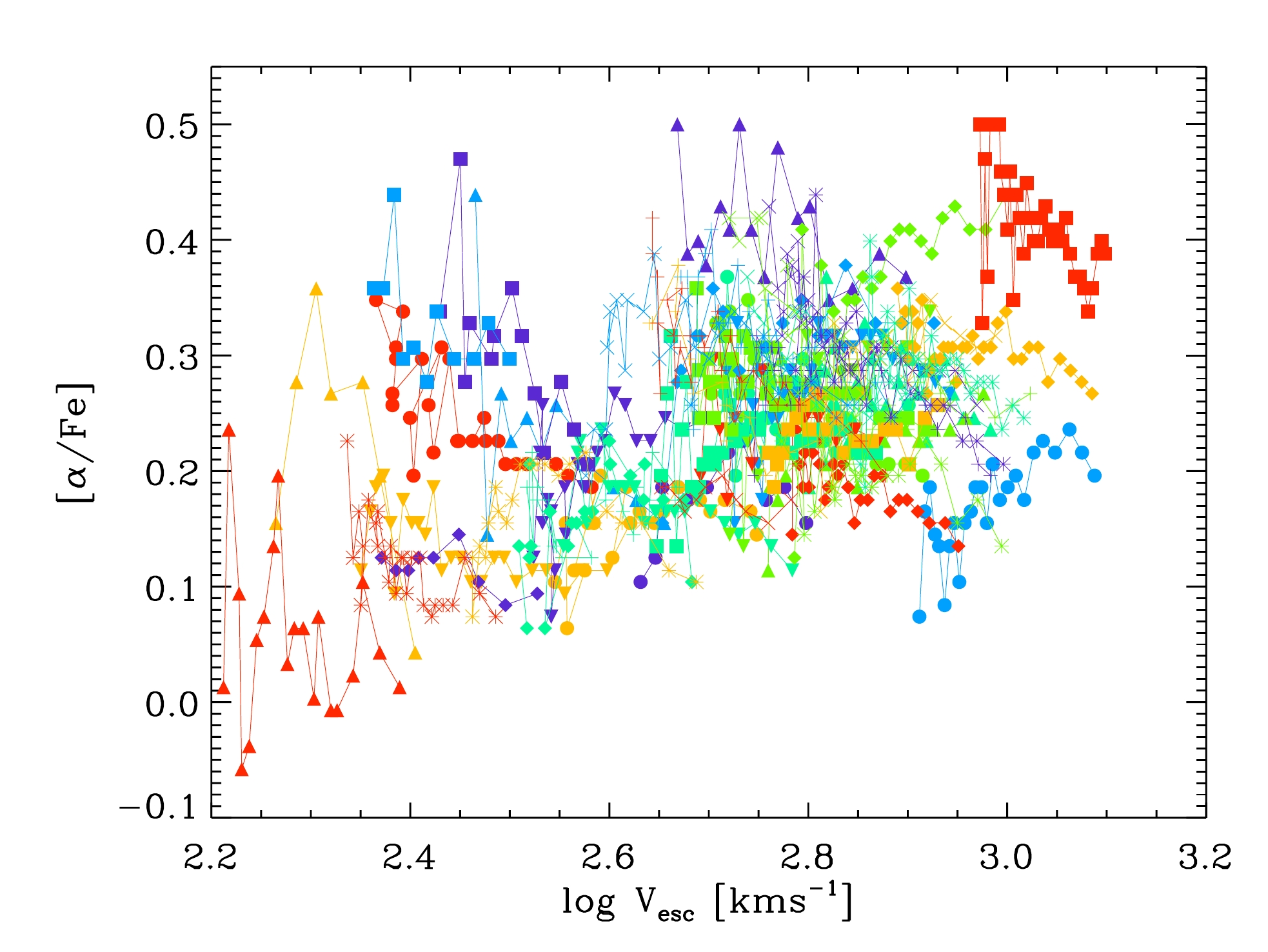}
\includegraphics[width=3.5in]{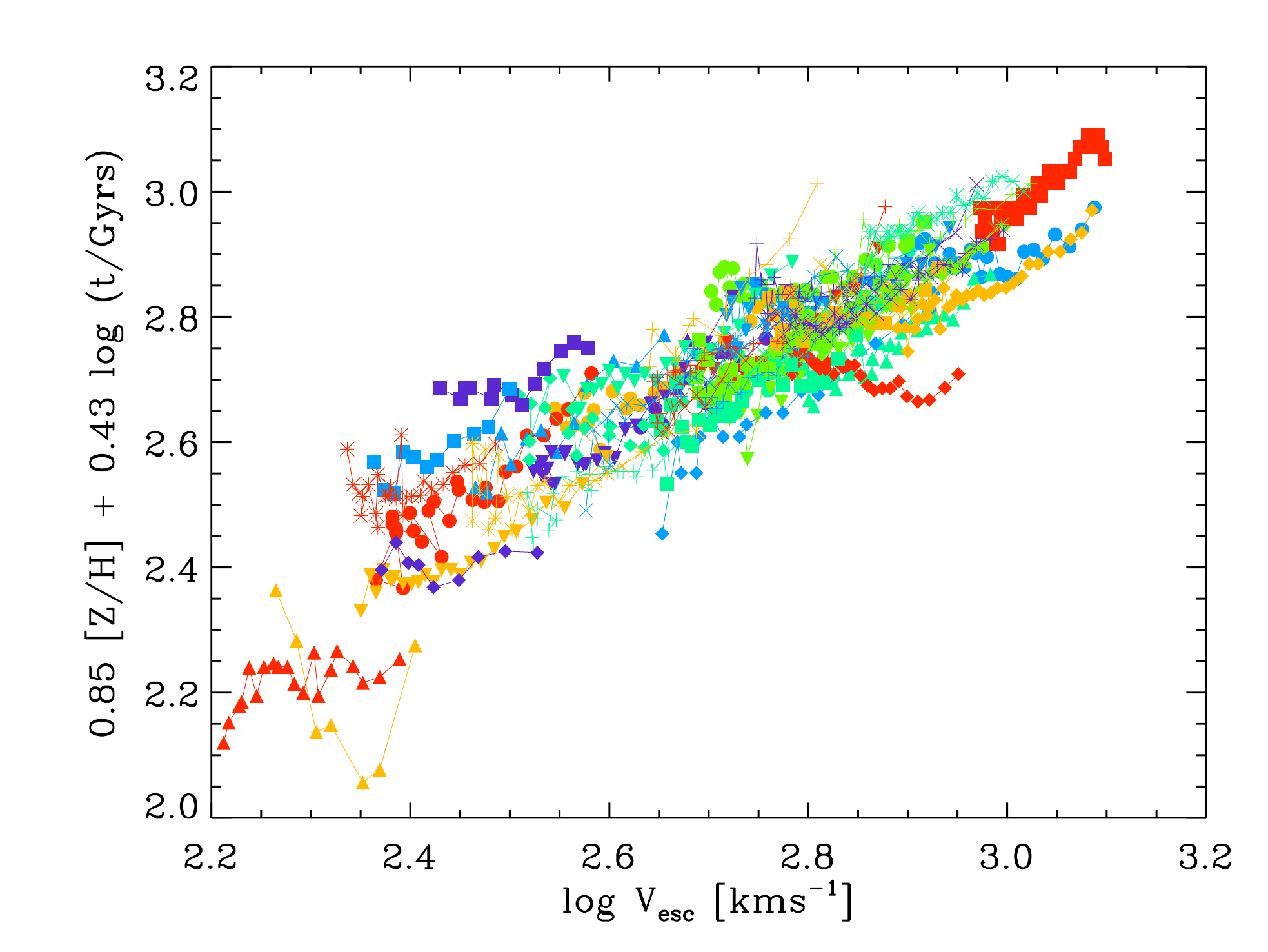}
\caption{Upper panels and bottom left: the V$_\mathrm{esc}$ vs. [Z/H], Age and [$\alpha$/Fe]. There is a strong correlation between [Z/H] and V$_\mathrm{esc}$, though not as tight as with the individual line strength indices. There is a weaker correlation with age and essentially no correlation with [$\alpha$/Fe]. None of the figures exhibit the local and global correlation observed in the Mgb- and Fe5015-V$_\mathrm{esc}$ relations. The sharp cutoff at the top of the upper right and lower left figures is due to the limited range of the SSP model. Lower right panel: Edge-on view of the plane connecting [Z/H], age and V$_\mathrm{esc}$, derived from a linear fit to the three variables. The relationship between the combination of these three variables is much tighter than in the other three panels. Colours and symbols are as described in Fig. \ref{Fig:Index_Vesc}.}
\label{Fig:Vesc_Plane}
\end{minipage}
\end{figure*}

\begin{figure*}
\begin{minipage}{7in}
\includegraphics[width=3.5in]{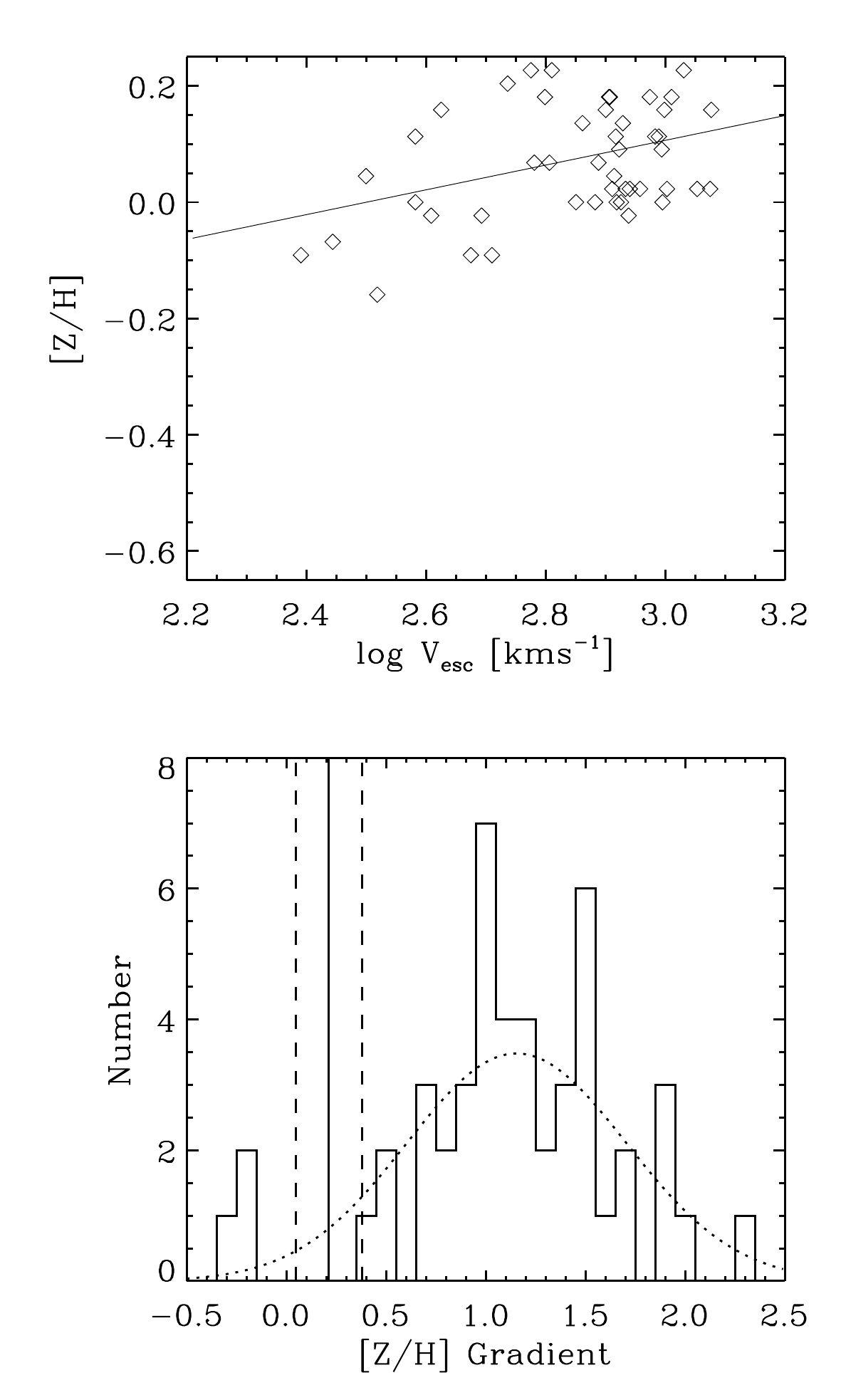}
\includegraphics[width=3.5in]{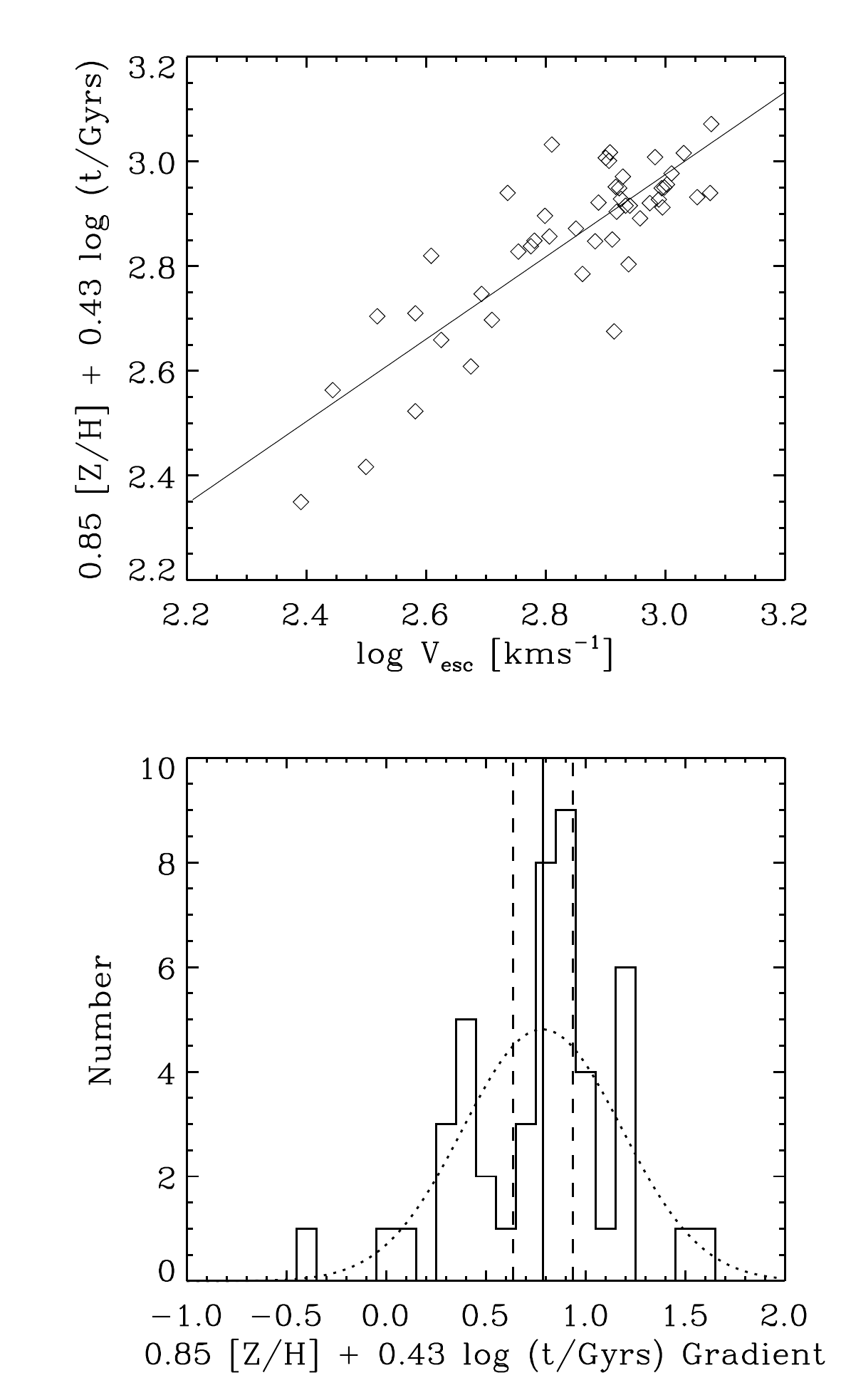}
\caption {Local gradients for [Z/H] and 0.85 [Z/H] + 0.43 $\log t$. The histograms show the individual galaxy gradients determined from a linear fit to each galaxy's profile. The dotted line shows a distribution with the same mean, $\sigma$ and total area as the individual gradients. The vertical solid line shows the global gradient determined from fitting to Re/8 values only, with the dashed lines indicating the 2 $\sigma$ error. As can be seen the local and global gradients are the same for our combination of Z and age but not for Z alone.}
\label{Fig:SSPgrads}
\end{minipage}
\end{figure*}

We search for correlations in this four-dimensional V$_{esc}$, age, [Z/H], [$\alpha$/Fe] space using principal components analysis \citep[PCA; see e.g.][]{PCA1,PCA2} the results of which are shown in Table \ref{Tab:PCA}. As can be seen the first two principal components  account for 90 per cent of the variance. The properties of local ellipticals are therefore confined to a two-dimensional hyperplane, similar to the result found by \citet{Trager2} but for $\sigma$ instead of V$_\mathrm{esc}$. Face-on and edge-on views of this hyperplane are shown in Fig. \ref{Fig:PCA_Plane}. There is a lack of points in the bottom right quadrant of the face-on view of the plane, due to the upper cut-off in age of 18 Gyrs imposed by the SSP model. We checked this result by using the models of \citet{TMB03} to calculate the SSP parameters of our sample and while the precise values in Table \ref{Tab:PCA} change by $\sim 5$ percent the conclusion that galaxies are confined to a hyperplane is independent of the SSP model used.

Assuming the hyperplane to be infinitely thin (i.e. the contributions from PC3 and PC4 are zero) then we can express two of our variables in terms of the other two variables. The choice of dependent and independent variables is entirely arbitrary, but in the interests of physical insight we choose [Z/H] and age as our two independent variables and seek to express V$_{esc}$  in terms of them. Performing a linear fit to the three-dimensional age, [Z/H], V$_\mathrm{esc}$ space we find that the variables are related by:
\begin{equation}
\log\left(\frac{\mathrm{V}_{\mathrm{esc}}}{500\mathrm{kms}^{-1}}\right)= 0.85 \left[\frac{\mathrm{Z}}{\mathrm{H}}\right] + 0.43 \log \left(\frac{\mathrm{t}}{\mathrm{Gyrs}}\right) - 0.29 \label{Eq:SSP}
\end{equation}
It is important to bear in mind that these are not true ages and metallicities but SSP-equivalent values. This combination of variables is shown in the lower panel of Fig. \ref{Fig:Vesc_Plane}. The scatter in this relation is greatly reduced from that of any relation between just two of the four variables we are considering here as shown in Fig. \ref{Fig:Vesc_Plane}. In Fig. \ref{Fig:SSPgrads} we show that the local gradients within a galaxy again follow the global gradient, though this result is not as tight as the local-and-global relation for Mgb-V$_\mathrm{esc}$. The global gradient, determined from fitting to Re/8 values is $0.79 \pm 0.08$. The mean of the individual gradients is 0.78, with the width of the distribution given by a $\sigma$ of 0.40. The typical error on the individual gradients is 0.11. The global gradient is consistent with the local gradient well within the errors. The local gradients again show a broader distribution, implying an intrinsic scatter of 32 per cent. This is not the case for [Z/H] alone; here the local gradients are significantly steeper than the global one. The specific combination of age and [Z/H] depends on the SSP model used, but the tightness of the plane and the local and global connection do not. 

\section{Discussion}
\label{Sec:Discussion}
\subsection{Caveats}
We have already mentioned a few caveats to consider when analysing our results. While these points have been discussed more fully elsewhere in the text we summarise them here in the interest of showing that none of these issues threaten our conclusions. The four principal caveats in this work are: triaxiality, inclination, dark matter and the shape of the Mgb isophotes. The first three of these affect our determination of the V$_\mathrm{esc}$ of our galaxies whereas the fourth affects the extraction of our Mgb-V$_\mathrm{esc}$ profiles.

\begin{figure*}
\begin{minipage}{7in}
\centering
\includegraphics[height=2.5in]{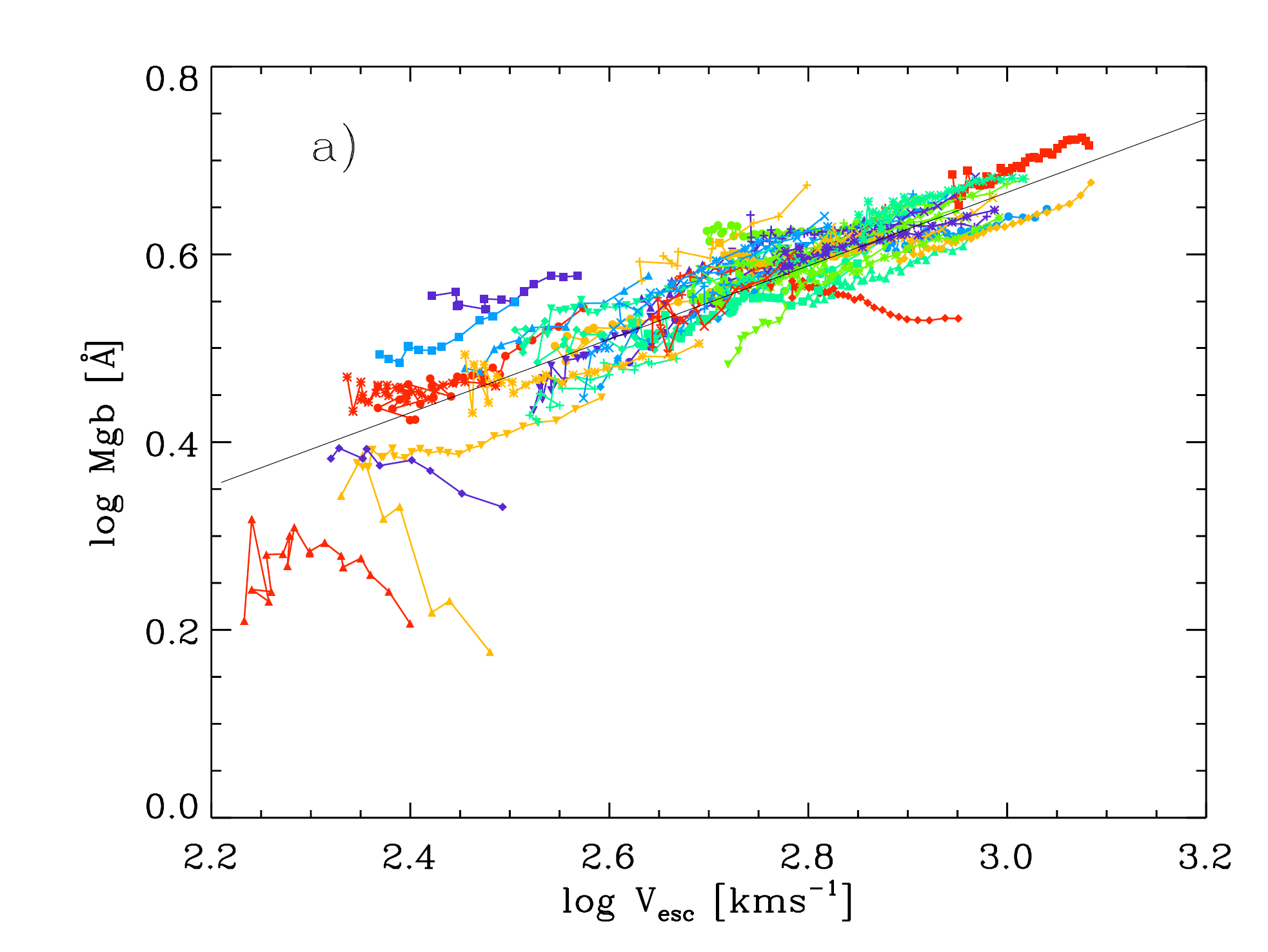}
\includegraphics[height=2.5in]{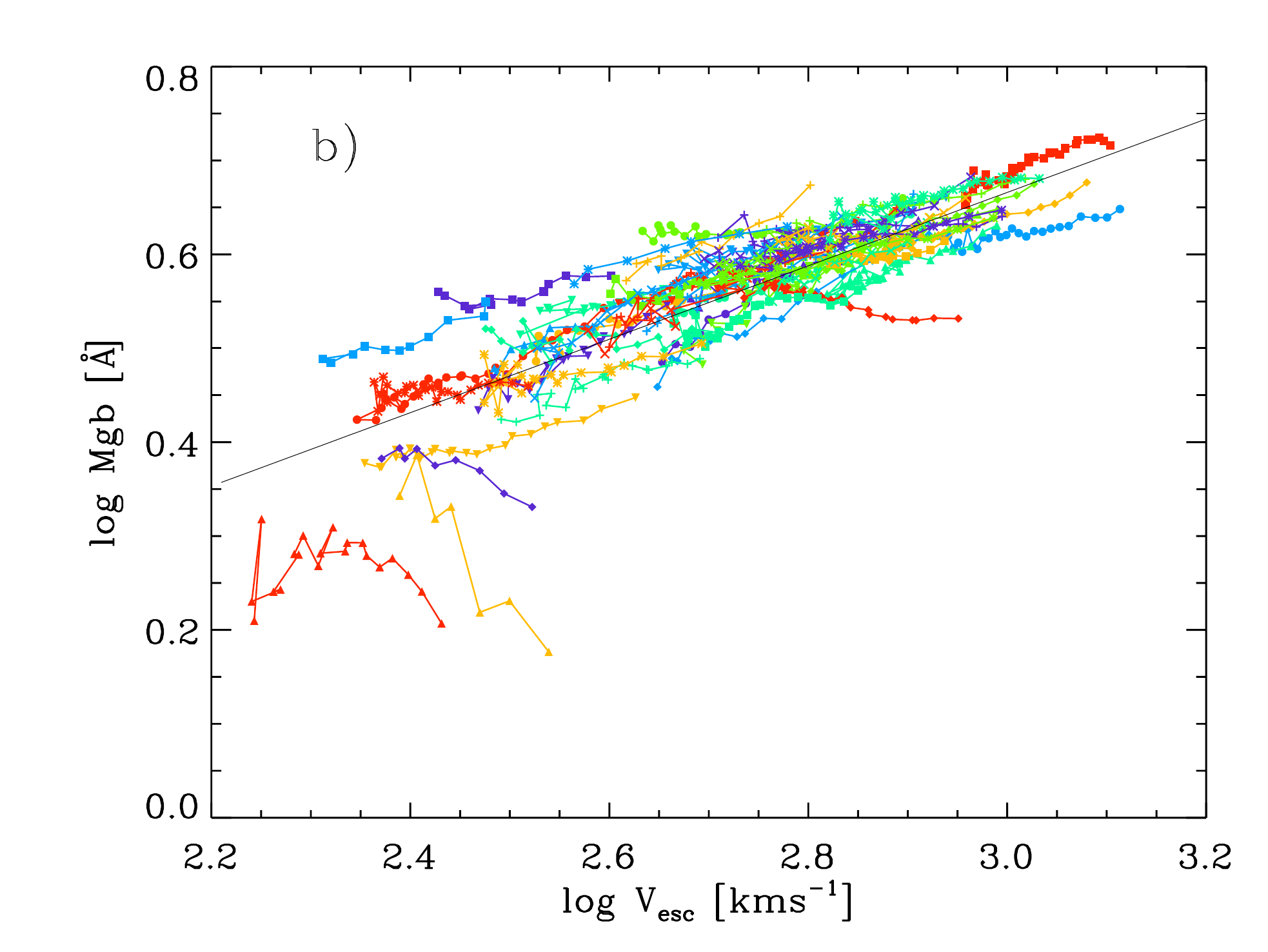}
\caption{The Mgb-V$_\mathrm{esc}$ relation derived under three different assumptions. Panel a) shows the relation derived assuming all our galaxies are seen with an inclination $i=90^{\circ}$. Panel b) shows the results of assuming that all the absorption arises from the equatorial plane of our galaxies, i.e. in a disc. The solid line in both panels is taken from Table \ref{Tab:Vesc1}, using the complete sample. As can be seen when comparing this figure with the top panel of Fig. \ref{Fig:Index_Vesc} the assumptions made in deriving panels a) and b) do not significantly affect our results.}
\label{Fig:Caveats}
\end{minipage}
\end{figure*}

\newcounter{Lcount}
\begin{list}{\bf \roman{Lcount})}
{\usecounter{Lcount}}
\item \textbf{Bars and triaxiality:} Several of our galaxies are triaxial or barred systems. While triaxial objects can have very different orbit families to axisymmetric systems the distribution of the matter and hence $\Phi$ will not be significantly different. Furthermore triaxial systems tend to be rounder so the deviations in shape are typically small. The M/L is also relatively robust against the assumption of axisymmetry, which is expected due to the Virial Theorem and the tight scaling relations followed by fast and slow rotators, e.g. the Fundamental Plane \citep{DD}. Because of this we are able to produce reasonable values for the second moments and velocity fields and hence the V$_\mathrm{esc}$ of triaxial or barred objects under the assumption of axisymmetry. Therefore we do not expect that more detailed triaxial modelling \citep{Lorenzi,KDC}  will significantly alter our conclusions. There is no systematic dependence of the residuals from the Index-V$_\mathrm{esc}$ relations on bar strength (determined qualitatively by eye for our sample) and that our `clean' axisymmetric sample discussed in Section \ref{Sec:Results} shows no significant improvements in the tightness of the relations. 
\item \textbf{Dark matter:} Our determination of V$_\mathrm{esc}$ is based on modelling of the photometry and so we are assuming that mass follows light, while allowing for a constant dark matter fraction. We note that dark matter makes up only a small fraction of the total density in the central regions and hence doesn't significantly affect the potential in the region we are studying. There is much observational evidence from dynamical studies and from lensing to support this view. In Section \ref{Sec:DM} we consider the effect of a dark halo and note that while the local gradients in V$_\mathrm{esc}$ do change the effect is modest over the SAURON field of view.

\item \textbf{Inclination: }For many of our galaxies it was possible to estimate the inclination from methods other than our \textsc{JAM} modelling (6 with embedded discs, 16 edge-on objects) and in these cases our \textsc{JAM} inclinations fall within the errors on our independent estimates. For our other galaxies, while we expect that our inclination estimates are accurate in most cases \citep[see also][]{newJeans} we also show in Fig. \ref{Fig:Caveats} that our V$_\mathrm{esc}$ values do not depend strongly on inclination. In panel a) of this figure we show the Mgb-V$_\mathrm{esc}$ relation for our galaxies under the assumption that they are all edge-on.  As can be seen there is little difference between this panel and the top panel of Fig. \ref{Fig:Index_Vesc} which shows our Mgb-V$_\mathrm{esc}$ relation using our best estimates for the inclinations. 
\item \textbf{Mgb discs:} Finally, as noted in Paper VI, the Mgb isocontours do not always follow the isophotes which we are using to extract our Mgb-V$_\mathrm{esc}$ profiles. We investigated this issue by considering the idea that our Mgb absorption comes entirely from a disc and re-calculated our V$_\mathrm{esc}$ based on this assumption. In panel b) of Fig. \ref{Fig:Caveats} we show the results of this, again, there is little difference between panel b) and the upper panel in Fig. \ref{Fig:Index_Vesc}, though the scatter is slightly larger. We also investigated the effect of using different apertures to extract V$_\mathrm{esc}$ and Index profiles from the maps by varying the ellipticity of our apertures. While the scatter increased slightly when circular apertures were used the relations did not significantly change.
\end{list}  

\subsection{The line strength-V$_\mathrm{esc}$ relations}
It is clear from the tightness of the correlations shown in Fig. \ref{Fig:Index_Vesc} that the stellar populations of early-type galaxies are closely linked with the depth of the local potential they reside in (characterised in this study by V$_\mathrm{esc}$). That this should be the case is by no means obvious. While we might expect that the formation of a star is influenced by the potential that it forms in it is perfectly possible for that potential to have changed significantly between the star's formation and the present day. We expect that the availability of gas and it's ability to cool will also play a role. The tight correlation observed is more easily accommodated in a monolithic collapse scenario for star formation in which the gravitational potential $\Phi$ is largely unchanged, but it is clear that galaxies do not form in this way - we need to address how mergers fit into this picture.


While in the monolithic collapse scenario the potential does not change the same is not true of mergers; in this case the potential can be significantly altered as more mass is added to the galaxy and the distribution of that mass can also be changed. In this case it is clear that the potential a star forms in and the potential we observe it in several billions years later are different. It has been known for some time \citep{White,Barnes} that during a merger the stars are preserved in their rank-order of binding energy, in the sense that the most deeply bound stars before the merger are also most deeply bound after the merger. This suggests a possible link between the potential a star formed in and the potential it finds itself in after the merger. More recent work by \citet{Hopkins_P} has argued that in both wet and dry mergers the radial gradients of metallicity are preserved, again suggesting that the tightness of the observed line strength-V$_\mathrm{esc}$ relationships can be compatible with hierarchical merging. However it is the detail that the {\it local and global} relations are the same that is our key result and more detailed modelling is required before we can properly compare model predictions to our result.

Four galaxies in our sample show signs of recent star formation, and these galaxies are significant outliers in the Mgb- and H$\beta$-V$_\mathrm{esc}$ relations. But, when we convert our line strength indices into SSP-equivalent parameters we find that these four galaxies are confined to the same region in the four-dimensional parameter space of V$_\mathrm{esc}$, t, [Z/H] and [$\alpha$/Fe] as those galaxies which lie on the Index-V$_\mathrm{esc}$ relations. As these objects all have unusually high H$\beta$ they are all likely to be relatively young objects. This suggests that even recently disrupted objects have some regular properties connected with the potential $\Phi$ that survive whatever process moved the galaxy off the Index-V$_\mathrm{esc}$ relation. It seems likely that as these objects age they will return to the Index-V$_\mathrm{esc}$ relations. 


\begin{figure}
\includegraphics[width=3.5in]{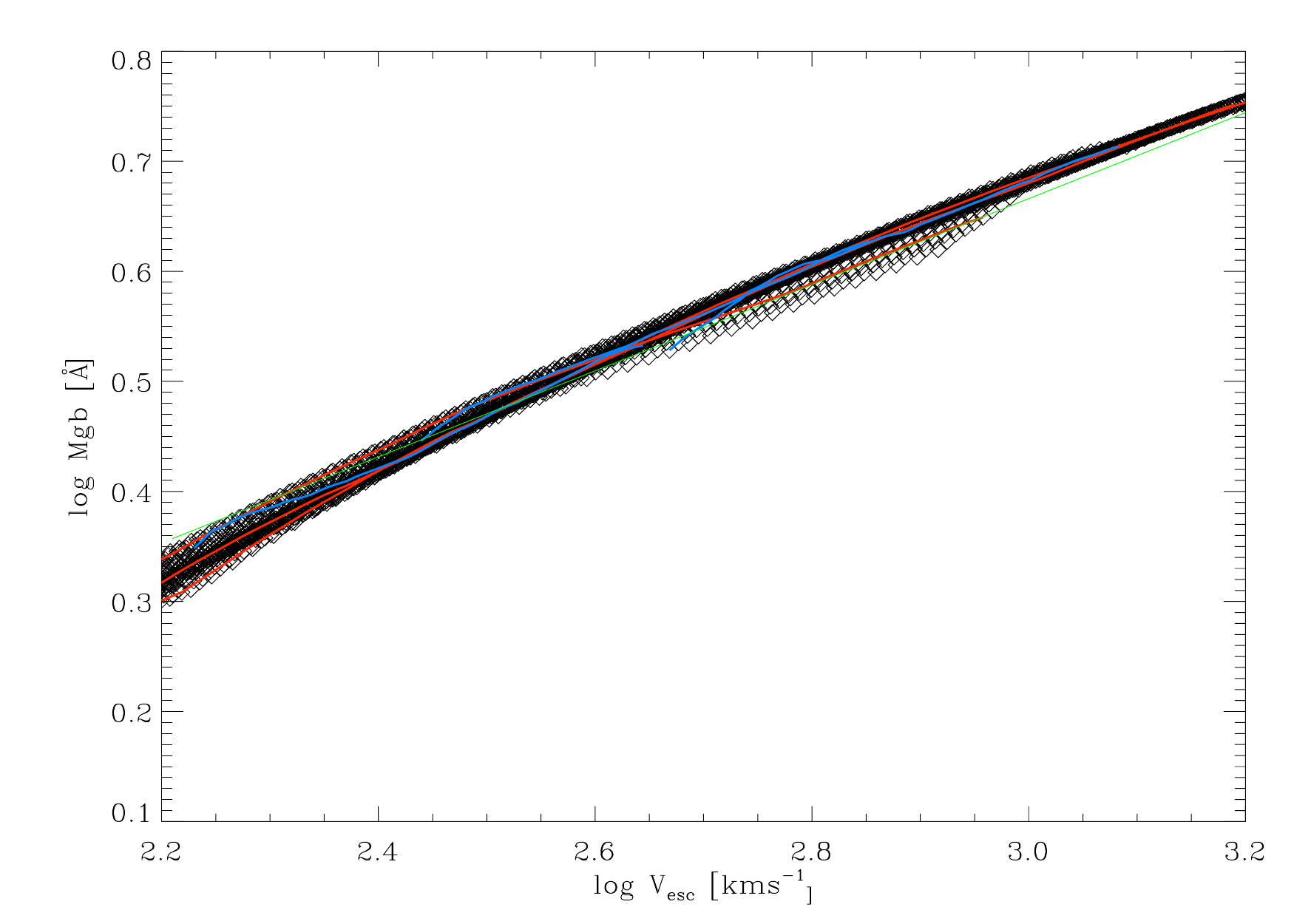}
\includegraphics[width=3.5in]{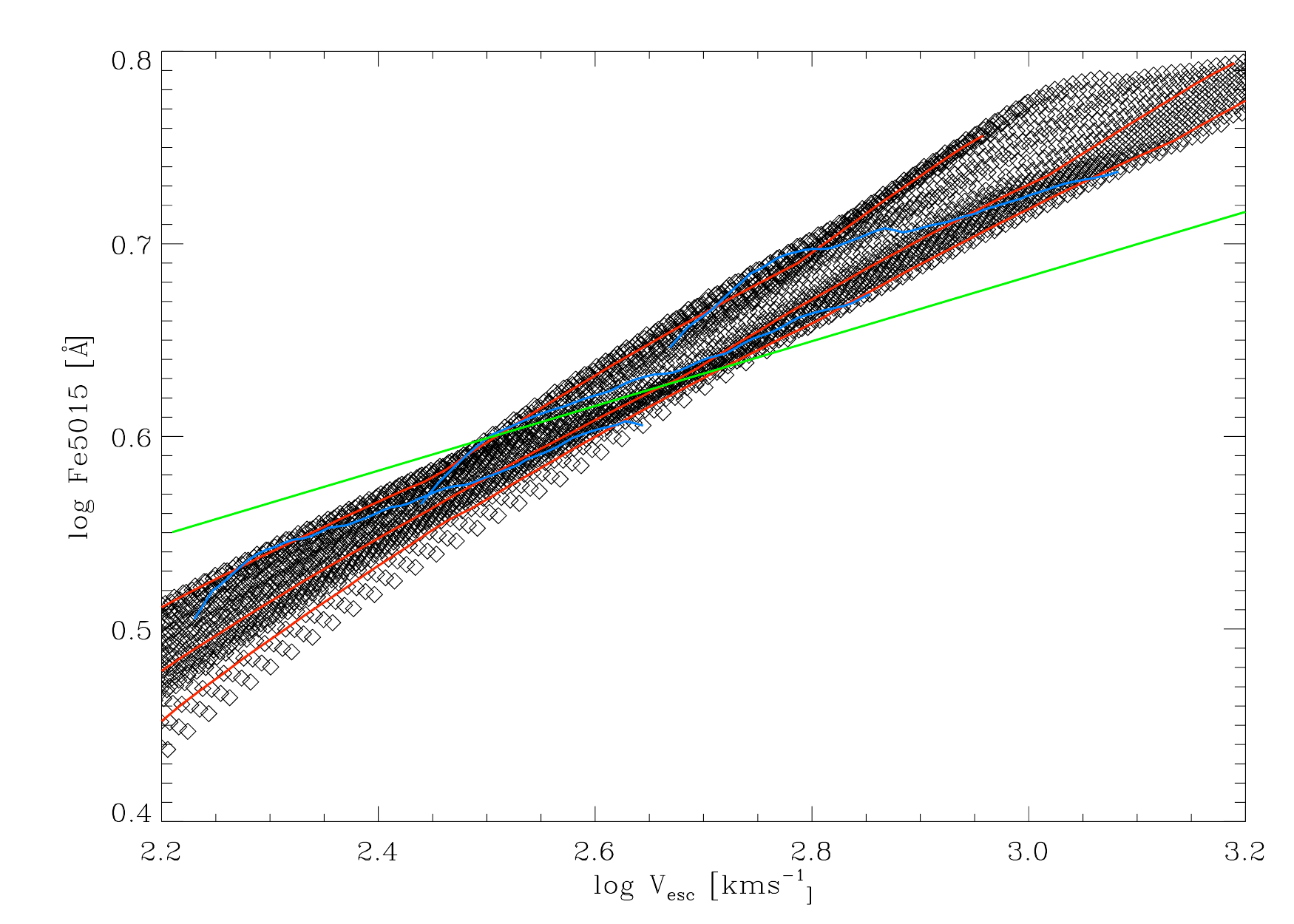}
\includegraphics[width=3.5in]{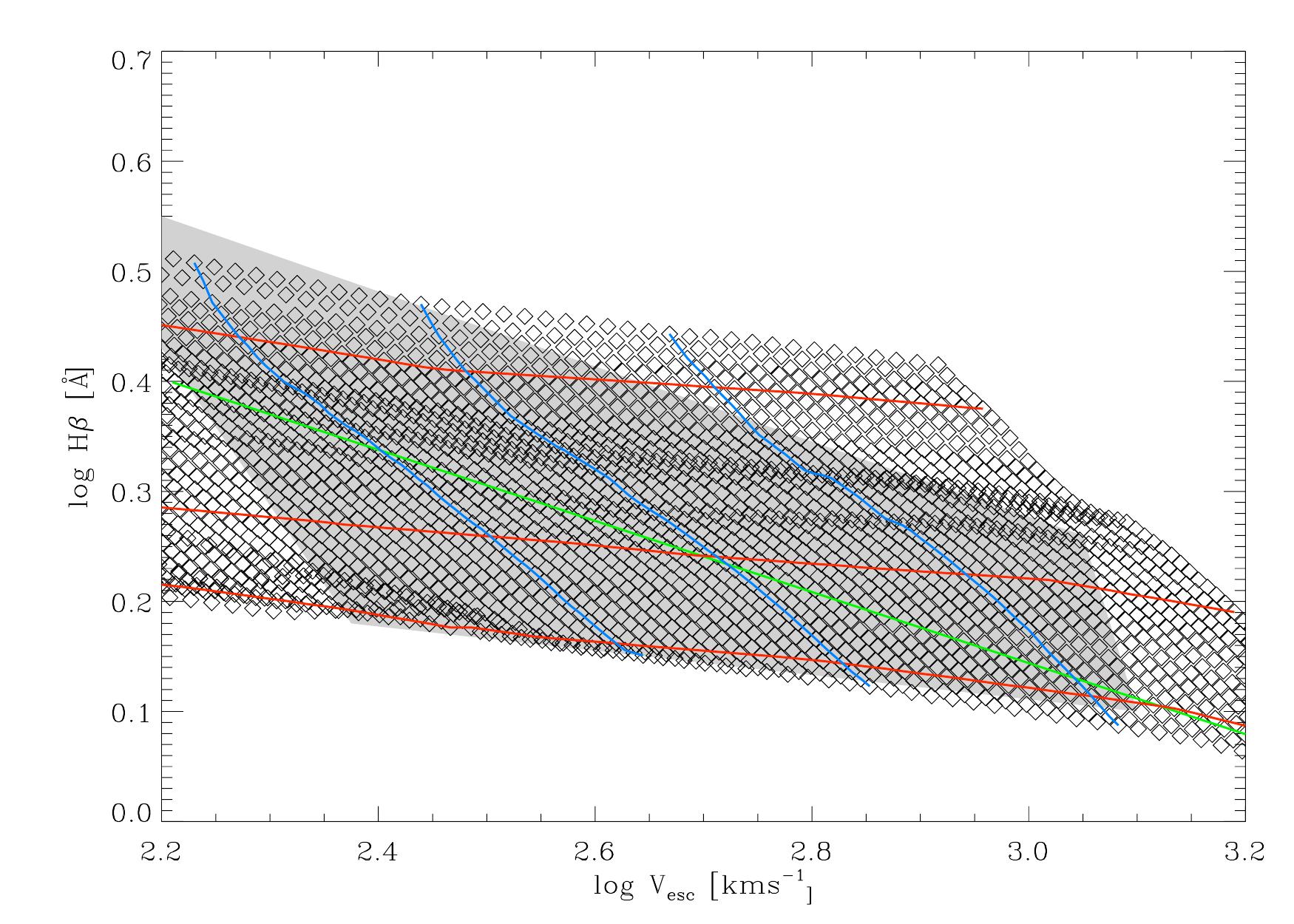}
\caption{The line strength - V$_\mathrm{esc}$ relations predicted by re-inverting the SSP model of \citet{Schiavon} and applying the constraint of Equation \ref{Eq:SSP} at constant [$\alpha$/Fe/] = 0.33 and minimum age, t $>$ 2 Gyrs. The grid of points show the predictions of the SSP model. The red and blue lines show lines of constant age and [Z/H] respectively. The green line is the best fit to the observations, given in Table \ref{Tab:Vesc1}. In the lower panel the grey shaded area shows the region occupied by the observations. With these constraints the SSP model is able to reproduce the observed Mgb and H$\beta$-V$_\mathrm{esc}$ relations but the condition of fixed [$\alpha$/Fe] is required to reproduce the Fe5015-V$_\mathrm{esc}$ relation.}
\label{Fig:SSP_inverted}
\end{figure}

It is interesting to ask why we observe a tight local and global relation in the Mgb-V$_\mathrm{esc}$ relation but not in the case of the other two indices. In order to investigate this issue we begin with our SSP hyperplane equation, Equation \ref{Eq:SSP}. We re-invert the SSP model grid of \citet{Schiavon} and consider the constraints on the line strength-V$_\mathrm{esc}$ relations implied by Equation \ref{Eq:SSP}, using a constant [$\alpha$/Fe] of 0.33, the mean value for galaxies in our sample. We also impose a minimum age of t $>$ 2 Gyrs.  This is shown in Fig. \ref{Fig:SSP_inverted}. By design the Mgb-V$_\mathrm{esc}$ relation produced in this way tightly follows the observations. In the case of H$\beta$ the region allowed by the model (the grid of points) is well matched to the region occupied by the observations (shaded area). The observed local gradients broadly follow the lines of constant age in the modeled region, which may suggest why the local and global connection is not observed between H$\beta$ and V$_\mathrm{esc}$. In the case of Fe5015 the model predicts a tight correlation with V$_\mathrm{esc}$ but one that is somewhat steeper than the observed relation. Allowing other values of [$\alpha$/Fe] is required to match the observations; this is consistent with the SSP hyperplane having a small thickness. We again note that the steeper local gradients in Fe5015 appear to follow lines of constant age. These results suggest two possible conclusions; either the local and global correlation in Mgb is a conspiracy of the interaction between age and metallicity in producing stellar absorption indices or that there still remain weaknesses in the SSP models used to derive Equation \ref{Eq:SSP}.

We find essentially no difference between the relations for fast- and slow-rotators, which are thought to have significantly different assembly histories. There are clear differences between the two classes of galaxies in many of their properties \citep[see][for a discussion of these differences]{Paper IX} but not in the Index-V$_\mathrm{esc}$ relation. What does this tell us about the assembly of these galaxies? If fast rotators are the progenitors of slow rotators and we assume that fast rotators lie naturally upon the Index-V$_\mathrm{esc}$ relations we have observed whatever process leads to these differences between fast- and slow-rotators must preserve the links between stellar population properties and the gravitational potential. If, as \citet{Di Matteo} suggest, equal mass dry mergers between giant elliptical galaxies can significantly alter the metallicity gradients of the remnant, the lack of a difference between the relations for fast- and slow-rotators may provide a significant constraint on the modelling of the formation of these galaxies.
\section{Conclusions}
In this work we have examined the link between the local escape velocity, V$_\mathrm{esc}$ (determined from photometric observations and dynamical modelling) and the local line strength indices. We discuss the impact on our results of non-axisymmetry, dark matter, inclination and substructure within the line strength maps. Single stellar population models were used to convert our line strength measurements into representative values for the age t, metallicity [Z/H], and alpha enhancement [$\alpha$/Fe]. We then used some simple models to explore the impact of the observed correlations on the formation history of early-type galaxies. The main findings of this work are as follows:
\newcounter{Lcount2}
\begin{list}{\roman{Lcount2})}
{\usecounter{Lcount2}}
\item The line strength indices Mgb and Fe5015 are correlated with V$_\mathrm{esc}$ (both with rms scatters of 0.033) while H$\beta$ is anti-correlated with V$_\mathrm{esc}$ (with an rms scatter of 0.049). Using the models of \citet{Schiavon} the scatter in the Mgb relation corresponds to a spread $\Delta \mathrm{[Z/H]} \sim 0.2$ at a fixed age of 9 Gyrs. The scatter in the H$\beta$ relation corresponds to a spread $\Delta \mathrm{t}  \sim 4.7$ Gyrs with [Z/H] fixed at solar metallicity. The tightness of these relations provide an important check for simulations of early-type galaxy formation. (In comparison the index - $\sigma_e$ relations have rms scatters of 0.028, 0.030 and 0.046 for Mgb, Fe5015 and H$\beta$).
\item For Mgb the correlation within a galaxy (the local relation) is the same as that between the central values of different galaxies (the global relation). This is the key difference when considering V$_\mathrm{esc}$ compared to using $\sigma$. 
\item For outliers characterised by high H$\beta$ the residuals in the Mgb-V$_\mathrm{esc}$ relation correlate with H$\beta$. We use this correlation to modify our Index-V$_\mathrm{esc}$ relation such that  $\log (\mathrm{V}_{\mathrm{esc}}/500 \mathrm{kms}^{-1}) = 0.16 + 3.57\log(\mathrm{Mgb}/4\mathrm{\AA}) + 1.29\log(\mathrm{H}\beta/1.6\mathrm{\AA})$. The scatter of this corrected relation is consistent with the measurement errors.
\item We divided our sample into several sub-populations: S0s and ellipticals, field and group/cluster objects and fast- and slow-rotators. The Index-V$_\mathrm{esc}$ relations for each of these sub-populations are consistent with the relations for the entire sample. We find no dependence on these simple divisions into morphological type and environment. We also find no significant difference between fast- and slow-rotators.
\item When converting these line strength measurements to SSP parameters we find that all the galaxies are confined to a two-dimensional plane within the four-dimensional space of V$_\mathrm{esc}$, age [Z/H] and [$\alpha$/Fe]. This plane is described by the equation: $\log \mathrm({V}_{\mathrm{esc}}/500\mathrm{kms}^{-1}) = 0.85 \mathrm{[Z/H]} + 0.43 \log (\mathrm{t}/\mathrm{Gyrs})$ - 0.29. Those galaxies that were outliers in the Index-V$_\mathrm{esc}$ relations do not stand out in this SSP-hyperplane.
\item We find that in the Z-V$_\mathrm{esc}$ diagram the local gradients are significantly steeper than the global relation. When we consider the above combination of Z and age we recover the local and global relation, in that the local gradients are the same as the global one. This tight relation does not depend on the SSP model used. \end{list} 

How the connection between stellar populations and the gravitational potential, both locally and globally, is preserved as galaxies assemble hierarchically presents a major challenge to models.

\section{Acknowledgements}
We thank Anne-Marie Weijmans for useful discussion on the influence of dark matter. The SAURON project is made possible through grants 614.13.003, 781.74.203, 614.000.301 and 614.031.015 from NWO and financial contributions from the InstitŸt National des Sciences de l'Univers, the UniversitŽ Lyon I, the Universities of Durham, Leiden and Oxford, the Programme National Galaxies, the British Council, PPARC grant 'Observational Astrophysics at Oxford 2002Ð2006' and support from Christ Church Oxford, and the Netherlands Research School for Astronomy NOVA. NS is grateful for the support of an STFC studentship. MC acknowledges support from a STFC Advanced Fellowship (PP/D005574/1). RLD is grateful for the award of a PPARC Senior Fellowship (PPA/Y/S/1999/00854), postdoctoral support through PPARC grant PPA/G/S/2000/00729, STFC grant PP/E001114/1 and from the Royal Society through a Wolfson Merit Award. The PPARC Visitors grant (PPA/V/S/2002/00553) to Oxford also supported this paper. GvdV acknowledges support provided by NASA through Hubble Fellowship grant HST-HF-01202.01-A awarded by the Space Telescope Science Institute, which is operated by the Association of Universities for Research in Astronomy, Inc., for NASA, under contract NAS 5-26555. This paper is based on observations obtained at the William Herschel Telescope, operated by the Isaac Newton Group in the Spanish Observatorio del Roque de los Muchachos of the Instituto de Astrof'sica de Canarias. It is also based on observations obtained at the 1.3m Mcgraw-Hill Telescope at the MDM observatory on Kitt Peak, which is owned and operated by the University of Michigan, Dartmouth College, the Ohio State University, Columbia University and Ohio University. This project made use of the HyperLeda and NED data bases. Part of this work is based on HST data obtained from the ESO/ST-ECF Science Archive Facility.

\appendix
\section{Mass-to-light ratio correlations}

\begin{figure}
\includegraphics[width=3.25in]{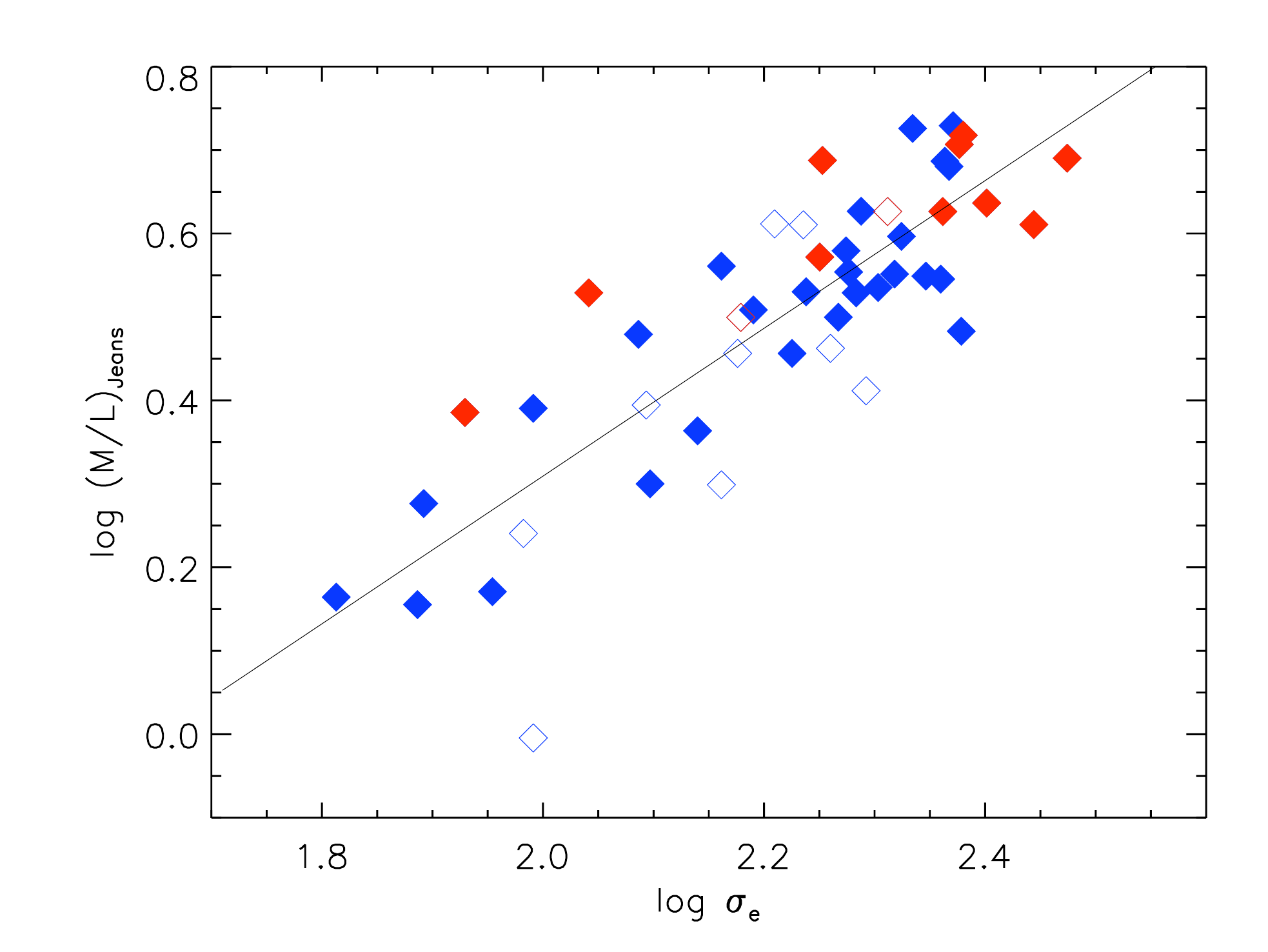}
\caption{M/L - $\sigma_e$ relation. Solid symbols are those galaxies ranked as having a good fitting MGE in Table \ref{table:sample} (rank 1), open symbols represent those galaxies having discrepancies between the MGE models and the photometry, typically due to bars (rank 2 or 3). Blue symbols indicate fast rotators and red symbols indicate slow rotators. The solid line is a fit to the data, excluding the galaxy with the lowest M/L, NGC3489. }
\label{Fig:M/L_sigma}
\end{figure}
\begin{figure}
\includegraphics[width=3.25in]{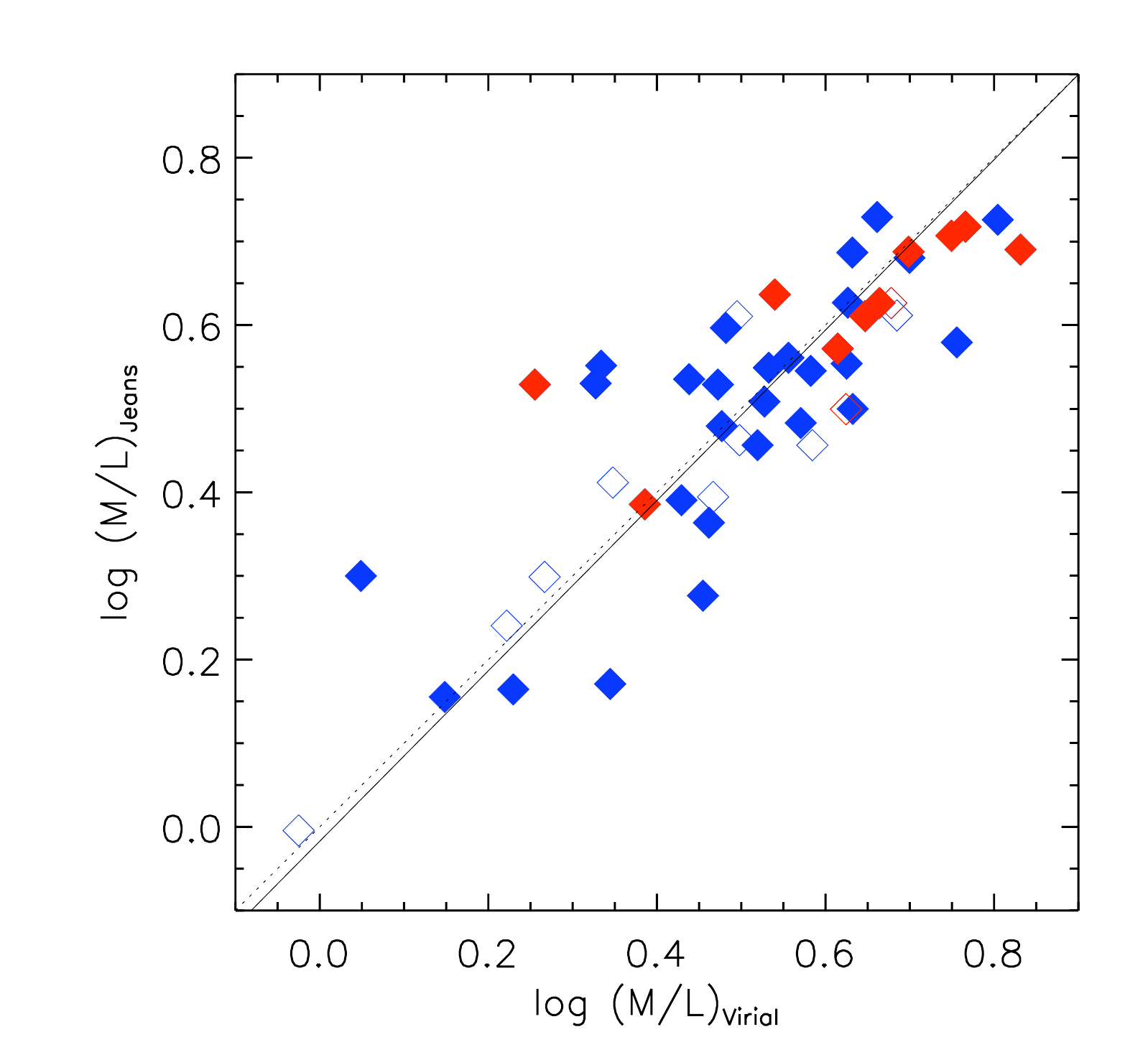}
\caption{(M/L)$_\mathrm{Jeans}$ vs (M/L)$_\mathrm{Virial}$. The solid line is a fit to the data while the dotted line shows a one-to-one correlation. Symbols as in Fig. \ref{Fig:M/L_sigma}.}
\label{Fig:M/L_Virial}
\end{figure}
\begin{figure}
\includegraphics[width=3.25in]{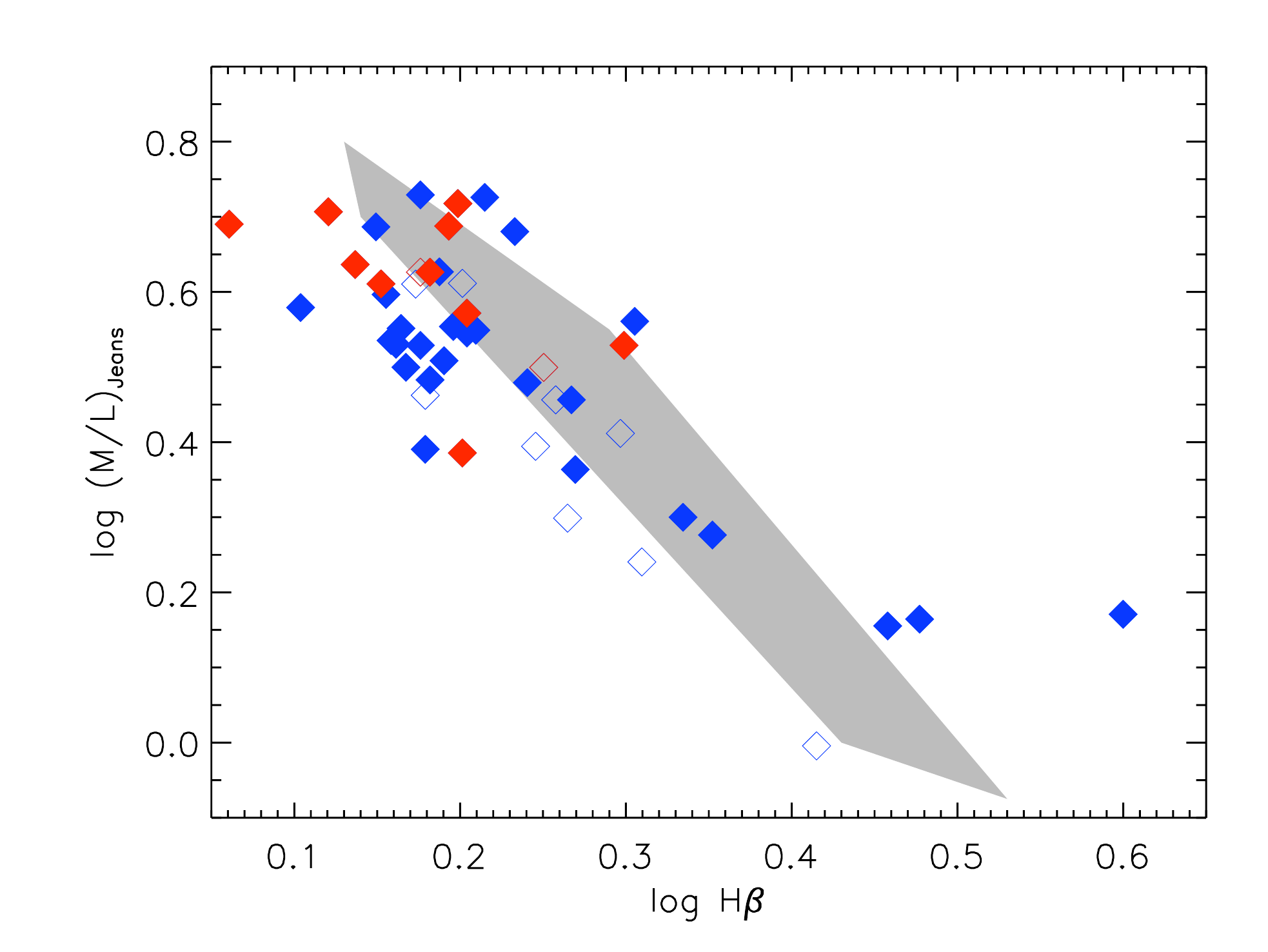}
\label{Fig:M/L_Hbeta}
\caption{Dynamical M/L versus the observed line strength index, H$\beta$. Symbols as in Fig. \ref{Fig:M/L_sigma}. The shaded region indicates the predictions of the SSP models of \citet{Vazdekisa} and \citet{Vazdekisb} using a Saltpeter IMF.}
\end{figure}
In Paper IV we examined the correlation between the dynamical mass-to-light ratio (M/L)$_\mathrm{Jeans}$, and several other properties of local early-type galaxies, including $\sigma_e$, H$\beta$ and the virial mass-to-light ratio, (M/L)$_\mathrm{Virial}$. Here we revisit those correlations with the additional 24 galaxies contributed by this work. The analysis was carried out precisely as described in Section 4.2 of Paper IV, using mass-to-light ratios derived from JAM modelling rather than Schwarzschild modelling. Where available we used surface brightness fluctuation distances, otherwise distances are redshift only and have a significantly larger error \citep{Tonry,Paturel,Mei}. All mass-to-light ratios were converted to {\it I}-band using galaxy colours obtained from \citet{Tonry} and \citet{Prugniel}. As in Paper IV we adopted an error in $\sigma_e$ of 5 percent and a 6 percent modelling error in (M/L)$_\mathrm{Jeans}$ to which we quadratically co-added the distance errors. The fit was carried out by quadratically adding an intrinsic error to make $\chi^2/\nu = 1$, where $\nu$ is the degrees of freedom. To preempt our conclusions we find no significant change from the results presented in Paper IV.

We find a tight correlation with $\sigma_e$, with an observed rms scatter of 30 percent, shown in Fig. \ref{Fig:M/L_sigma}. This implies an intrinsic scatter of 11 percent. The best fitting relation has the form:
\begin{equation}
\mathrm{(M/L)_{Jeans}}  = (3.77 \pm 0.14) \left (\frac{\sigma_e}{200\ \mathrm{km\ s}^{-1}} \right )^{0.89 \pm 0.09}
\end{equation}
We also compared our dynamical (M/L)$_\mathrm{Jeans}$ to the observed virial (M/L)$_\mathrm{Virial} = \beta R_e \sigma_e^2 / G \mathrm{L}$ (where we adopt $\beta$ = 5 as found in Paper IV), shown in Fig. \ref{Fig:M/L_Virial}. With our adopted modelling error of 6 percent in the (M/L)$_\mathrm{Jeans}$ the scatter in (M/L)$_\mathrm{Virial}$ required to make $\chi^2/\nu = 1$ is 26 percent. The best fitting relation has the form:
\begin{equation}
\mathrm{(M/L)_{Jeans}} \propto \mathrm{(M/L)}_\mathrm{Virial}^{1.02 \pm 0.10}
\end{equation}
We also present the relation with H$\beta$ (see Fig. \ref{Fig:M/L_Hbeta}). This is closely related to the population mass-to-light ratio (M/L)$_\mathrm{Pop}$, which is largely driven by variations in H$\beta$. We again reproduce the trend found in Paper IV. We also indicate the predictions of the SSP models of \citet{Vazdekisa} and \citet{Vazdekisb} as shown in fig. 16 of Paper IV.

\section{PSF parameters used in the model MGEs}
Table \ref{Tab:PSFs} contains the details of the point spread functions used when evaluating our model MGEs. These PSFs were calculated by fitting circular Gaussians to PSFs produced using the TinyTim software. The PSFs have the form: \newline $\mathrm{PSF} = \Sigma^n_{k=1} G_k \exp [ -R^2 / 2\sigma_k^2] / 2\pi \sigma_k^2$.
\begin{table}
\caption{Parameters of the model MGE PSFs}
\label{Tab:PSFs}
\begin{center}
\begin{tabular}{l c c c}
\hline
Instrument/Filter & k & $G_k$ & $\sigma_k$\\
& & & (arcsec)\\
\hline
&1& 0.226 & 0.02\\
&2& 0.573 & 0.05\\
WFPC2/F606W &3& 0.092 & 0.14\\
&4& 0.071 & 0.33\\
&5& 0.038 & 0.88\\
\hline
&1& 0.254 & 0.02\\
&2& 0.560 & 0.06\\
WFPC2/F702W &3& 0.083 & 0.16\\
&4& 0.070 & 0.38\\
&5& 0.033 & 1.05\\
\hline
&1& 0.280 & 0.02\\
&2& 0.546 & 0.06\\
WFPC2/F814W &3& 0.073 & 0.18\\
&4& 0.071 & 0.45\\
&5& 0.030 & 1.39\\
\hline
&1& 0.061 & 0.03\\
&2& 0.089 & 0.09\\
WFPC1/F555W &3& 0.145 & 0.19\\
&4& 0.637 & 1.00\\
&5& 0.066 & 1.91\\
\hline
&1& 0.445 & 0.05\\
&2& 0.303 & 0.15\\
ACS/F475W &3& 0.101 & 0.37\\
&4& 0.081 & 1.04\\
&5& 0.061 & 3.38\\
\hline
\end{tabular}
\end{center}
Notes: The model PSFs are formed from the sum of circular Gaussians fitted to TinyTim PSFs and have the form: $\mathrm{PSF} = \Sigma^n_{k=1} G_k \exp [ -R^2 / 2\sigma_k^2] / 2\pi \sigma_k^2$. The numerical weights are normalised such that $\Sigma^n_{k=1} = 1$
\end{table}
\section{Further examples of velocity fields calculated from \textsc{JAM} modelling}
\begin{figure*} 
\centering  
\includegraphics[width=1.5in,height=4in]{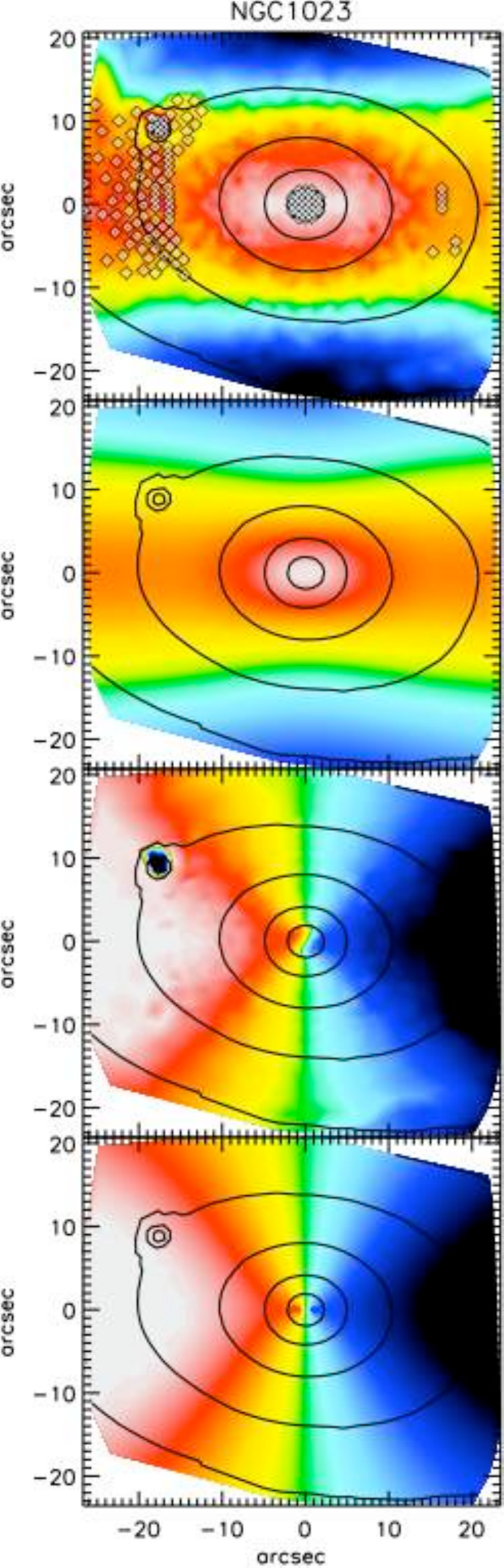}
\includegraphics[width=1.5in,height=4in]{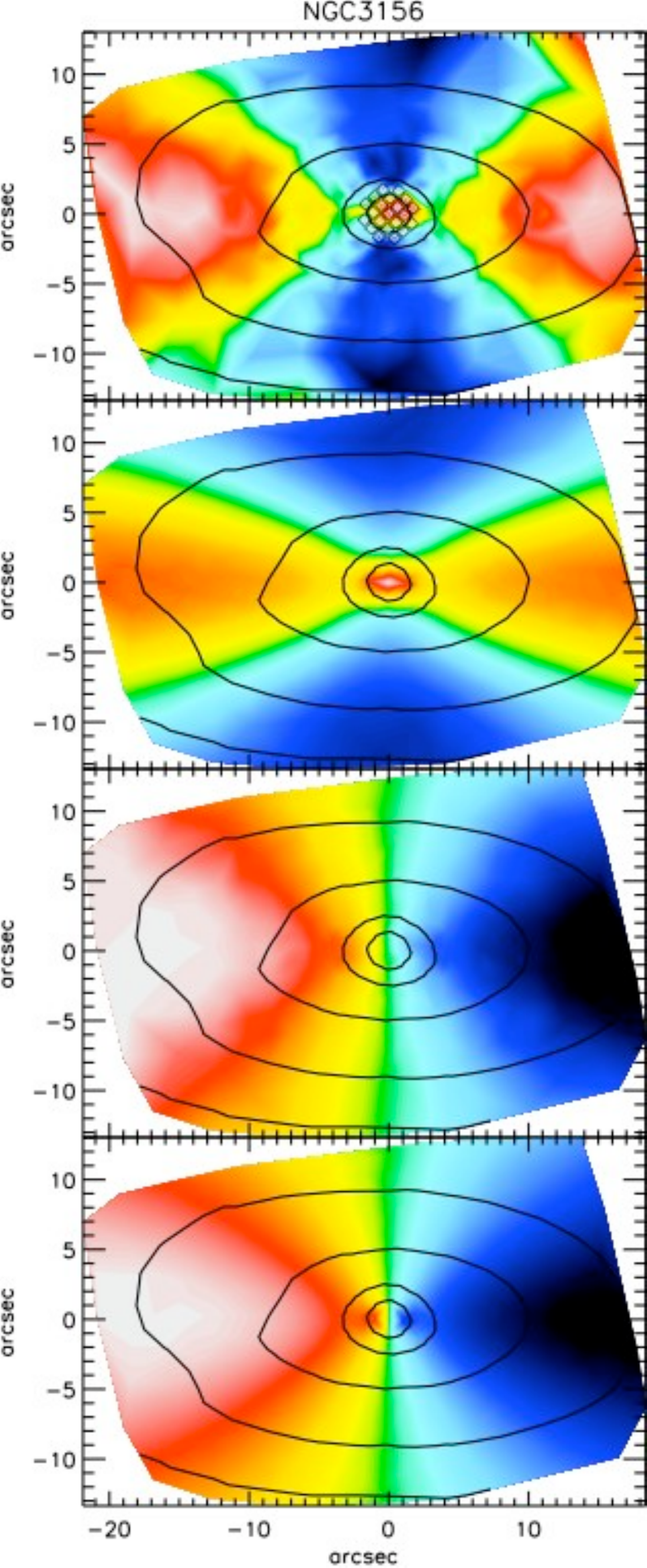}  
\includegraphics[width=1.5in,height=4in]{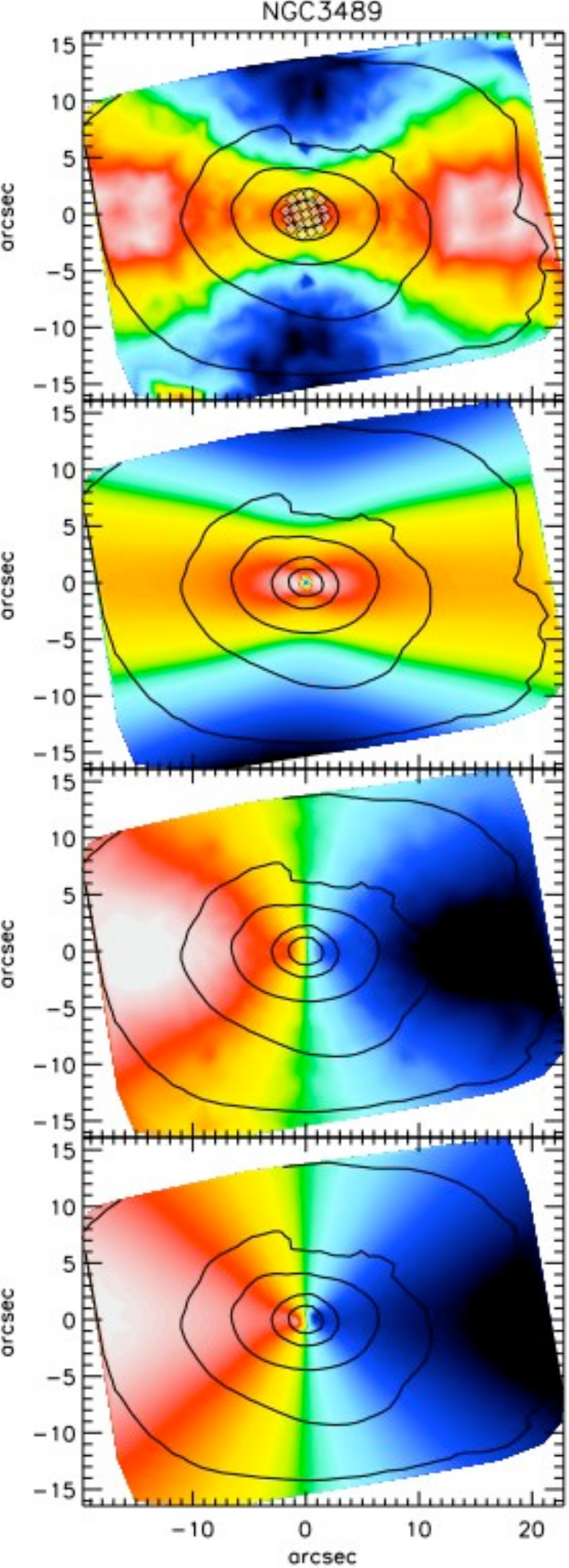}  
\includegraphics[width=1.5in,height=4in]{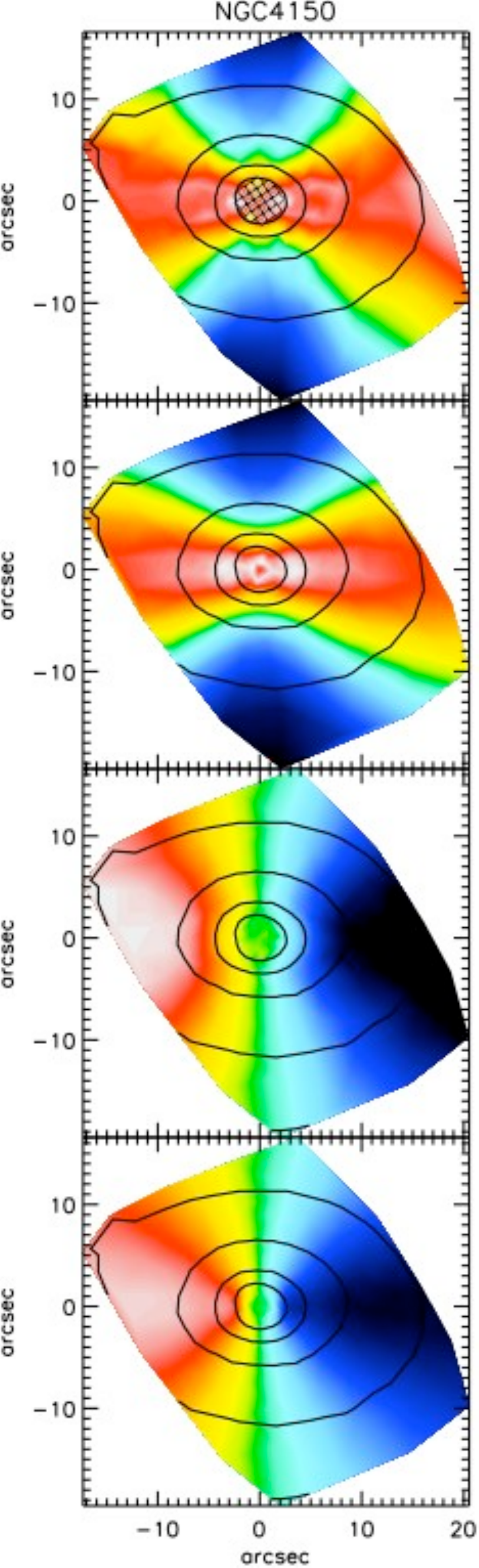}  
\includegraphics[width=1.5in,height=4in]{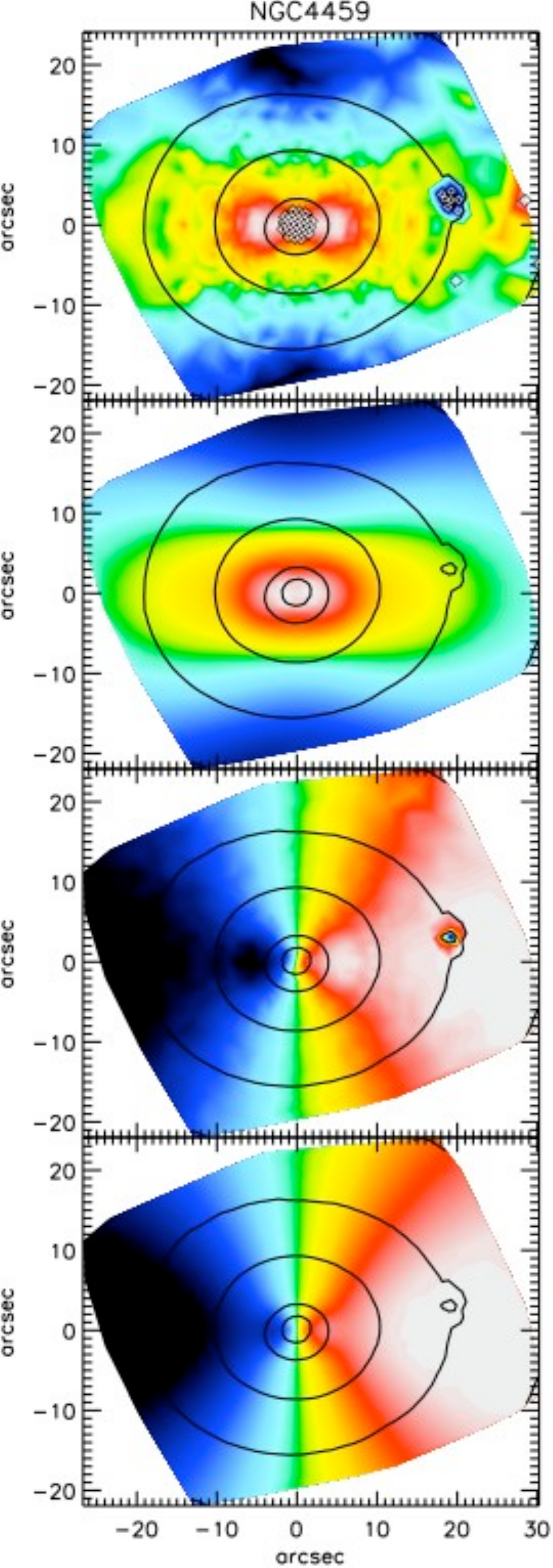}  
\includegraphics[width=1.5in,height=4in]{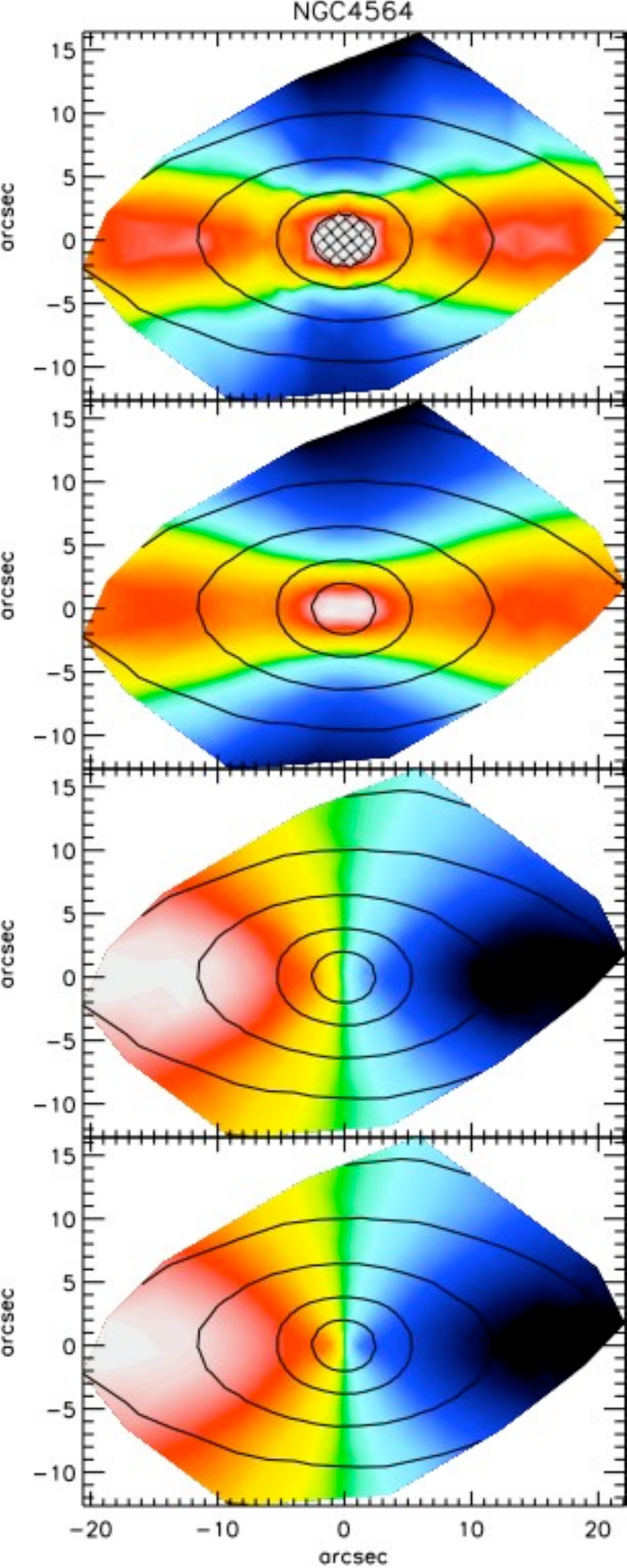}
\includegraphics[width=1.5in,height=4in]{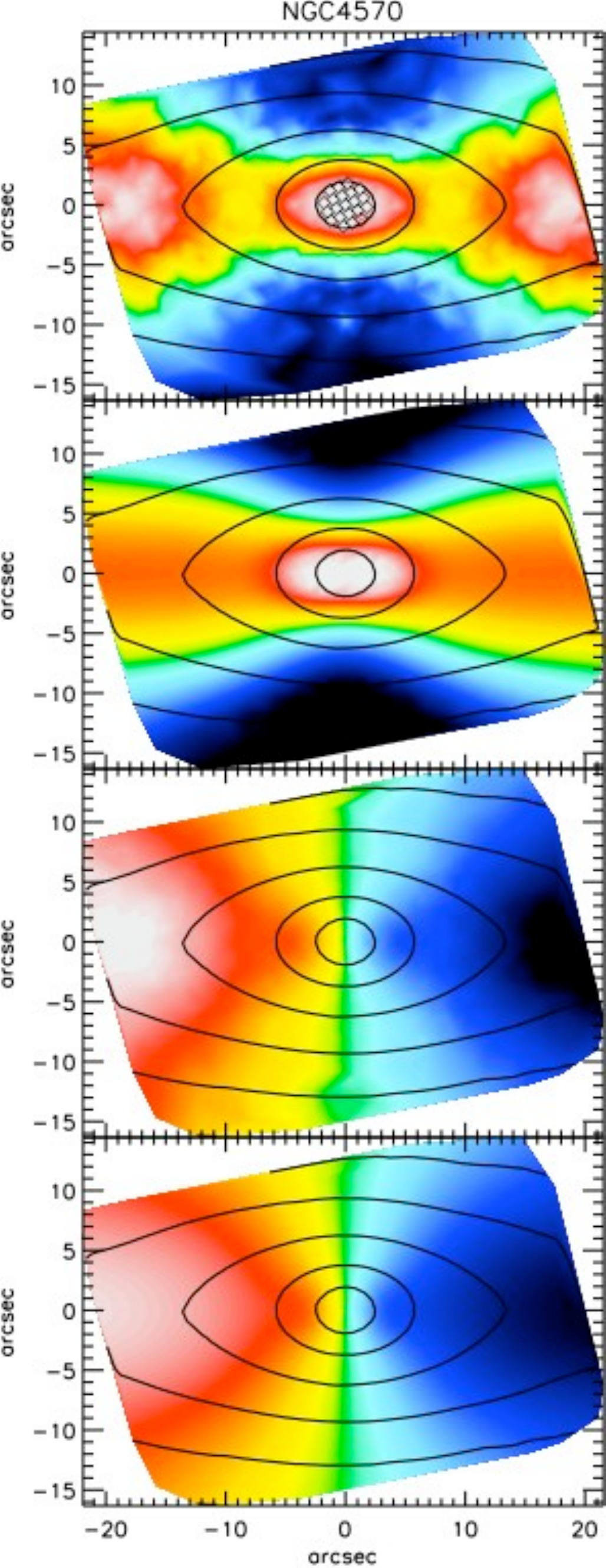}
\includegraphics[width=1.5in,height=4in]{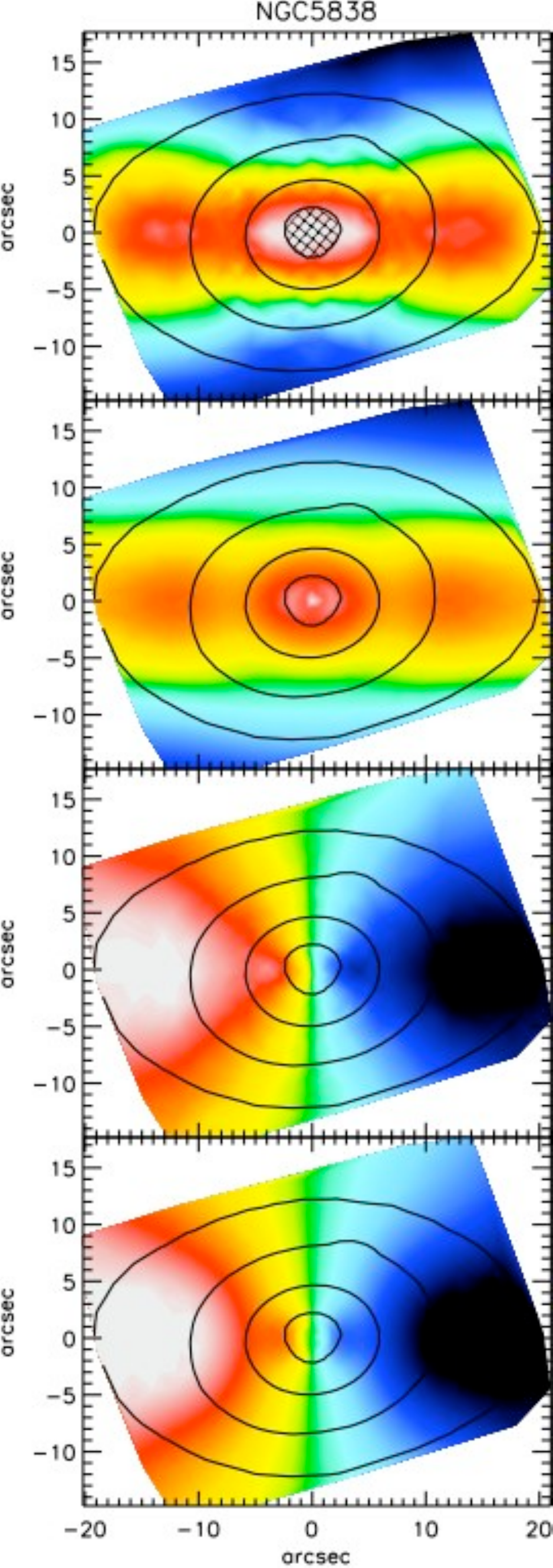}  
\label{Fig:JAM_examples}
\caption{Examples of the first and second moments of the velocity fields, v ($\mu_1$) and $\sqrt{v^2 + \sigma^2}$ ($\mu_2$) derived from our \textsc{JAM} models alongside the bi-symmetrised measured $\mu_1$ and $\mu_2$ from the SAURON kinematics. From top to bottom for each galaxy the figures show: observed $\mu_2$, model $\mu_2$, observed $\mu_1$, model $\mu_1$. We show here a selection of galaxies from our sample to illustrate the range in the quality of fits we achieved.}
\end{figure*}

In Fig. \ref{Fig:JAM_examples} we show the results of \textsc{JAM} modelling for eight galaxies from our sample.  Further examples are shown in \citet{newJeans}. Details of the \textsc{JAM} modelling can be found in Section \ref{Sec:Jeans}. We show the observed and modeled results for both the first and second moments of the velocity for each galaxy. As can be seen in all cases we achieve good agreement between the observations and our models. 
\section{Multi-Gaussian Expansion parameters for 24 galaxies in our sample}
In Appendix B of Paper IV we presented the (distant-independent) MGE parameters for 23 galaxies from the SAURON sample\footnote{The MGE parameters for NGC 2974 were given in \citet{Davor}. Those for NGC 3032 were given in \citet{Young}.}. In Table \ref{Tab:MGEs} and \ref{Tab:MGEs2} we present the MGE parameters for the remaining 23 galaxies. The constant-PA models were obtained by fitting the HST photometry (where available) at small radii and the ground-based MDM photometry at large radii. They provide an accurate description of the surface brightness of the galaxies from $R \approx 0.01$ arcsec to about twice the maximum $\sigma_j$ used in each galaxy (usually corresponding to $5-10 R_e$). Dust and bright foreground stars were excluded from the fit. The matching of the different images is described in Section \ref{Sec:MGEs}. The deconvolved surface brightness $\Sigma$ is defined as follows:
\begin{equation}
\Sigma (x',y') =  \sum_{j=0}^N \frac{L^\prime_j}{2\pi\sigma^{\prime2}_jq^\prime_j} \exp \left \{ - \frac{1}{2\sigma^{\prime2}_j} \left ( x^{\prime2}_j + \frac{y^{\prime2}_j}{q^{\prime2}_j}\right ) \right \}
\end{equation}
where the model is composed of $N$ Gaussian components of dispersion $\sigma_j$, axial ratio q$_j$ and peak intensity $I_j$. The coordinates ($x',y'$) are measured on the sky plane, with the $x'$-axis corresponding to the galaxy major axis. The total luminosity of each Gaussian component is given by $L_j = 2\pi I_j \sigma_j^2 q_j$. See \citet{MGE II} for details. The obscured areas in each figure were masked in the fitting process.
\begin{table*}
\caption{MGE parameters for the deconvolved {\it I}-band surface brightness}
\label{Tab:MGEs}
\begin{center}
\begin{tabular}{c c c c c c c c c c c c c}
\hline
j & $\log I_j$ & $\log \sigma_j$ & q$_j$ & $\log I_j$ & $\log \sigma_j$ & q$_j$ & $\log I_j$ & $\log \sigma_j$ & q$_j$& $\log I_j$ & $\log \sigma_j$ & q$_j$\\
& (L$_{\sun}$ pc$^{-2}$) & (arcsec) &  & (L$_{\sun}$ pc$^{-2}$) & (arcsec) &  & (L$_{\sun}$ pc$^{-2}$) & (arcsec) & & (L$_{\sun}$ pc$^{-2}$) & (arcsec) &\\
\hline
& \multicolumn{3}{c}{NGC 474} & \multicolumn{3}{c}{NGC 1023} & \multicolumn{3}{c}{NGC 2549}& \multicolumn{3}{c}{NGC 2685}\\\
&&&&&&&&&&&&\\
      1 &       5.224 &      -1.762 &      0.922 &       6.084 &      -1.645 &      0.800 &       5.775 &      -1.593 &      0.753 &       5.278 &      -1.363 &      0.500 \\
       2 &       4.697 &      -1.097 &      0.989 &       5.214 &      -1.012 &      0.747 &       4.760 &     -0.877 &      0.646 &       4.657 &     -0.845 &      0.500 \\
       3 &       4.089 &     -0.652 &      0.978 &       4.837 &     -0.637 &      0.613 &       4.260 &     -0.363 &      0.811 &       4.261 &     -0.335 &      0.392 \\
       4 &       4.005 &     -0.192 &      0.965 &       4.288 &     -0.383 &      0.800 &       4.078 &     0.062 &      0.732 &       4.110 &      0.145 &      0.345 \\
       5 &       3.595 &     0.043 &      0.992 &       4.366 &     -0.152 &      0.474 &       3.512 &      0.538 &      0.554 &       3.689 &      0.450 &      0.422 \\
       6 &       3.391 &      0.283 &      0.922 &       4.328 &    -0.015 &      0.800 &       2.981 &      0.778 &      0.742 &       2.949 &      0.684 &      0.600 \\
       7 &       3.028 &      0.435 &      0.997 &       4.180 &      0.261 &      0.800 &       3.239 &      0.801 &      0.233 &       3.101 &      0.903 &      0.293 \\
       8 &       3.029 &      0.621 &      0.900 &       3.869 &      0.580 &      0.800 &          5.666* &       1.248 &      0.434 &       2.609 &       1.033 &      0.600 \\
       9 &       2.680 &      0.857 &      0.900 &       3.539 &      0.896 &      0.757 &       5.666 &       1.248 &      0.434 &       2.653 &       1.308 &      0.261 \\
       10 &       2.161 &       1.322 &      0.931 &       3.091 &       1.320 &      0.634 &       3.122 &       1.595 &      0.421 &       1.950 &       1.340 &      0.600 \\
      11 &       1.566 &       1.661 &       1.00 &       1.637 &       1.655 &      0.800 &          3.186* &       1.602 &      0.430 &       1.902 &       1.599 &      0.600\\
      12 & - & - & - &       2.441 &       1.827 &      0.300 &       2.549 &       1.702 &      0.409 &      0.985 &       1.914 &      0.600 \\
      13 & - & - & - &       2.061 &       2.033 &      0.3828 &          2.865* &       1.911 &      0.403 & - & - & - \\
      14 & - & - & - &      0.648 &       2.255 &      0.800 &       2.829 &       1.920 &      0.404 & - & - & - \\
&&&&&&&&&&&&\\
      & \multicolumn{3}{c}{NGC 2695} & \multicolumn{3}{c}{NGC 2699} & \multicolumn{3}{c}{NGC 2768}& \multicolumn{3}{c}{NGC 3384}\\\
&&&&&&&&&&&&\\
       1 &       3.870 &   -0.009 &      0.710 &       5.388 &      -1.762 &      0.810 &       4.801 &      -1.337 &      0.750 &       5.093 &      -1.762 &      0.700 \\
       2 &       3.575 &      0.293 &      0.710 &       4.549 &      -1.241 &      0.900 &       4.532 &     -0.721 &      0.482 &       5.203 &      -1.222 &      0.700 \\
       3 &       3.103 &      0.586 &      0.710 &       4.072 &     -0.484 &      0.900 &       4.370 &     -0.544 &      0.750 &       5.064 &     -0.884 &      0.700 \\
       4 &       2.673 &      0.948 &      0.701 &       3.541 &     -0.323 &      0.866 &       4.084 &     -0.178 &      0.750 &       4.640 &     -0.486 &      0.665 \\
       5 &       1.940 &       1.375 &      0.710 &       3.775 &    -0.067 &      0.884 &       3.654 &      0.182 &      0.750 &       4.071 &     -0.170 &      0.700 \\
       6 &      0.921 &       1.720 &      0.710 &       3.347 &      0.338 &      0.700 &       3.458 &      0.512 &      0.674 &       4.094 &      0.261 &      0.700 \\
       7 & - & - & - &       2.893 &      0.573 &      0.798 &       3.204 &      0.862 &      0.689 &       4.113 &      0.273 &      0.436 \\
       8 & - & - & - &       2.704 &      0.815 &      0.900 &       2.792 &       1.225 &      0.504 &       3.909 &      0.513 &      0.700 \\
       9 & - & - & - &       1.603 &       1.152 &      0.900 &       2.545 &       1.449 &      0.553 &       3.657 &      0.827 &      0.700 \\
       10 & - & - & - &       1.384 &       1.241 &      0.700 &       2.021 &       1.797 &      0.305 &       3.019 &       1.162 &      0.700 \\
      11 & - & - & - &       1.324 &       1.357 &      0.900 &       1.923 &       1.910 &      0.523 &       1.927 &       1.775 &      0.696 \\
      12 & - & - & - &      0.657 &       1.878 &      0.724 &       1.182 &       2.139 &      0.544 &       2.300 &       1.781 &      0.415 \\
      13 & - & - & - & - & - & - & - & - & - &       1.04279 &       2.127 &      0.537 \\
 &&&&&&&&&&&&\\
      & \multicolumn{3}{c}{NGC 3489} & \multicolumn{3}{c}{NGC 4262} & \multicolumn{3}{c}{NGC 4270}& \multicolumn{3}{c}{NGC 4382}\\\
&&&&&&&&&&&&\\
       1 &       6.070 &      -1.624 &      0.500 &       5.410 &      -1.422 &      0.910 &       5.049 &      -1.589 &      0.692 &       4.148 &     -0.728 &      0.770 \\
       2 &       5.466 &     -0.950 &      0.500 &       4.459 &     -0.903 &      0.910 &       3.961 &     -0.721 &      0.700 &       4.195 &     -0.344 &      0.780 \\
       3 &       4.790 &     -0.573 &      0.625 &       4.119 &     -0.570 &      0.910 &       3.630 &     -0.301 &      0.606 &       4.337 &  -0.001 &      0.780 \\
       4 &       4.726 &     -0.181 &      0.500 &       4.006 &     -0.235 &      0.910 &       3.551 &    -0.025 &      0.682 &       3.994 &      0.383 &      0.780 \\
       5 &       4.363 &      0.109 &      0.700 &       3.780 &      0.251 &      0.906 &       3.145 &      0.250 &      0.700 &       3.591 &      0.707 &      0.770 \\
       6 &       3.831 &      0.397 &      0.700 &       3.312 &      0.518 &      0.910 &       2.839 &      0.550 &      0.676 &       3.194 &      0.962 &      0.780 \\
       7 &       3.702 &      0.722 &      0.700 &       2.266 &      0.747 &      0.910 &       2.570 &      0.877 &      0.500 &       2.802 &       1.252 &      0.780 \\
       8 &       3.010 &       1.103 &      0.700 &       2.259 &       1.024 &      0.910 &       2.412 &       1.151 &      0.500 &       2.303 &       1.460 &      0.770 \\
       9 &       2.623 &       1.515 &      0.505 &       1.858 &       1.358 &      0.910 &       1.693 &       1.444 &      0.500 &       2.551 &       1.752 &      0.780 \\
       10 &       1.445 &       1.876 &      0.694 &      0.439 &       1.743 &      0.910 &      0.636 &       1.750 &      0.700 &       1.818 &       2.154 &      0.780 \\
  \hline
  \end{tabular}
\end{center}
Note: * indicates where a negative Gaussian was used to achieve a satisfactory fit. This only occurs in the most discy objects.
\end{table*}
\begin{table*}
\caption{MGE parameters for the deconvolved {\it I}-band surface brightness}
\label{Tab:MGEs2}
\begin{center}
\begin{tabular}{c c c c c c c c c c c c c}
\hline
j & $\log I_j$ & $\log \sigma_j$ & q$_j$ & $\log I_j$ & $\log \sigma_j$ & q$_j$ & $\log I_j$ & $\log \sigma_j$ & q$_j$& $\log I_j$ & $\log \sigma_j$ & q$_j$\\
& (L$_{\sun}$ pc$^{-2}$) & (arcsec) &  & (L$_{\sun}$ pc$^{-2}$) & (arcsec) &  & (L$_{\sun}$ pc$^{-2}$) & (arcsec) && (L$_{\sun}$ pc$^{-2}$) & (arcsec) &\\
\hline
       & \multicolumn{3}{c}{NGC 4387} & \multicolumn{3}{c}{NGC 4477} & \multicolumn{3}{c}{NGC 4546}& \multicolumn{3}{c}{NGC 4564}\\\
&&&&&&&&&&&&\\
1 &       4.793 &      -1.422 &      0.750 &       4.785 &      -1.337 &      0.910 &       6.012 &      -1.707 &      0.770 &       5.304 &      -1.482 &      0.800 \\
       2 &       4.120 &     -0.959 &      0.804 &       4.203 &     -0.918 &      0.910 &       5.028 &      -1.063 &      0.773 &       4.876 &     -0.993 &      0.800 \\
       3 &       3.274 &     -0.679 &      0.750 &       4.126 &     -0.377 &      0.900 &       4.639 &     -0.682 &      0.732 &       4.521 &     -0.586 &      0.800 \\
       4 &       3.311 &     -0.492 &      0.740 &       3.441 &      0.141 &      0.910 &       4.251 &     -0.397 &      0.753 &       4.224 &     -0.242 &      0.800 \\
       5 &       3.201 &     -0.261 &      0.718 &       3.453 &      0.431 &      0.910 &       4.107 &     -0.160 &      0.592 &       3.873 &     0.097 &      0.712 \\
       6 &       3.166 &    -0.037 &      0.850 &       3.154 &      0.779 &      0.910 &       3.940 &     0.050 &      0.780 &       3.521 &      0.348 &      0.741 \\
       7 &       2.980 &      0.279 &      0.773 &       2.567 &       1.081 &      0.910 &       3.750 &      0.304 &      0.812 &       3.270 &      0.638 &      0.701 \\
       8 &       2.860 &      0.591 &      0.647 &       2.184 &       1.604 &      0.910 &       3.465 &      0.582 &      0.918 &       2.270 &      0.929 &      0.800 \\
       9 &       2.573 &      0.848 &      0.592 &      0.421 &       1.892 &      0.910 &       2.917 &       1.031 &      0.473 &       2.852 &       1.063 &      0.350 \\
       10 &       2.229 &       1.128 &      0.546 & - & - & - &       2.723 &       1.345 &      0.400 &       2.347 &       1.333 &      0.350 \\
      11 &       1.558 &       1.388 &      0.638 & - & - & - &       2.003 &       1.602 &      0.451 &       2.111 &       1.461 &      0.413 \\
      12 &      0.732 &       1.634 &      0.756 & - & - & - &       1.378 &       1.859 &      0.641 &       1.711 &       1.630 &      0.489 \\
      13 & - & - & - & - & - & - & - & - & - &       1.150 &       1.840 &      0.712 \\
&&&&&&&&&&&&\\
& \multicolumn{3}{c}{NGC 4570} & \multicolumn{3}{c}{NGC 5198} & \multicolumn{3}{c}{NGC 5308}& \multicolumn{3}{c}{NGC 5831}\\\
&&&&&&&&&&&&\\
 1 &       6.101 &      -1.762 &      0.550 &       4.387 &      -1.008 &      0.867 &       5.922 &     -0.931 &      0.405 &       4.843 &      -1.585 &      0.700 \\
       2 &       5.009 &     -0.944 &      0.580 &       4.196 &     -0.634 &      0.900 &          5.955* &     -0.917 &      0.411 &       4.518 &      -1.044 &      0.813 \\
       3 &       4.646 &     -0.585 &      0.537 &       3.860 &     -0.270 &      0.900 &       4.875 &     -0.654 &      0.605 &       4.314 &     -0.708 &      0.869 \\
       4 &       4.392 &     -0.327 &      0.752 &       3.525 &     0.050 &      0.900 &       4.337 &     -0.298 &     0.0869 &       3.714 &     -0.393 &      0.700 \\
       5 &       4.207 &     0.015 &      0.804 &       3.013 &      0.406 &      0.875 &       4.250 &     -0.218 &      0.649 &       3.779 &     -0.295 &      0.900 \\
       6 &       3.866 &      0.405 &      0.644 &       2.676 &      0.729 &      0.850 &       3.889 &      0.166 &     0.095 &       3.649 &    -0.020 &      0.700 \\
       7 &       3.522 &      0.638 &      0.665 &       2.494 &       1.051&      0.850 &       3.835 &      0.198 &      0.590 &       3.480 &      0.238 &      0.700 \\
       8 &       3.138 &       1.042 &      0.471 &       1.861 &       1.359 &      0.900 &       3.325 &      0.385 &      0.194 &       3.202 &      0.474 &      0.700 \\
       9 &       2.723 &       1.397 &      0.150 &       1.049 &       1.743 &      0.900 &       3.520 &      0.595 &      0.547 &       2.877 &      0.801 &      0.700 \\
       10 &       2.544 &       1.588 &      0.236 & - & - & - &          3.201* &      0.854 &      0.331 &       2.506 &       1.048 &      0.900 \\
      11 &       1.684 &       1.746 &      0.326 & - & - & - &       3.235 &      0.906 &      0.509 &       1.969 &       1.367 &      0.900 \\
      12 &      0.847 &       1.954 &      0.768 & - & - & - &       2.953 &       1.338 &      0.118 &       1.470 &       1.630 &      0.900 \\
      13 & - & - & - & - & - & - &       2.516 &       1.512 &      0.193 &      0.804 &       1.952 &      0.900 \\
      14 & - & - & - & - & - & - &       1.494 &       1.740 &      0.241 & - & - & - \\
      15 & - & - & - & - & - & - &      0.786 &       1.805 &      0.700 & - & - & - \\
&&&&&&&&&&&&\\
      & \multicolumn{3}{c}{NGC 5838} & \multicolumn{3}{c}{NGC 5982} & \multicolumn{3}{c}{NGC 7332}&\\\
&&&&&&&&&&&&\\
 1 &       5.583 &      -1.637 &      0.950 &       4.019 &     -0.719 &      0.850 &       6.192 &      -1.721 &      0.281 &        &      &     \\
       2 &       4.752 &      -1.091 &      0.950 &       4.192 &     -0.405 &      0.850 &       4.815 &      -1.054 &      0.589 &       &      &     \\
       3 &       5.075 &     -0.779 &      0.530 &       4.138 &    -0.080 &      0.850 &       4.580 &     -0.709 &      0.351 &      &    &       \\
       4 &       4.371 &     -0.205 &      0.950 &       3.647 &      0.319 &      0.650 &       4.337 &     -0.591 &      0.700 &        &     &      \\
       5 &       4.041 &      0.264 &      0.780 &       3.444 &      0.577 &      0.708 &       4.133 &     -0.263 &      0.700 &        &    &      \\
       6 &       3.556 &      0.625 &      0.890 &       2.837 &      0.806 &      0.706 &       3.798 &     0.041 &      0.602 &     &     &    \\
       7 &       2.954 &      0.989 &      0.466 &       2.728 &       1.039 &      0.650 &       3.551 &      0.254 &      0.700 &        &    &     \\
       8 &       2.886 &       1.087 &      0.835 &       2.346 &       1.345 &      0.650 &       3.245 &      0.479 &      0.246 &     &      &      \\
       9 &       1.091 &       1.450 &      0.793 &       1.396 &       1.685 &      0.709 &       3.125 &      0.684 &      0.665 &        &   &     \\
       10 &       2.286 &       1.684 &      0.350 &       1.282 &       1.843 &      0.850 &       2.606 &       1.144 &      0.516 &       &        &     \\
      11 &      0.998 &       2.018 &      0.350 & - & - & - &       2.395 &       1.512 &      0.152 &       &      &      \\
      12 & - & - & - & - & - & - &       1.197 &       1.516 &      0.700 &        &      &      \\
      13 & - & - & - & - & - & - &       1.845 &       1.656 &      0.232 &      &      &       \\
      14 & - & - & - & - & - & - &       1.026 &       1.821 &      0.274 &  &  &  \\
      15 & - & - & - & - & - & - &      0.160 &       1.906 &      0.700 &  &  &  \\
\hline
\end{tabular}
\end{center}
Note: * indicates where a negative Gaussian was used to achieve a satisfactory fit. This only occurs in the most discy objects.
\end{table*}

\label{lastpage}

\end{document}